\documentclass{article}

\usepackage{arxiv}

\usepackage[utf8]{inputenc} % allow utf-8 input
\usepackage[T1]{fontenc}    % use 8-bit T1 fonts
\usepackage{hyperref}       % hyperlinks
\usepackage{url}            % simple URL typesetting
\usepackage{booktabs}       % professional-quality tables
\usepackage{amsfonts}       % blackboard math symbols
\usepackage{nicefrac}       % compact symbols for 1/2, etc.
\usepackage{microtype}      % microtypography
\usepackage{lipsum}
\usepackage{graphicx}
\graphicspath{ {./images/} }
\usepackage{amsfonts}
\usepackage{amsmath}
\usepackage{latexsym}
\usepackage{listings}
\usepackage{lscape}
\usepackage{graphicx}
\usepackage{setspace}
\usepackage{color}
\usepackage{multirow}
\usepackage{rotating}
\usepackage{epsfig}
\usepackage{amssymb, amsmath}
\usepackage{subfig}
\def\wt{\widetilde}

\def\Qbp1{Q\left[(\wt X, \wt \Sigma )\rightarrow (\wt X \cup \{x\},\wt \Sigma \cup \{\Sigma\})\right]}
\def\Qdp1{Q\left[(\wt X \cup \{x\},\wt \Sigma \cup \{\Sigma\})\rightarrow (\wt X, \wt \Sigma )\right]}
\def\Qbm1{Q\left[(\wt X\backslash\{x_i\}, \wt \Sigma\backslash\{\Sigma_i\} )\rightarrow (\wt X,\wt \Sigma)\right]}
\def\Qdm1{Q\left[(\wt X,\wt \Sigma)\rightarrow (\wt X\backslash\{x_i\}, \wt \Sigma\backslash\{\Sigma_i\} )\right]}

\def\bX{\mathbf X}
\def\bH{\mathbf H}

\def\bC{\mathbf{C}}

\def\bV{\boldsymbol V}

\def\bB{\boldsymbol B}

\def\b0{\mathbf{0}}

\def\argmax{\operatornamewithlimits{argmax}}

\def\bB{\mathbf{B}}

\def\bV{\mathbf{V}}

\def\bX{\mathbf{X}}

\title{Minorization-Maximization-based Steepest Ascent for Large-scale Survival Analysis with Time-Varying Effects: Application to the National
 Kidney Transplant Dataset}

\author{
 Kevin He \footnote{Corresponding Author} \\
Department of Biostatistics\\ University of Michigan\\ Ann
Arbor, MI 48109\\
  \texttt{kevinhe@umich.edu} \\
   * Corresponding Author\\
   \And
Ji Zhu \\
Department of Statistics\\ University of Michigan\\ Ann
Arbor, MI 48109\\
  \texttt{jizhu@umich.edu} \\
  \And
Jian Kang \\
Department of Biostatistics\\ University of Michigan\\ Ann
Arbor, MI 48109\\
  \texttt{jiankang@umich.edu} \\
    \And
Yi Li \\
Department of Biostatistics\\ University of Michigan\\ Ann
Arbor, MI 48109\\
  \texttt{yili@umich.edu} \\
}

\begin{document}
\maketitle
\begin{abstract}
The time-varying effects model is a flexible and powerful tool for modeling the dynamic changes of
covariate effects. However, in survival analysis, 
its computational
burden increases quickly as the number of sample sizes or predictors grows.  Traditional
 methods that perform well for moderate sample sizes and low-dimensional data do not scale to massive
data. Analysis of national kidney transplant data with a massive sample size and large number of predictors defy any existing statistical methods and software.
 In view of these difficulties, we propose a Minorization-Maximization-based steepest ascent procedure for estimating the time-varying effects. Leveraging the block structure formed by the basis expansions, the proposed procedure iteratively updates the optimal block-wise direction along which the approximate increase in the log-partial likelihood is maximized. The resulting estimates ensure the ascent
property and serve as refinements of the previous step. The performance of the proposed method is examined by simulations and  applications to the analysis of national kidney transplant data.

\end{abstract}

% keywords can be removed

\section{Introduction}

End-stage renal disease (ESRD) is one of the most deadly and costly
diseases in the United States. Kidney transplantation is the preferred treatment for ESRD. However, despite aggressive efforts to increase the number of kidney donors, the demand far exceeds the supply, with fewer than $15\%$ of eligible patients likely to receive a transplant  \cite{r33}. 
To  optimize  treatment  strategies  for  ESRD  patients,  an  important  aspect  is   to  understand  why  the  outcome  is  worse  for  certain  patients.   Thus,  there  is  urgent  need  to   accurately identify risk factors associated with post-transplant mortality.

For this purpose, the proportional hazards model \cite{r5} has been widely employed.  However,
the Cox model assumes the covariate effects are constant over time,  which is often violated due to the complex relationships between baseline conditions and post-treatment outcomes.
One example is obesity, generally viewed as a risk factor for mortality; however, previous studies \cite{r6} showed improved survival in obese kidney dialysis patients, which has been labeled as reverse epidemiology. One possible explanation is that obesity has a protective effect in the short run (Figure 1a), but becomes a risk factor after long-term exposure. Another example is core muscle size. Englesbe et al. \cite{r7} found that  large core muscle size has a strong protective effect in the short term after surgery, with a weakening  association over time (Figure 1b). In contrast, the constant estimate provided by the Cox model is close to zero. Thus, accounting for time-varying effects provides
valuable clinical information that could be obscured otherwise.

\begin{figure}[h]
 \caption{Example time-varying effects in clinical studies. The solid lines are B-spline based estimates. The constant estimates are provided by the Cox model.}
\centering
\subfloat[Obesity for Dialysis Outcomes]{
\includegraphics[width=2.5in, height=2.5in]{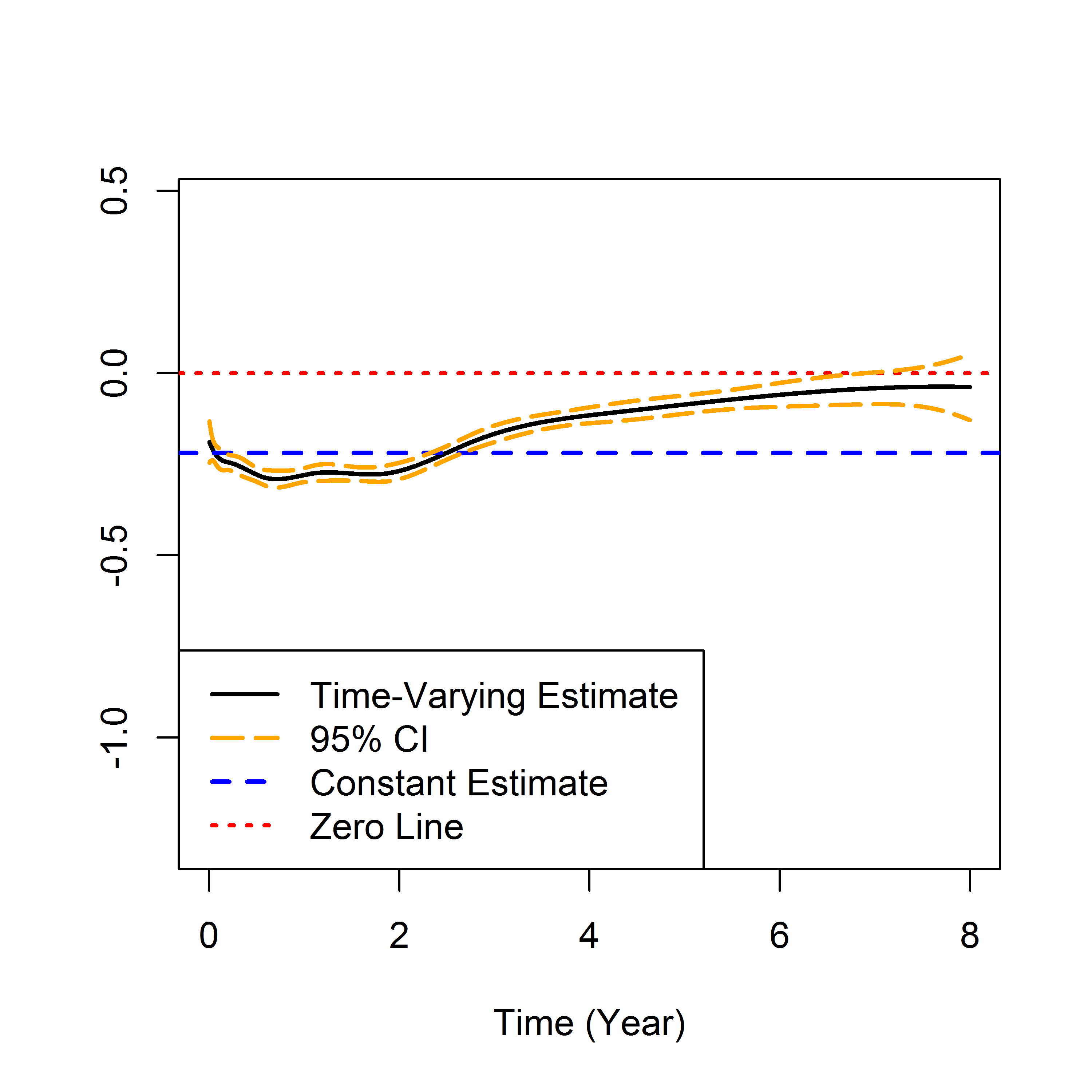}
  }
    \centering
\hspace{0.1pt}
\subfloat[Core Muscle Size for Surgical Outcomes]{
\includegraphics[width=2.5in, height=2.5in]{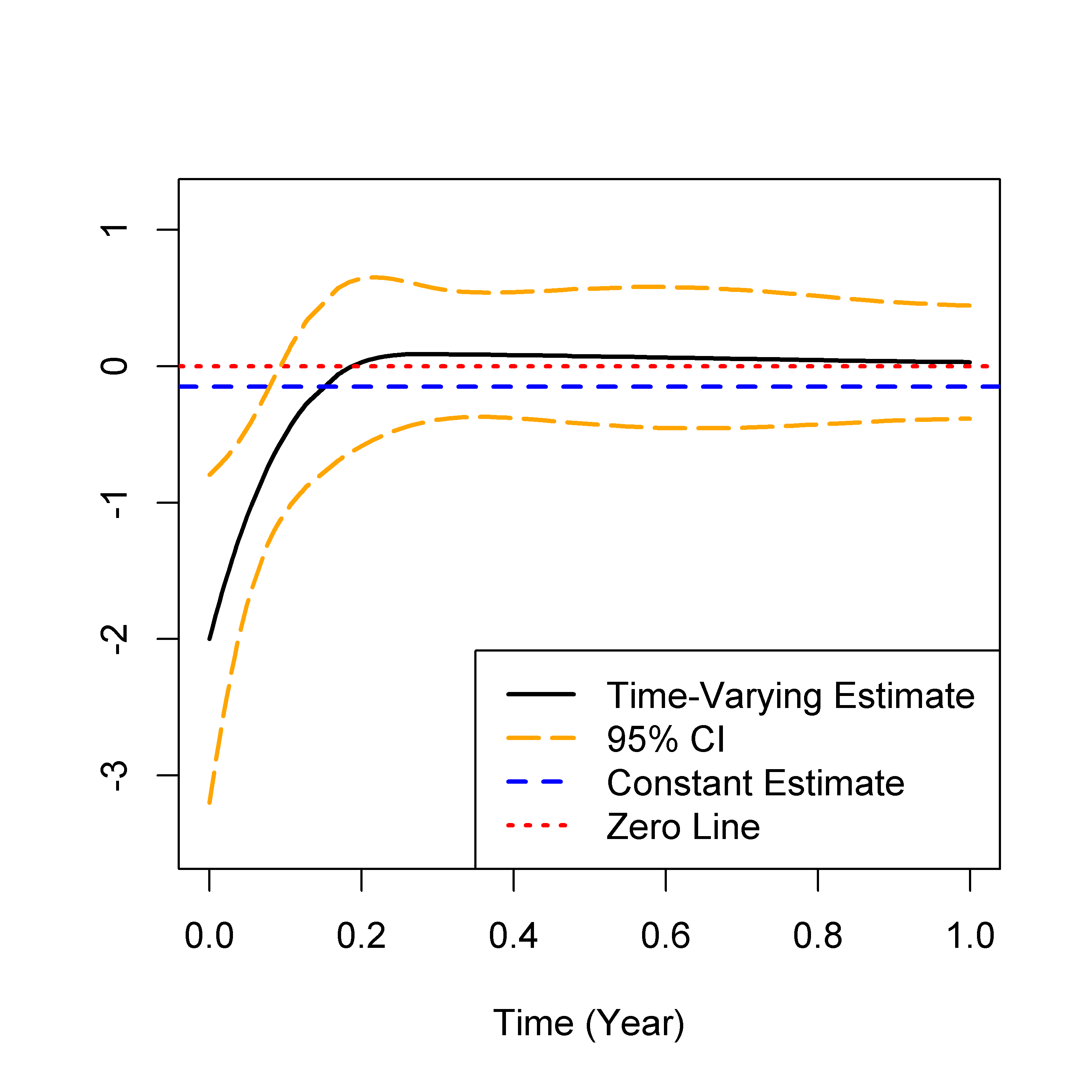}
  }
\end{figure}

\vspace{-3pt}
\begin{figure}[h]
 \caption{Computation time and estimation error.}% Figures (a) and (c) compare these methods until the algorithms converge. Figures (b) and (d) take a close look at the first two minutes. }
\centering
\subfloat[P=5]{
\includegraphics[width=2.75in, height=2.25in]{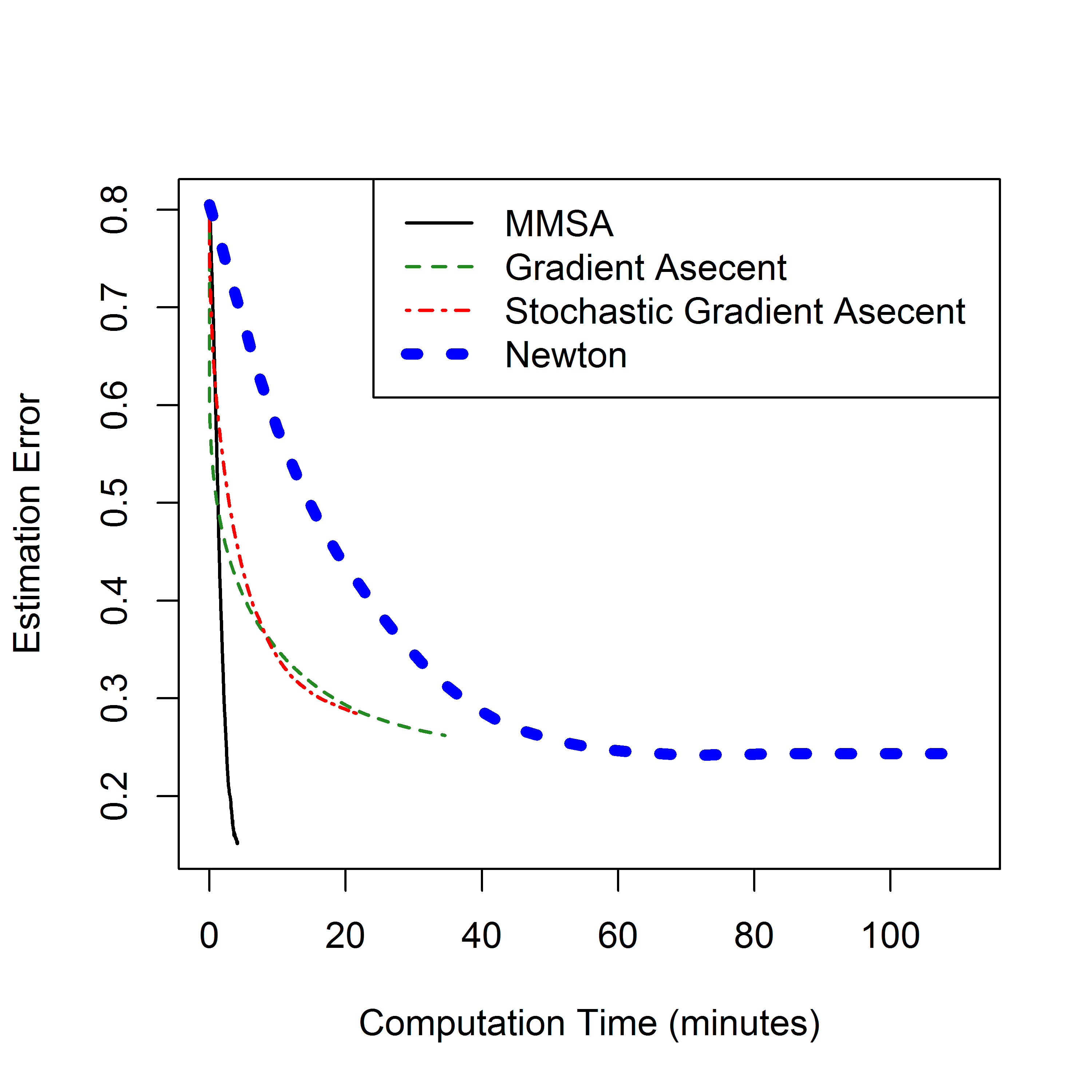}
  }
    \centering
\hspace{0.1pt}
\subfloat[P=20]{
\includegraphics[width=2.75in, height=2.25in]{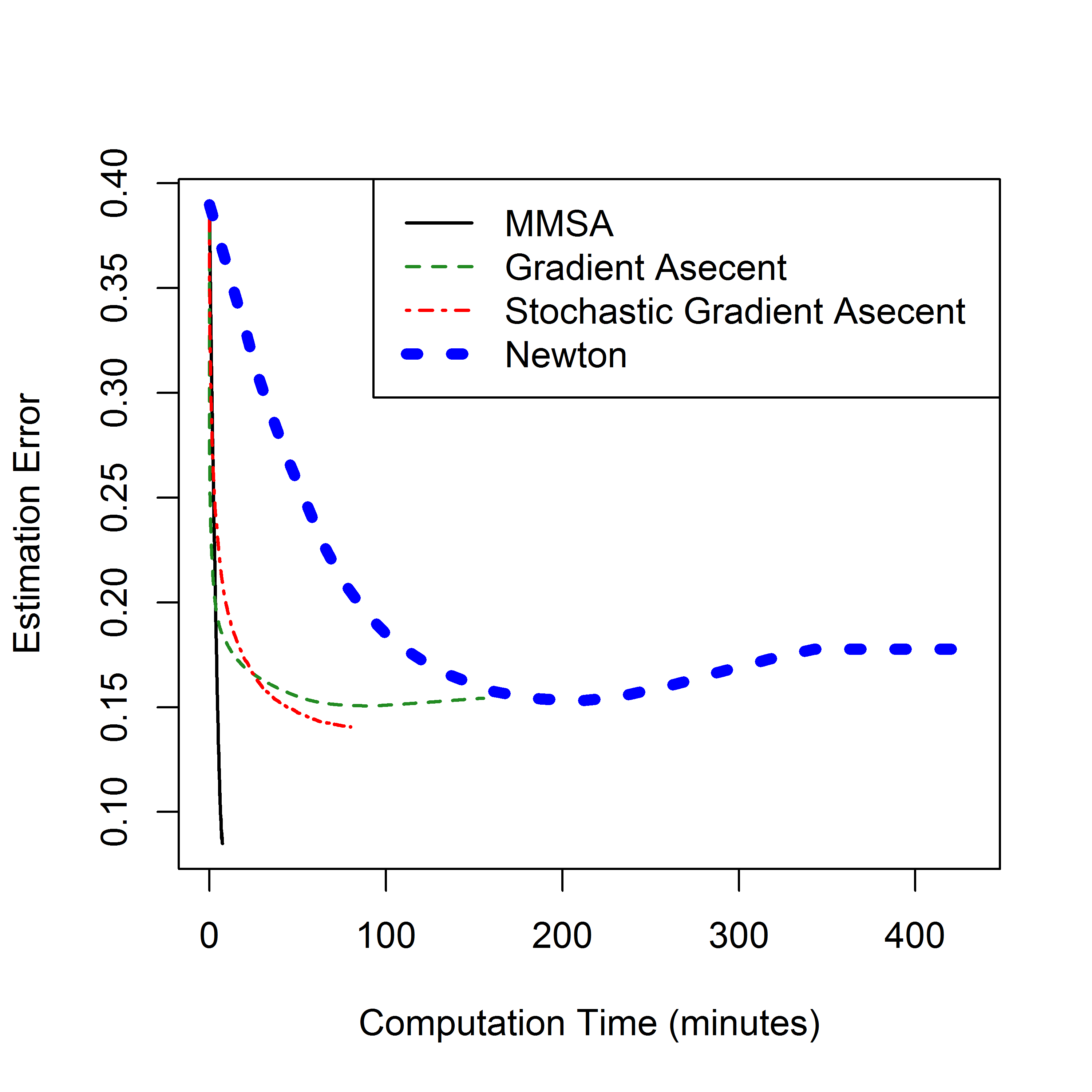}
  }
\end{figure}
\vspace{-3pt}

To extend the standard Cox model, time-varying effects have been widely studied.
Zucker and Karr \cite{r42}
 utilized a penalized partial likelihood approach
 and proposed nonparametric estimation of the time-varying effects. A specialized algorithm for this problem was then provided by Hastie and Tibshirani \cite{r18}. Alternatively,
Gray \cite{r15,r16} proposed using fixed knots spline functions.
 Kernel-based partial likelihood approaches have also been developed  \cite{r34,r24}. Some recent studies \cite{r20,r39} have proposed selection
of time-varying effects using penalized methods such as adaptive Lasso \cite{r40,r41}. Xiao, Lu, and Zhang \cite{r38} combined the ideas of local polynomial smoothing and group nonnegative garrote to achieve these goals. He et al. \cite{r43} considered a frailty model with time-varying effects.

While successful, these methods present challenges for large-scale studies.
To estimate time-varying effects in survival analysis, datasets are usually expanded in a repeated measurement format, where the time is divided into small intervals of a single event. Within each interval, the covariate values and outcomes for at-risk subjects are stacked to a large working dataset, which becomes infeasible for large sample sizes.
To avoid the data expansion, a routine based on Kronecker product has been suggested \cite{r28}. However, in our motivating setting, the pertinent analysis file is extremely large because there are over 300,000 transplant patients. The algorithm,
which involves iterative computation and inversion of the observed information matrix, can easily overwhelms a computer with an 32G memory.

%To avoid the intensive data expansion problem, a routine based on Kronecker product was suggested by \citet{Perperoglou2006}. However,

Moreover, to estimate time-varying effects, one may represent the coefficients using basis expansions such as B-splines. Thus, the parameter vector carries block structures, for which extra parameters are created and
 the computational burden increases quickly as the number of predictors grows. In particular, the kidney transplant database includes more than 160 predictors and many of which are comorbidities with rare frequencies. The inversion
of the observed information matrix leads to unstable estimations, especially in the right-tail of the follow-up period, because the data tends to be sparse there due to the censoring.

To exemplify this issue, we conducted a simulated example to compare the proposed method (termed MMSA) with Newton approach, gradient ascent, and stochastic gradient ascent with  step size determined by Adagrad algorithm \cite{r32}.
 Detailed simulation set-up is provided in Setting 2 of Section 3. 
 Figure 2 compares the computation times and average estimation errors.
 When the number of parameters is large, the Newton approach introduces large estimation biases. Gradient-based methods also face serious limitations by overlooking useful information from the Hessian matrix. In contrast, the proposed approach is computationally efficient and substantially improves the estimation error.

%Moreover, in classical gradient-based procedures, the updates at each iteration are computed based on gradient information only. Empirically we find that its performance is more sensitive to the choice of learning rate. For example, a sufficiently small learning rate is needed to ensure that the estimates in each iteration of gradient-based procedure serve as refinements of the previous step, which requires a large number of iterations and thus more computation time. In contrast, the proposed MMSA substantially improves the computational efficiency.

Our proposed solution is motivated by boosting \cite{r9}, which was originally introduced for classifying
binary outcomes.
Breiman \cite{r2} formulated boosting as a gradient descent approach with a special loss function. Mason et al. \cite{r25} developed a related algorithm, which was mainly acknowledged in the machine learning community. Friedman, Hastie and Tibshirani \cite{r10} and Friedman \cite{r11} laid out a gradient boosting framework to handle a variety of loss functions. B\"{u}hlmann and Yu \cite{r3} proposed a novel component-wise boosting procedure
based on  $\ell_2$ loss functions, and B\"{u}hlmann and Yu  \cite{r4} further demonstrated that the
component-wise procedure works well in high-dimensional linear models. Wolfson \cite{r37} developed a modification of gradient boosting under the estimation equation settings.

While demonstrating promising performance for proportional hazards model \cite{r45}, as illustrated by Hofner, Hothorn, and Kneib \cite{r19}, conventional gradient boosting cannot accommodate time-varying effects in survival analysis. Alternatively, likelihood-based boosting was considered  \cite{r19}. However, this approach is computationally intensive,  preventing its use in large-scale settings.

To fill in these gaps, we propose a new steepest ascent procedure based on a Minorization-Maximization (MM) algorithm \cite{r23}. Our proposed approach 
converts a difficult optimization problem into a sequence of simpler ones. Simplicity is achieved by avoiding iterative computation and inversion of large-scale observed information matrix, which is especially important  for large-scale analysis. Leveraging the block structure formed by the basis expansions, the proposed procedure iteratively updates the optimal block-wise direction along which the directional derivative is
maximized and, hence, the approximate increase in log-partial likelihood is greatest. The resulting estimates ensure the ascent
property and serve as refinements of the previous step.
As exemplified in Figure 2, our procedure provides
 well-behaved results, achieving less estimation error and improved computational efficiency.

%The remainder of this article is organized as follows: we describe the proposed  method in Section 2. Numerical properties are examined in Section $3$ through simulations. We then apply it to national kidney transplant data in Section $4$.  The article concludes with a discussion.

 % for simultaneously selecting and automatically determining potential time-varying effects
% The proposed method improves the flexibility of the Cox-based risk-adjustment model currently used in the literature.

\section{Method}\label{model}
\subsection[]{Model}
In our motivating example, patients came from multiple transplant centers. %across the United States.  
In the absence of adjustment for center effects, the estimation of covariate effects may be substantially biased \cite{r22}. %due to uncontrolled confounding by centers 
To avoid this issue, we adopt a stratified model with center-specific baseline hazards. Another advantage of using a stratified model is that it greatly reduces the calculations  across the partial likelihood contributions, which is especially important for the large-scale data  exemplified in our study. %Thus, our proposed approach is based on the stratified model.
 %To incorporate the multi-center data structure in the kidney transplant database, 
 
Let $D_{ij}$ denote the death  time and $C_{ij}$ be the
censoring time for patient $i$ in center $j$,
 $i=1,\ldots, n_j$, and $j=1, \ldots, J$. Here $n_j$ is the sample size in center $j$, and $J$ is the number of centers. %The total number of patients is denoted by $n=\sum_{j=1}^Jn_j$.
 The observed
time is denoted as $T_{ij} = \min\{D_{ij},C_{ij}\}$, and
 the death indicator is given by $\delta_{ij} = I(D_{ij} < C_{ij})$.  Let
  $\textbf{X}_{ij}=(X_{ij1}, \ldots, X_{ijP})^T$ be a $P$-dimensional covariate
vector.  We assume that, upon conditioning on $\textbf{X}_{ij}$, $D_{ij}$ is independently censored by $C_{ij}$.
Consider a center-specific hazard function
\begin{eqnarray}
   \lambda(t|\bX_{ij}) = \lambda_{0j}(t)\exp(\bX_{ij}^T {\boldsymbol\beta}(t)), \nonumber
\end{eqnarray}
where $\lambda_{0j}(t)$ is the center-specific baseline hazard. 
 To estimate the time-varying coefficients ${\boldsymbol\beta}(t)=(\beta_{1}(t),\ldots,  \beta_{P}(t))$, we span
 $\boldsymbol\beta(\cdot)$  by a set of  B-splines on a fixed grid of knots: 
\begin{eqnarray}
   \beta_{p}(t)=\boldsymbol\theta_{p}^T  \bB(t)=\sum_{k=1}^K \theta_{pk} B_k(t), ~~ p=1, \ldots, P,   \nonumber
\end{eqnarray}
where   $\bB(t)=(B_1(t), \ldots,
B_K(t))^T$ forms a basis,  and
 $\boldsymbol\theta_{p}=(\theta_{p1}, \ldots, \theta_{pK})^T$ is a vector of coefficients with
 $\theta_{pk}$ being the coefficient for the $k$-th basis of the $p$-th
covariate.  Considering a length-$PK$  parameter vector
$\boldsymbol\theta=vec(\boldsymbol\Theta)$,  the vectorization of the coefficient matrix
$\boldsymbol\Theta=(\boldsymbol\theta_{1}, \ldots, \boldsymbol\theta_{P})^T$ by row,
 the log-partial likelihood function is
   \begin{align}
\ell(\boldsymbol\theta)=\sum_{j=1}^J \sum_{i=1}^{n_j} \delta_{ij} \left [\bX_{ij}^T \boldsymbol\Theta  \bB(T_{ij})
-\log
\left\{\sum_{i' \in R_{ij}}  \exp \left(\bX_{i' j}^T \boldsymbol\Theta  \bB(T_{ij}) \right) \right \} \right ],
 \label{eq:logpartial}
\end{align}
 where $R_{ij}=\{i': 1 \leq i' \leq n_j, ~ T_{i' j}\geq T_{ij}\}$ is the center-specific at-risk set. For small-sized problems, maximizing \eqref{eq:logpartial} can be achieved by a Newton's approach, which, however, becomes impractical for large-scale problems.
%It is worth noting that, in generalized linear models, time-varying effects can be estimated by including interactions between $\bX_i$ and $\bB(T_i)$. In survival analysis, however, the computation is much more intensive due to the cross-terms between $\bX_{\ell}$ and $\bB(T_i)$ for all $\ell \in R_i$, $i=1, \ldots, n$. %The massive $n$ and $p$ of large-scale time-to-event data introduce unique computational and statistical challenges.

\subsection[]{Motivation}

To improve computational efficiency and numerical stability,
we consider 
 a first-order approximation of $\ell(\boldsymbol\theta)$ around the current estimate $\widehat{\boldsymbol\theta}$: 
 \begin{eqnarray}
 \ell(\boldsymbol{\widehat
\theta}+\alpha \boldsymbol \mu)=\ell(\boldsymbol{\widehat
\theta})+\alpha \triangledown \ell(\widehat{\boldsymbol\theta})^T\boldsymbol \mu
 + o(\alpha),
 \nonumber
\end{eqnarray}
where $\boldsymbol \mu$ is the update direction of $\boldsymbol\theta$, $\alpha$ is a small positive value,  $\triangledown \ell(\widehat{\boldsymbol\theta})$ is the gradient, and
the term $\triangledown \ell(\widehat{\boldsymbol\theta})^T\boldsymbol \mu$  is the directional derivative along the direction $\boldsymbol \mu$:
   \begin{eqnarray}
\triangledown \ell(\widehat{\boldsymbol\theta})^T\boldsymbol \mu =\frac{\partial}{\partial
\alpha}\ell(\boldsymbol{\widehat
\theta}+\alpha \boldsymbol \mu)\bigg \rvert_{\alpha=0}
=\lim_{\alpha \rightarrow 0} \frac
 { \ell(\boldsymbol{\widehat
\theta}+\alpha \boldsymbol \mu)-\ell(\boldsymbol{\widehat
\theta}) }{\alpha}.
 \nonumber
\end{eqnarray}
If $\triangledown \ell(\widehat{\boldsymbol\theta})^T\boldsymbol \mu >0$, $\boldsymbol \mu$ is an ascent direction driving  $\ell(\boldsymbol\theta)$ uphill.
Intuitively, we wish to identify a unit norm update
direction such that
$\ell(\boldsymbol{\widehat
\theta}+ \alpha
\boldsymbol
\mu)$
 ascends most rapidly.
This motivates a steepest ascent direction that maximizes the direction derivative
  \begin{eqnarray}
 \boldsymbol \mu^{\star}= \argmax_{\boldsymbol \mu} \{ \triangledown \ell(\widehat{\boldsymbol\theta})^T\boldsymbol \mu ~\big |~ ||\boldsymbol \mu||_{\dagger}=1\},
 \label{eq:SD}
\end{eqnarray}
where $||\cdot||_{\dagger}$ is a norm on $\mathbb{R}^{PK}$. %The dual norm \cite{r23} of $||\boldsymbol \mu||_{\dagger}$ is given by
% \begin{align*}
%||\triangledown\ell(\widehat{\boldsymbol\beta})||_{\star}=\sup \{\triangledown\ell(\widehat{\boldsymbol\beta})^T\boldsymbol \mu~\big |~ ||\boldsymbol \mu||_{\dagger}=1\}. 
%\end{align*}
%It follows that the generalized Cauchy-Schwarz inequality holds such that$\triangledown\ell(\widehat{\boldsymbol\beta})^T\boldsymbol \mu \leq ||\triangledown\ell(\widehat{\boldsymbol\beta})||_{\star} ||\boldsymbol \mu||_{\dagger}.$

\subsection[]{Example Norms}

The choice of norm $||\boldsymbol \mu||_{\dagger}$  plays a critical role in the performance of  \eqref{eq:SD}. 
If we choose a $\ell_2$ norm, the corresponding dual norm \cite{r1} is the $\ell_2$ norm itself, which leads to a gradient ascent method. As illustrated in Figure 2, its  convergence can be very slow. %, especially when the condition numbers of the observed information matrix are large. %Figure 3 further compares the average estimated coefficients across various iterations of the proposed method and the gradient ascent. Detailed simulation set-up is provided in Section 3.2.
Alternatively,  if we consider a $\ell_1$ norm,  the corresponding dual norm is the $\ell_{\infty}$ norm, which leads to  a coordinate-wise gradient boosting procedure. However, this approach is not suitable for our motivating setting, in which the parameter vector carries a group structure formed by  representing the time-varying coefficients using basis expansions.

\subsection[]{Block-Wise Steepest Ascent}

To leverage the block structure of our parameter vectors, we
consider a $\ell_1$/quadratic norm, 
  \begin{eqnarray}
||\boldsymbol \mu||_{\dagger}=\sum_{p=1}^P\big|\big|\boldsymbol \mu_p\big|\big|_{\bH_p(\boldsymbol{\widehat
\theta})}, ~~\mbox{where}~~ ||\boldsymbol \mu_p||_{\bH_p(\boldsymbol{\widehat
\theta})}=\left(\boldsymbol \mu_p^T \bH_p(\boldsymbol{\widehat
\theta}) \boldsymbol \mu_p\right)^{1/2}.
\label{eq:optimum}
\end{eqnarray}
Here $\boldsymbol\mu_p$ is a $K$-dimensional vector corresponding to the $p$-th block of $\boldsymbol\mu$, and $\bH_p(\boldsymbol{\widehat
\theta})$ is a $K \times K$-dimensional matrix. Specifically, for $p=1, \ldots, P$, we choose 
\begin{align*}
\bH_p(\widehat{\boldsymbol\theta})=-\left(\triangledown \ell(\widehat{\boldsymbol\theta})_p^T (-\triangledown^2 \ell(\widehat{\boldsymbol\theta})_p)^{-1}  \triangledown \ell(\widehat{\boldsymbol\theta})_p\right)
 \triangledown^2 \ell(\widehat{\boldsymbol\theta})_p,
%\label{eq:H}
\end{align*}
where $\triangledown \ell(\widehat{\boldsymbol\theta})_p$ is a $K$-dimensional gradient vector and $\triangledown^2 \ell(\widehat{\boldsymbol\theta})_p$ is the $p$-th block diagonal of the Hessian matrix, corresponding to the $p$-th variable. The scalar $\triangledown \ell(\widehat{\boldsymbol\theta})_p^T (-\triangledown^2 \ell(\widehat{\boldsymbol\theta})_p)^{-1}  \triangledown \ell(\widehat{\boldsymbol\theta})_p$ plays a role as a normalization factor. 
  Apply the Cauchy-Schwarz inequality,
\begin{eqnarray}
\triangledown \ell(\widehat{\boldsymbol\theta})^T\boldsymbol \mu
 \leq \sum_{p=1}^P\big|\big|\triangledown \ell(\widehat{\boldsymbol\theta})_p\big|\big|_{\bH_p^{-1}(\boldsymbol{\widehat
\theta})}
  \big|\big|\boldsymbol \mu_p\big|\big|_{\bH_p(\boldsymbol{\widehat
\theta})}  \leq \max_p \left(\big|\big|\triangledown \ell(\widehat{\boldsymbol\theta})_p\big|\big|_{\bH_p^{-1}(\boldsymbol{\widehat
\theta})} \right)\sum_{p=1}^P
  \big|\big|\boldsymbol \mu_p\big|\big|_{\bH_p(\boldsymbol{\widehat
\theta})}.
 \nonumber
\end{eqnarray}
Thus, the dual norm of a $\ell_1$/quadratic norm is the $\ell_{\infty}$/quadratic norm. The resulting $\boldsymbol \mu^{\star}$ is the normalized direction in the unit ball of $\ell_1$/quadratic norm that extends farthest in the direction of gradient $\triangledown \ell(\widehat{\boldsymbol\theta})$, and is given by
\begin{eqnarray}
 \boldsymbol \mu^{\star}=(0, \ldots, 0, \widetilde{\boldsymbol\mu}_{p^{\star}}^T, 0, \ldots, 0)^T,
 \label{eq:block}
\end{eqnarray}
 where $p^{\star}$ is the optimal block maximizing the block-wise directional derivative
\begin{align}
p^{\star}=\argmax_p \nabla \ell(\boldsymbol{\widehat\theta})_p^T \widetilde{\boldsymbol\mu}_{p},
 \label{eq:pStar}
 \end{align}
with $\widetilde{\boldsymbol\mu}_{p}$ being a $K$-dimensional update vector corresponding to the $p$-th variable
\begin{align}
\widetilde{\boldsymbol\mu}_{p}=  \left(\bH_p(\widehat{\boldsymbol\theta})\right)^{-1}\triangledown \ell(\widehat{\boldsymbol\theta})_{p}.
\label{eq:muStar}
 \end{align}
 This leads to a Minorization-Maximization-based steepest ascent procedure that iteratively pursues the optimal block-wise direction  maximizing the directional derivative. 

 \subsection[]{Minorization Step}
 
In the minorization step, the $\ell_1$/quadratic norm considered in \eqref{eq:optimum} leads to the following minority surrogate function
\begin{align*}
g(\boldsymbol\theta|\widehat{\boldsymbol\theta}) =  \ell(\widehat{\boldsymbol\theta}) + \triangledown \ell(\widehat{\boldsymbol\theta})^T (\boldsymbol\theta-\widehat{\boldsymbol\theta})-\frac{1}{2\nu}(\boldsymbol\theta-\widehat{\boldsymbol\theta})^T \bH(\widehat{\boldsymbol\theta}) (\boldsymbol\theta-\widehat{\boldsymbol\theta}),
%\label{eq:surrogate}
\end{align*}
where $\nu$ is a small positive value to be specified. Here $\bH(\widehat{\boldsymbol\theta})$ is a block-diagonal matrix, where the blocks correspond to the basis expansions for each variable. In particular, 
\[
\bH(\widehat{\boldsymbol\theta})
=
\begin{bmatrix}
    \bH_1(\widehat{\boldsymbol\theta}) & 0 & \dots  & 0 \\
    0 & \bH_2(\widehat{\boldsymbol\theta})  & \dots  & 0 \\
    \vdots & \vdots  & \ddots & \vdots \\
    0 & 0  & \dots  & \bH_P(\widehat{\boldsymbol\theta})
\end{bmatrix}
\]
with all non-block-diagonal elements being zeros. It is obvious that $g(\boldsymbol{\widehat\theta}|\boldsymbol{\widehat\theta})=\ell(\boldsymbol{\widehat\theta})$. Proposition 1 in Section 2.7 shows that, with a suitable choice of $\nu$,  $g(\boldsymbol\theta|\widehat{\boldsymbol\theta}) \leq  \ell(\boldsymbol\theta)$ for all $\boldsymbol\theta$.

 \subsection[]{Maximization Step}
In the maximization step, the block-wise update in \eqref{eq:pStar} and \eqref{eq:muStar} maximizes $g(\boldsymbol{\theta}|\boldsymbol{\widehat\theta})$ subject to a constraint that only one variable is updated at each iteration.  
The resulting estimates of $\boldsymbol{\theta}$ ensure the ascent property and thus serve as refinements of the previous step.
 We summarize the proposed algorithm as follows:
 
\hspace{-0.5cm}\emph{Algorithm} (MMSA)
 \begin{enumerate}
 \item[(a)] Initialize $\boldsymbol {\widehat{\theta}}^{(0)}=\textbf{0}$. For $m=1,2,3, \ldots$, identify $p^{\star}$ as in \eqref{eq:pStar}.  
  \item[(b)] Update the estimate  by
\begin{align*}
 \boldsymbol {\widehat  \theta}^{(m)}_{p^{\star}}= \boldsymbol {\widehat  \theta}^{(m-1)}_{p^{\star}}+\nu ~ \widetilde{\boldsymbol\mu}_{p^{\star}}.
%  &{} \boldsymbol {\widehat  \theta}^{(m)}_{p}= \boldsymbol {\widehat  \theta}^{(m-1)}_{p}~ &{}\mbox{if}~p\neq p^{\star}.
\end{align*}
 \item[(c)] The iteration continues until  $\max_p \nabla \ell(\boldsymbol{\widehat\theta}^{(m-1)})_p^T \widetilde{\boldsymbol\mu}_{p}$ or the relative change in the
log-partial likelihood is less than a convergence threshold (e.g. $10^{-6}$).
\end{enumerate}
%\vspace{-0.25cm}

%Empirically, the best strategy appears to be a two-stage approach. In the first stage (Steps (a)-(d)),
Remark 1: The proposed MMSA is a block-wise procedure. At each iteration, only one variable is updated. The corresponding block-wise update direction maximizes the block-wise directional derivative, $\nabla \ell(\boldsymbol{\widehat\theta}^{(m-1)})_p^T \widetilde{\boldsymbol\mu}_{p}$, which ranks the importance of each predictor and measures how fast the log-partial likelihood would increase by including each predictor. 

Remark 2: The proposed algorithm converts a difficult optimization problem into a simpler surrogate function. Simplicity is achieved by avoiding iterative computation and inversion of large-scale observed information matrix. Numerical results show that the proposed algorithm provides sufficient
and rapid updates, achieving much computational efficiency.

%Remark 3: To further improve computational efficiency, a stochastic procedure \cite{r12} can be applied. At each iteration, sample a fraction $\eta$ (e.g. $\eta=0.2$) of the observations (without replacement), and estimate the  update using the subsample. %Not only does the sampling reduce the computing time (which is especially important for large-scale data), but also it improves prediction performance.

Remark 3: The learning rate, $\nu$, can be chosen to be
a small positive value, e.g. 0.05. Further clarification for the choice of $\nu$ is provided in Section 2.7.

Remark 4: To further improve computational efficiency, a stochastic procedure \cite{r12} can be applied. At each iteration, we sample a fraction $\eta$ (e.g. $\eta=0.2$) of the observations (without replacement), and estimate the  update using the subsample. Not only does the sampling reduce the computing time (which is especially important for large-scale
data), but also it improves estimation performance.

\begin{figure}[h]
 \caption{Average estimated coefficients across various MMSA and gradient ascent (termed gradient) iterations; $\beta_2(t)=sin3(\pi t/4)$ and $\beta_4(t)=-(t/3)^2 \exp(t/2)$ are time-varying effects; based on  100 simulation replicates; m: number of iterations.}
    \centering
\subfloat[MMSA: m=500]{
\includegraphics[width=1.5in, height=1.75in]{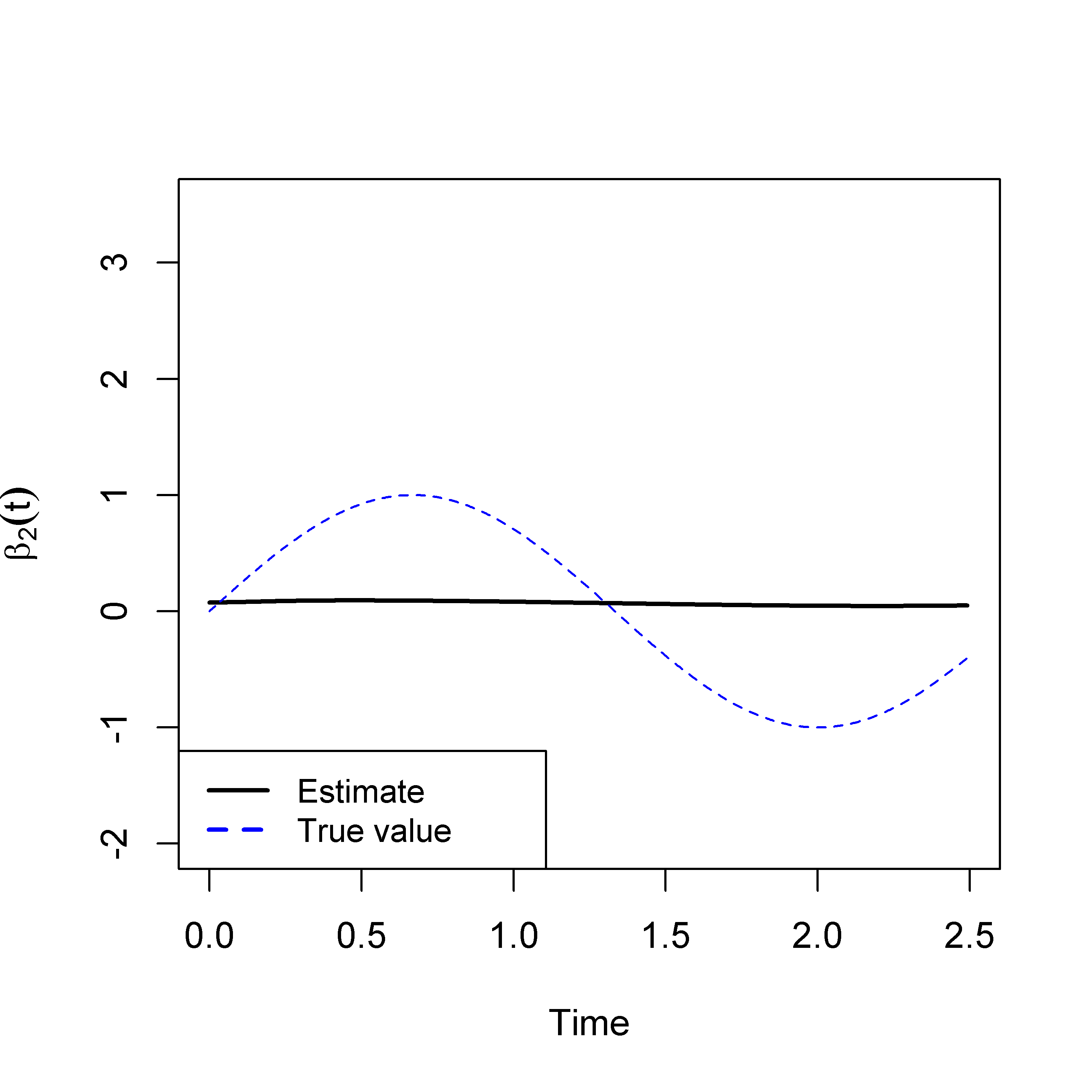}
  }
      \centering
      \hspace{0.1pt}
\subfloat[MMSA: m=1,000]{
\includegraphics[width=1.65in, height=1.75in]{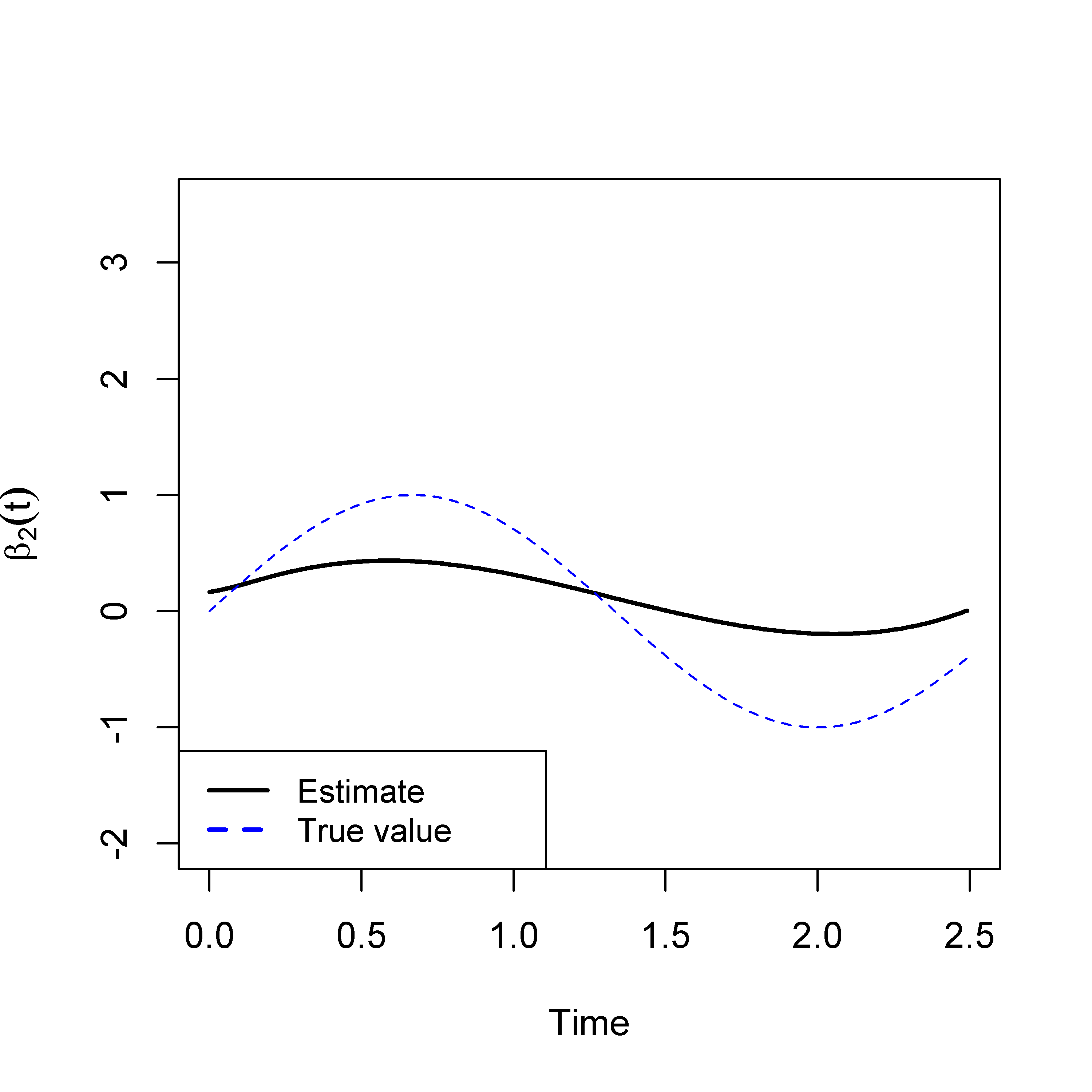}
  }
      \centering
\hspace{0.1pt}
\subfloat[MMSA: m=3,000]{
\includegraphics[width=1.65in, height=1.75in]{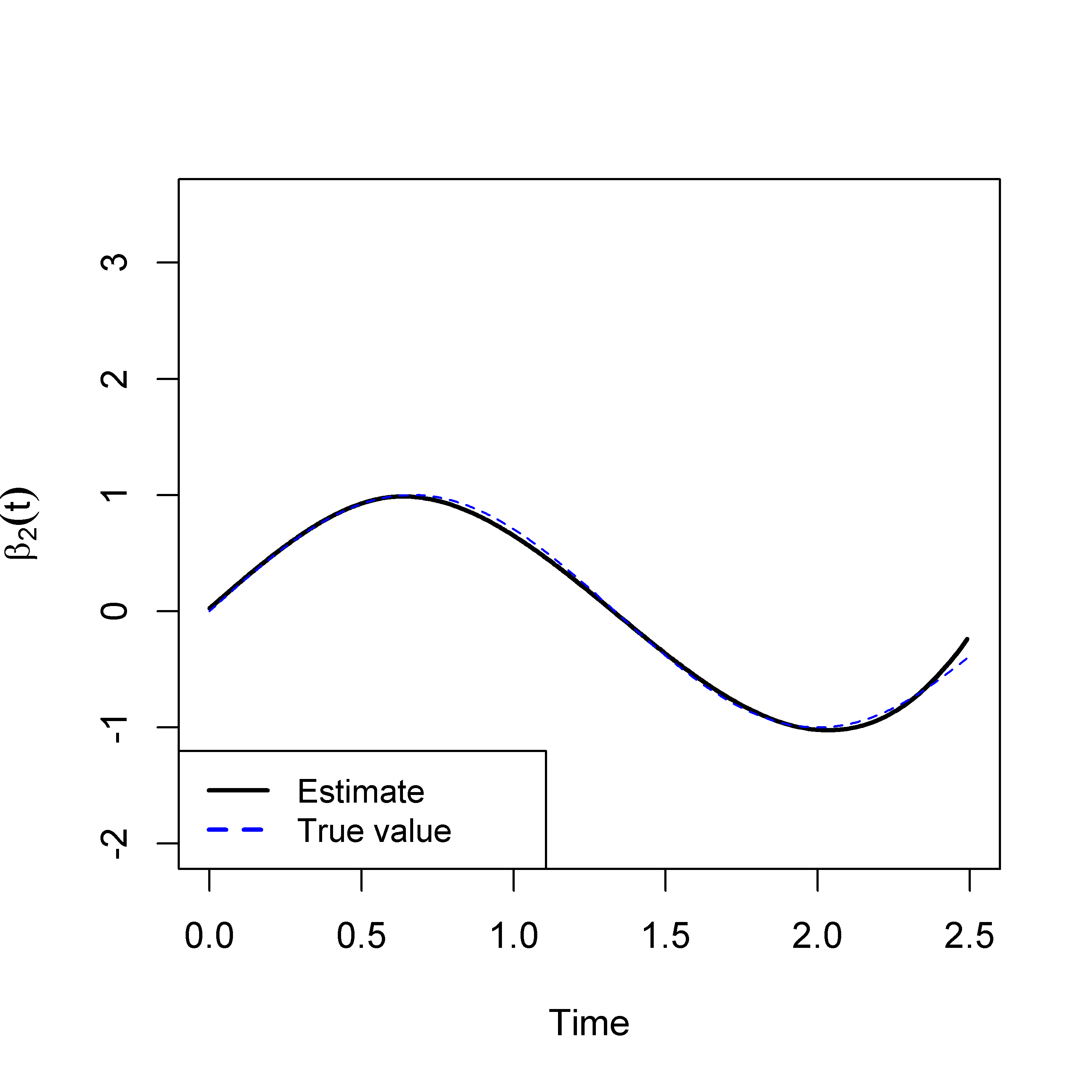}
  }
      \centering
\vspace{0.1pt}
\subfloat[MMSA: m=500]{
\includegraphics[width=1.65in, height=1.75in]{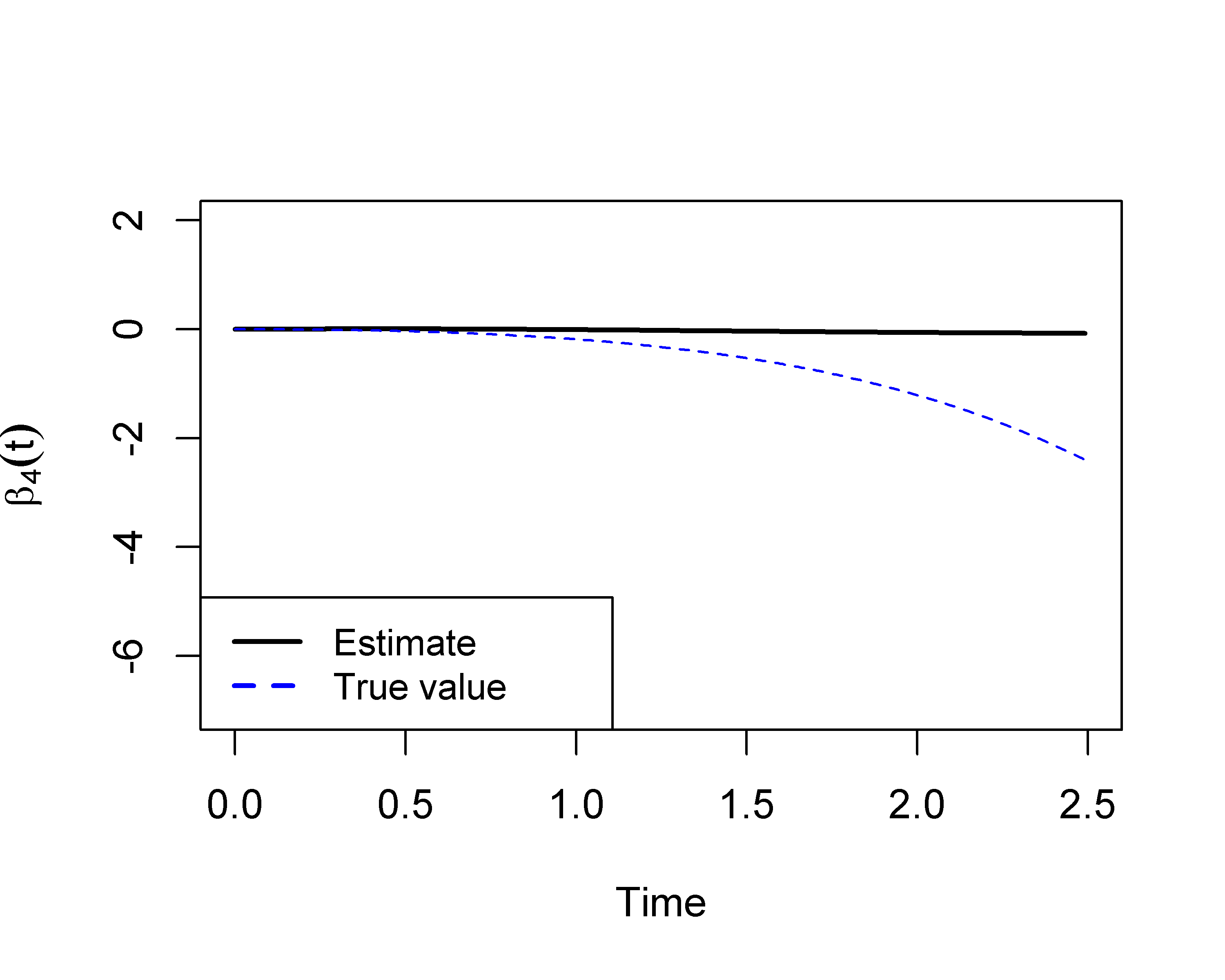}
  }
      \centering
\hspace{0.1pt}
\subfloat[MMSA: m=1,000]{
\includegraphics[width=1.65in, height=1.75in]{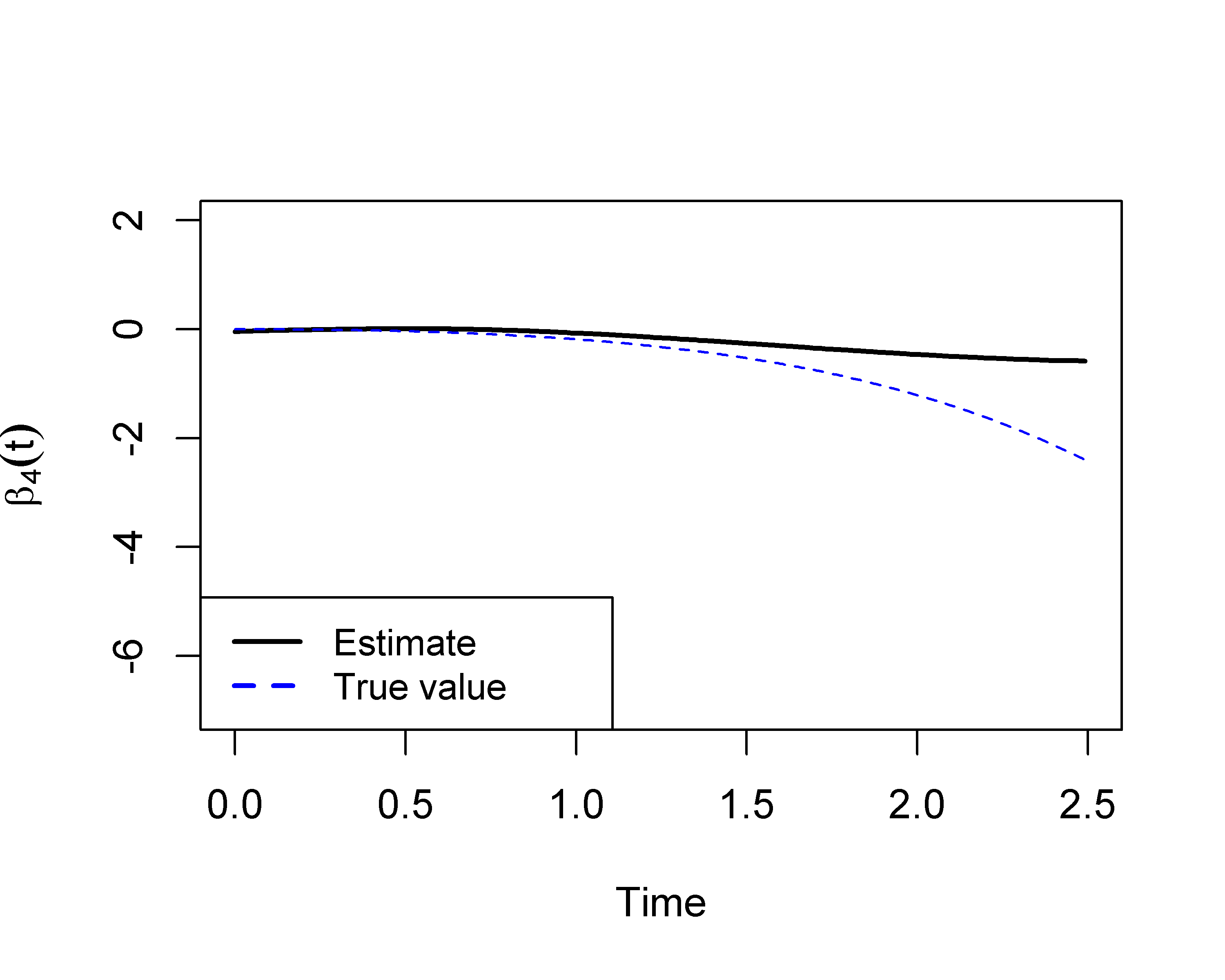}
  }
    \centering
\hspace{0.1pt}
\subfloat[MMSA: m=3,000]{
\includegraphics[width=1.65in, height=1.75in]{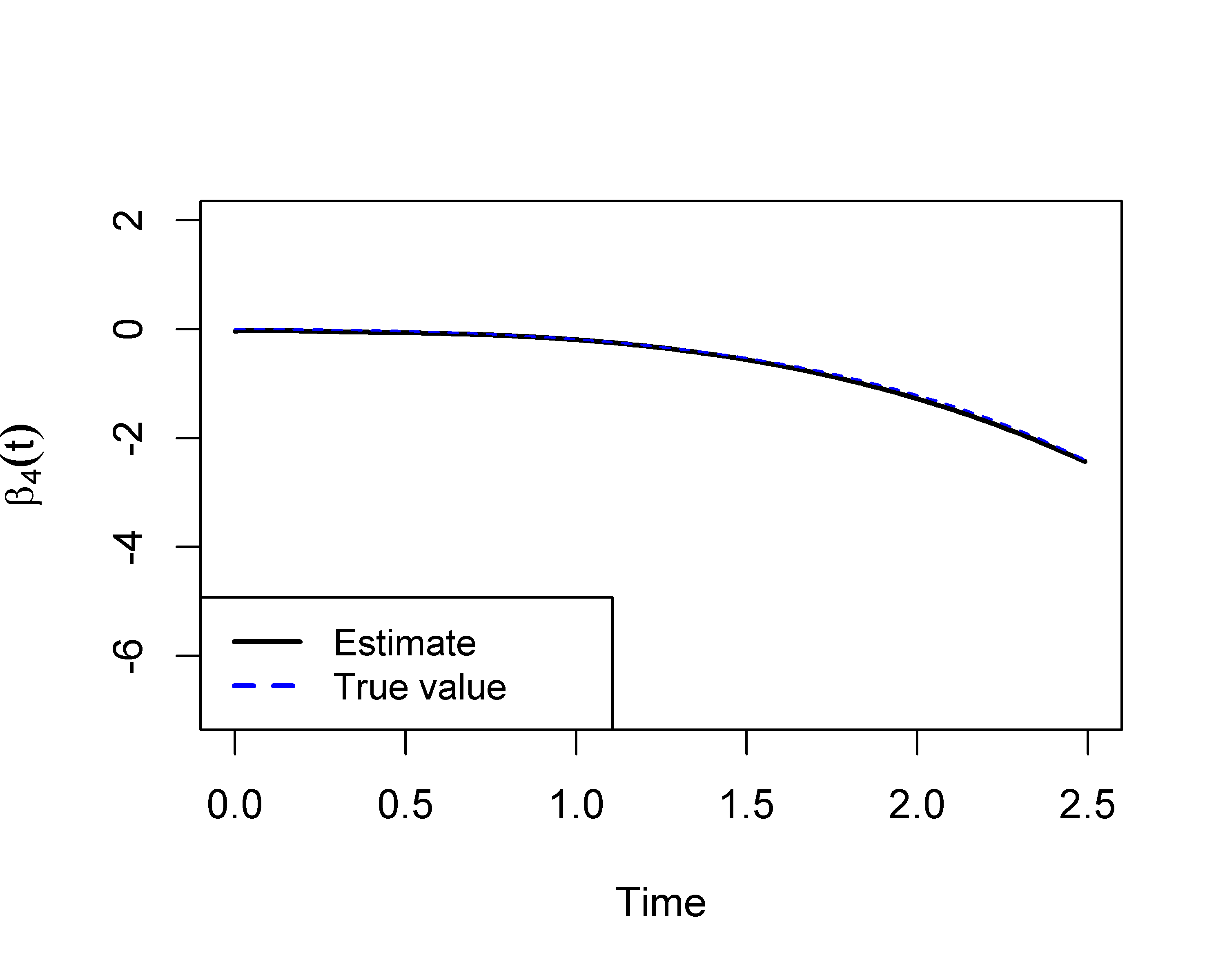}
  }
 \centering
 \vspace{0.1pt}
\subfloat[Gradient: m=500]{
\includegraphics[width=1.5in, height=1.75in]{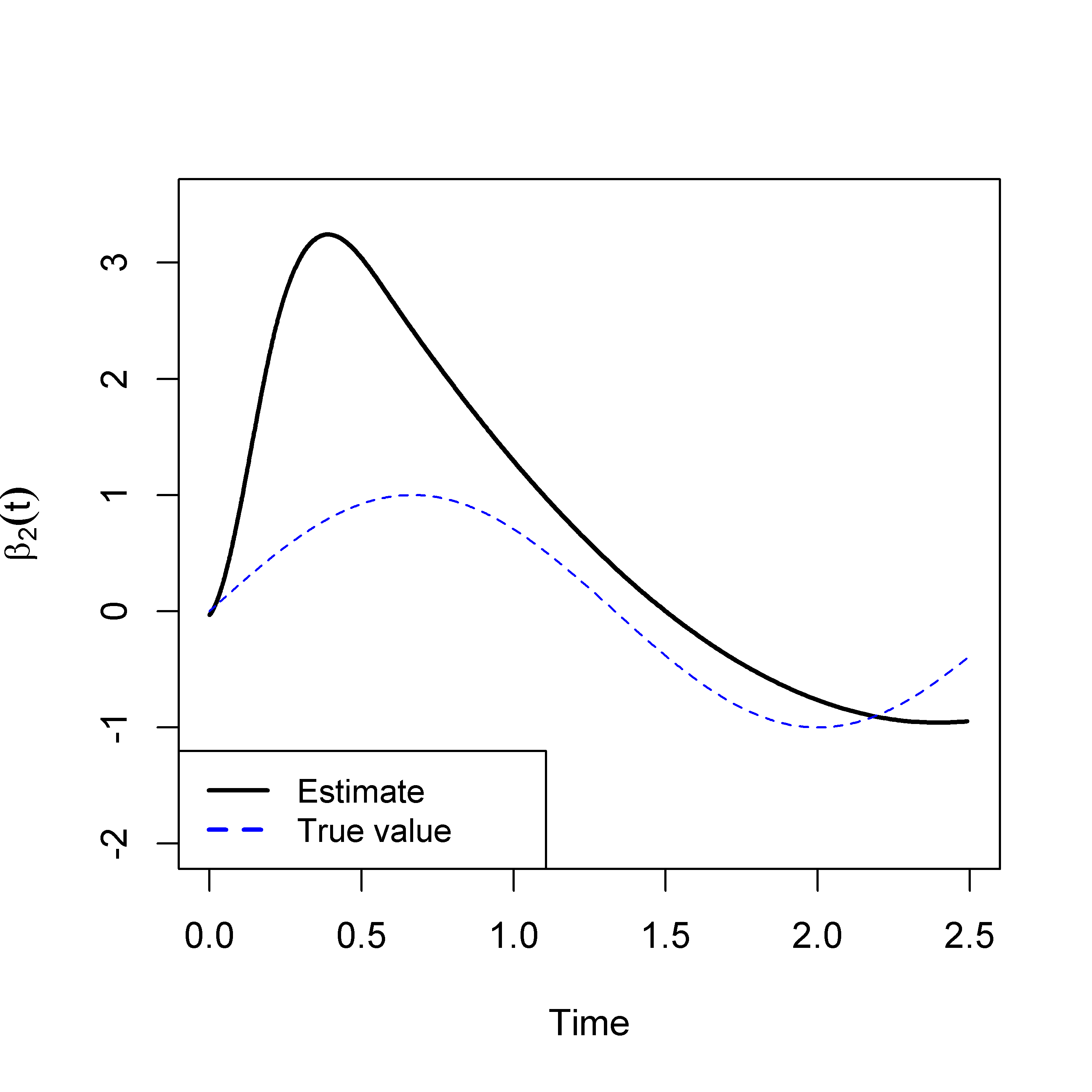}
  }
      \centering
      \hspace{0.1pt}
\subfloat[Gradient: m=1,000]{
\includegraphics[width=1.65in, height=1.75in]{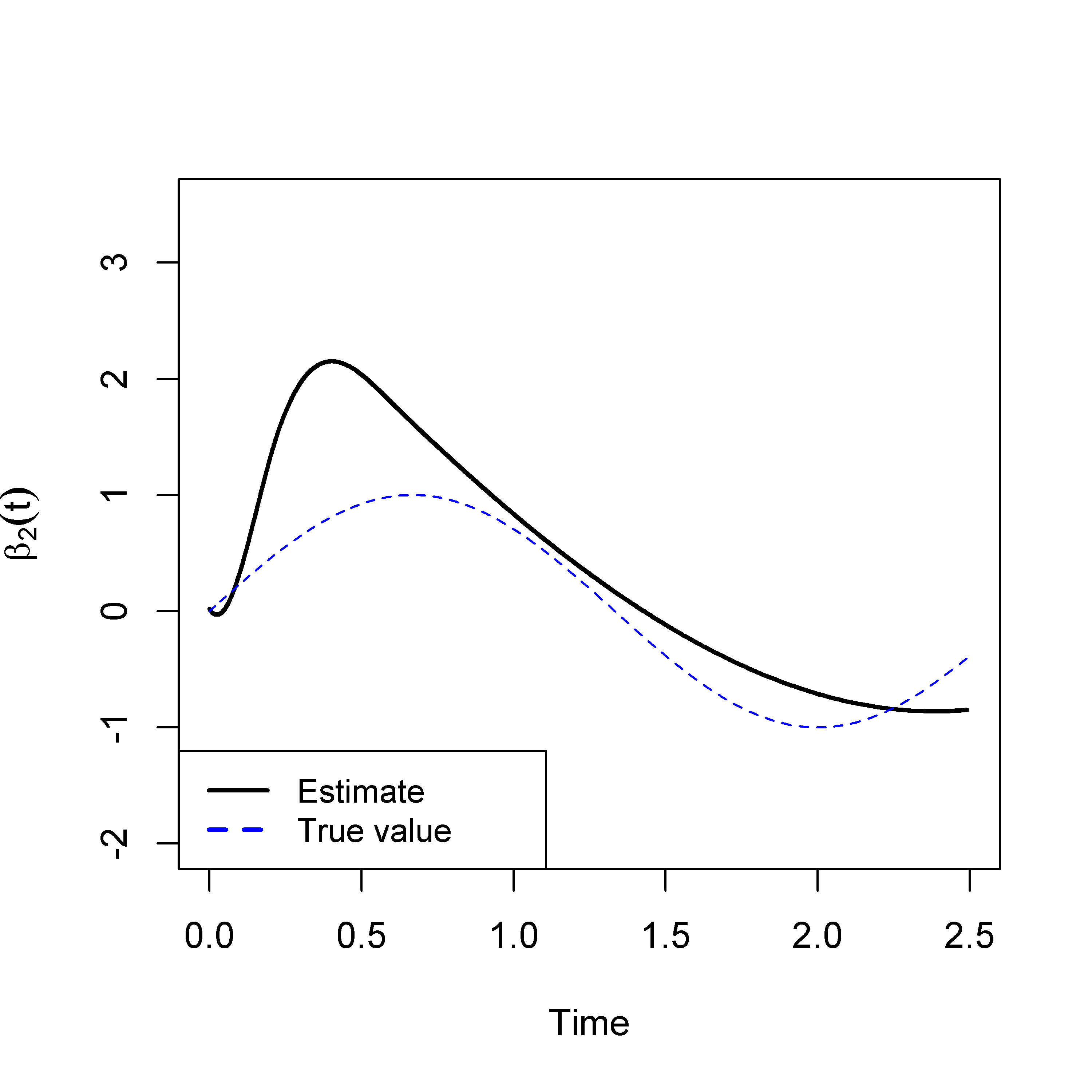}
  }
      \centering
\hspace{0.1pt}
\subfloat[Gradient: m=3,000]{
\includegraphics[width=1.65in, height=1.75in]{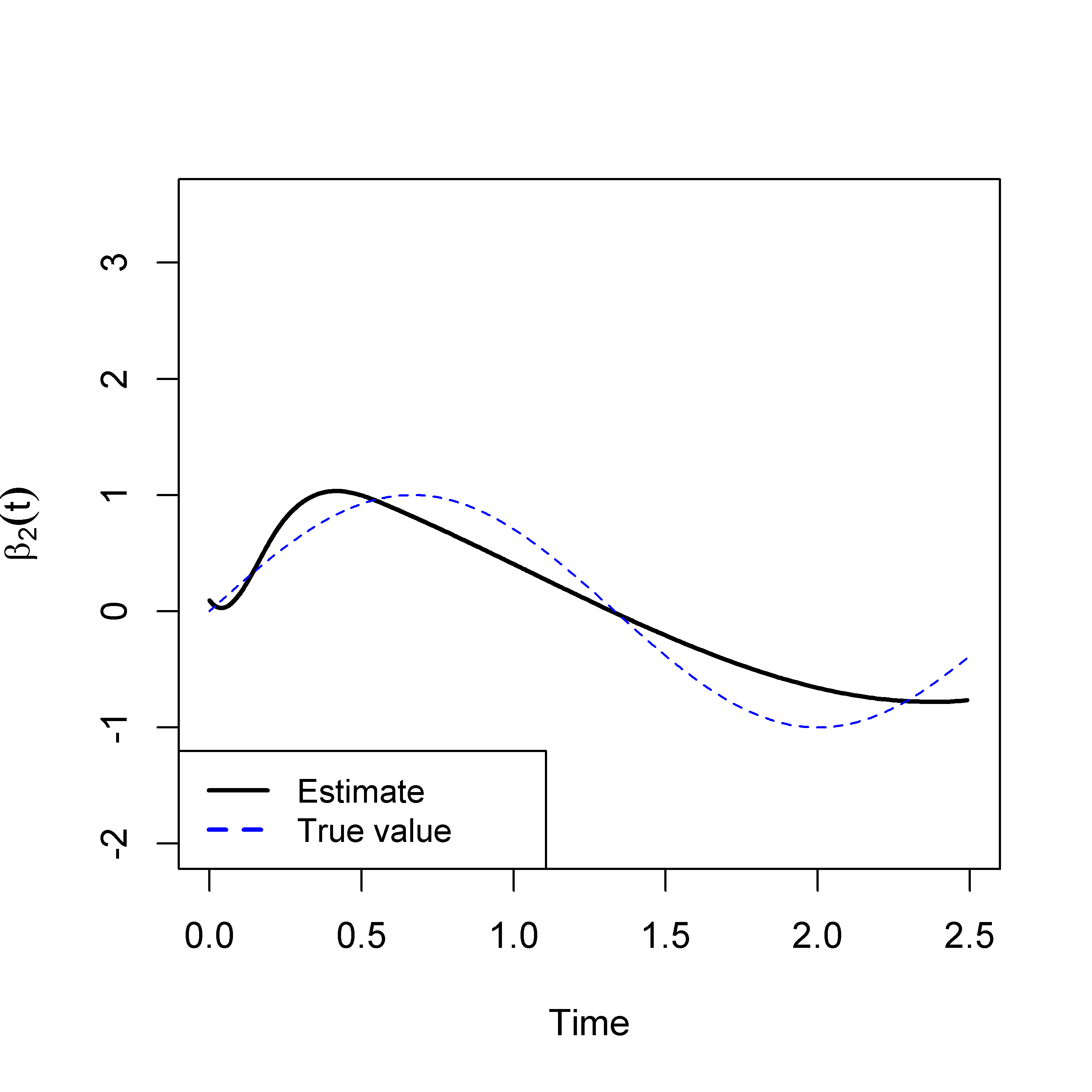}
  }
      \centering
\vspace{0.1pt}
\subfloat[Gradient: m=500]{
\includegraphics[width=1.65in, height=1.75in]{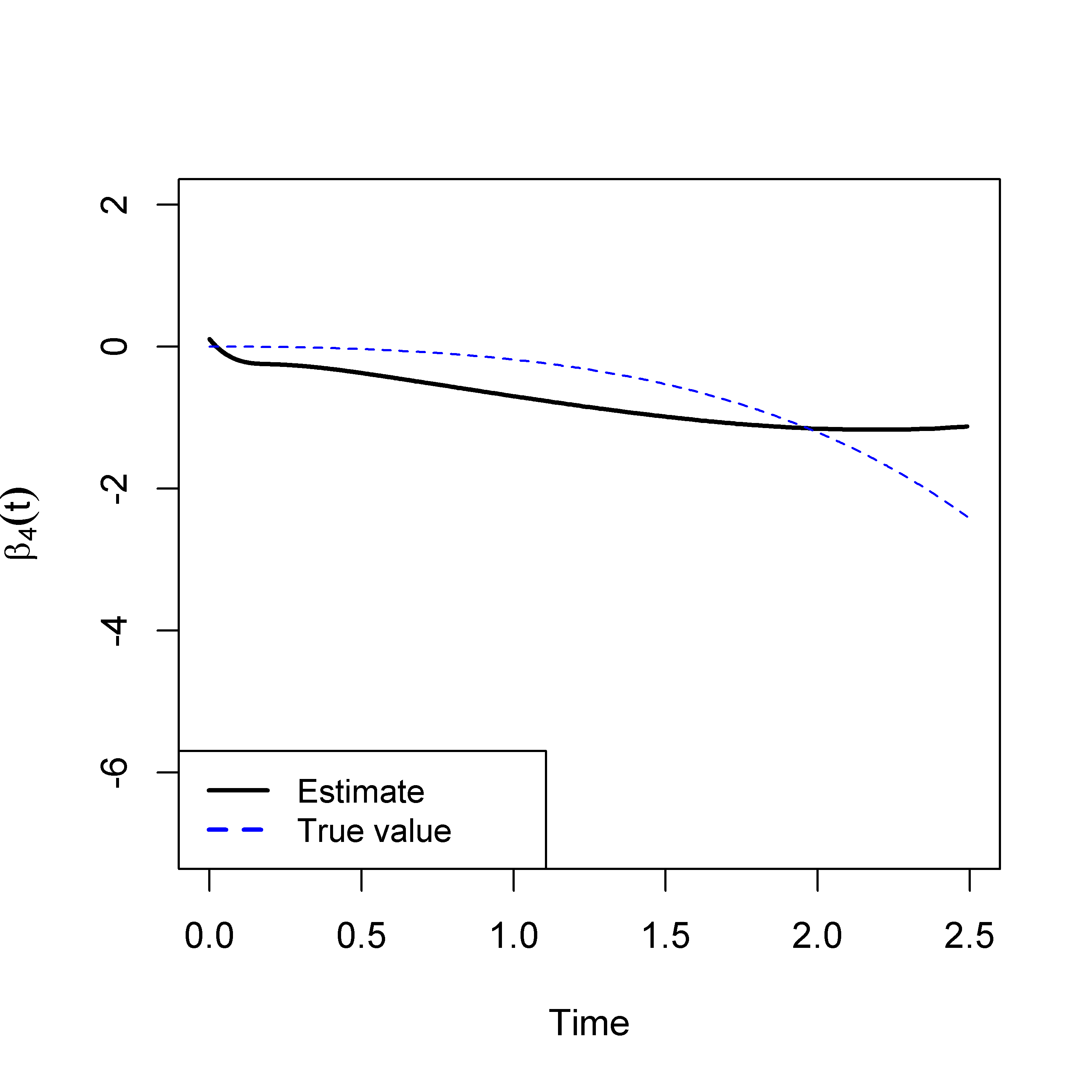}
  }
      \centering
\hspace{0.1pt}
\subfloat[Gradient: m=1,000]{
\includegraphics[width=1.65in, height=1.75in]{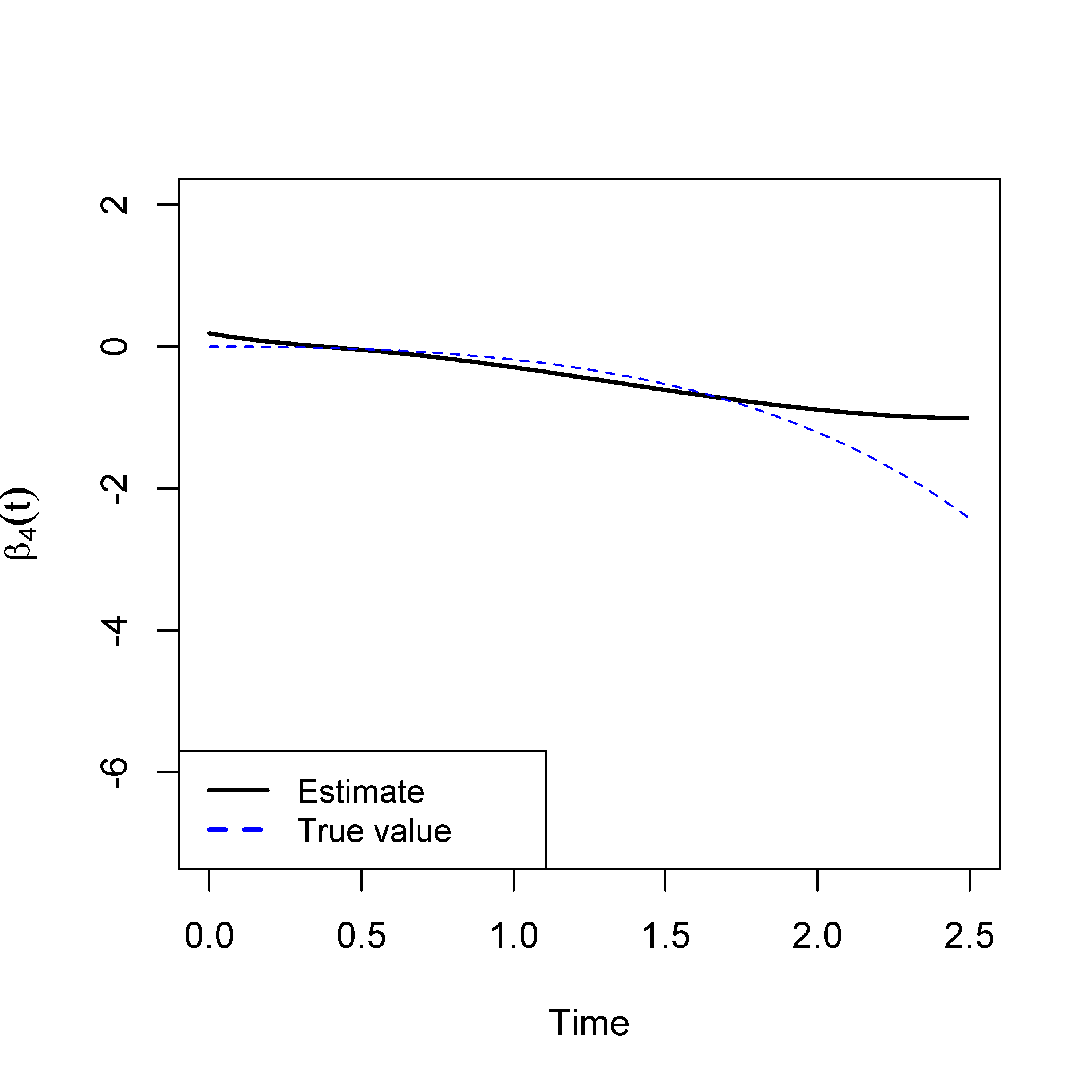}
  }
    \centering
\hspace{0.1pt}
\subfloat[Gradient: m=3,000]{
\includegraphics[width=1.65in, height=1.75in]{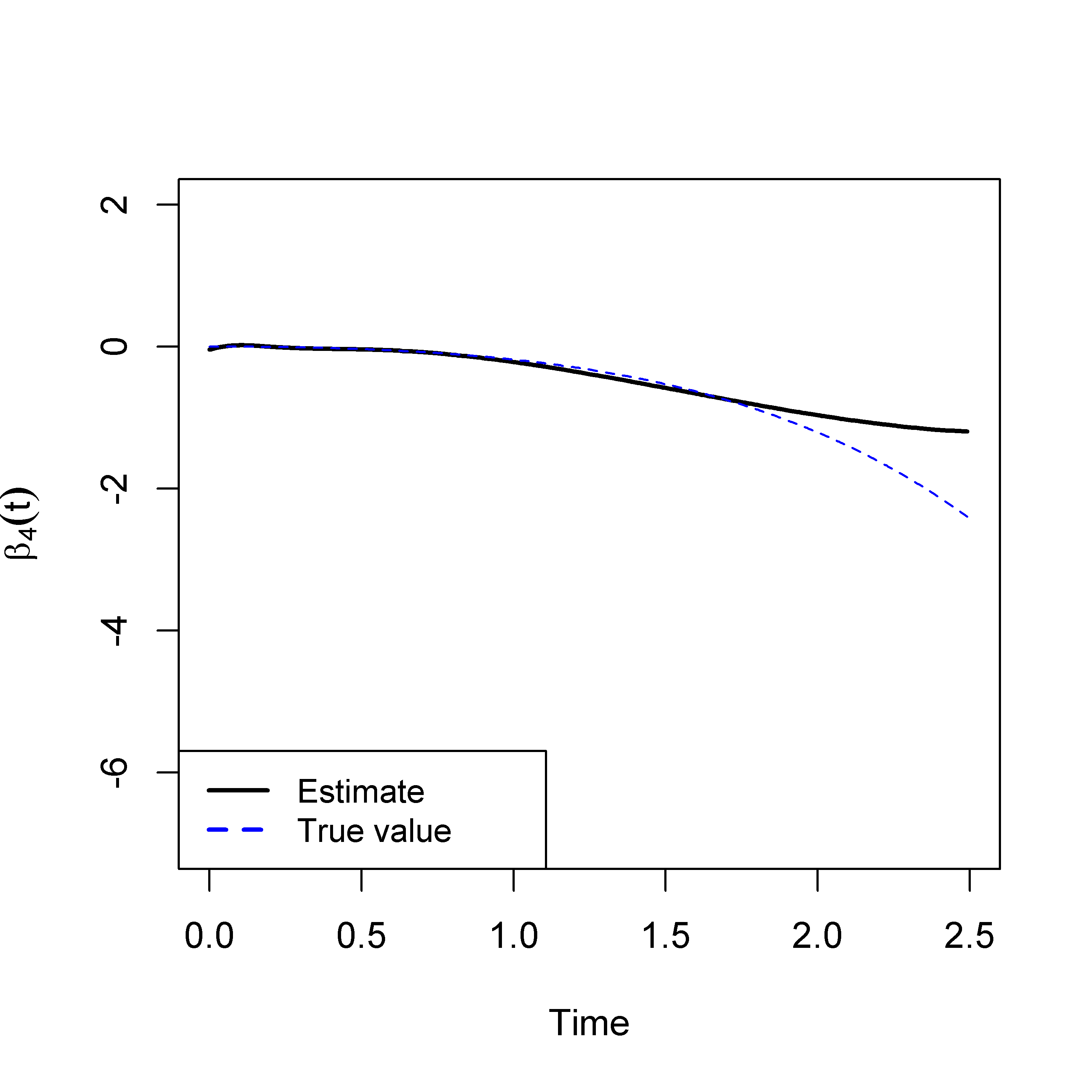}
  }
\end{figure}

\subsection{Numerical Properties}

To derive the numerical properties for the MMSA, we impose the
 following regularity conditions:
\begin{enumerate}
  \item[(A)] For any initial value $\boldsymbol\theta^{(0)}$, the block diagonal matrix, $\bH(\boldsymbol\theta)$, is positive definite in the super-level set
$
 \{\boldsymbol\theta: \ell(\boldsymbol\theta) \geq \ell(\boldsymbol\theta^{(0)})\}.
$
\item[(B)]  The negative log-partial likelihood function is coercive; i.e.,
  $
      \lim_{||\boldsymbol\theta||_2 \rightarrow \infty} - \ell (\boldsymbol\theta)=\infty. %\nonumber
$
\end{enumerate}

  Condition (A) guarantees the existence of the MMSA update and is satisfied in most practical applications.
The coercive assumption \cite{r23} defined in Condition (B) and the fact that the log-partial likelihood is upper bounded guarantee that
 the super-level set
  is compact. Therefore, the maximum value of $\ell(\boldsymbol\theta)$ is attained, e.g. Weierstrasss's theorem  \cite{r23}. The existence of a cluster point of the MMSA is also guaranteed by the compactness. 
  
  We now show that there exists a learning rate $\nu$ such that the proposed
algorithm satisfies the ascent property, which guarantees that the iterative estimates in each MMSA step serve as refinements of the previous step.
Let $\lambda_{max}(\cdot)$ represent the largest eigenvalues of an arbitrary non-negative definite matrix. At each MMSA iteration, given the current estimates $\boldsymbol{\widehat\theta}^{(m-1)}$, 
let
$\boldsymbol{\widetilde\theta}$ be a vector that lies between $\boldsymbol\theta$ and $\boldsymbol{\widehat\theta}^{(m-1)}$ such that 
\begin{align*}
\ell(\boldsymbol\theta) =  \ell(\widehat{\boldsymbol\theta}^{(m-1)}) + \triangledown\ell(\widehat{\boldsymbol\theta}^{(m-1)})^T (\boldsymbol\theta-\widehat{\boldsymbol\theta}^{(m-1)})+\frac{1}{2}(\boldsymbol\theta-\widehat{\boldsymbol\theta}^{(m-1)})^T \triangledown^2 \ell(\boldsymbol{\widetilde\theta}) (\boldsymbol\theta-\widehat{\boldsymbol\theta}^{(m-1)}).
\end{align*}

\textbf{Proposition 1 (Ascent Property)}
{\it
 Assume Condition (A) holds. For $\nu>0$ satisfying
\begin{align}
\lambda_{max}\left(\{\bH(\boldsymbol{\widehat\theta}^{(m-1)})\}^{-1/2}\{- \triangledown^2 \ell(\boldsymbol{\widetilde\theta})\}\{\bH(\boldsymbol{\widehat\theta}^{(m-1)})\}^{-1/2}\right) < 1/\nu,
\label{eq:rate}
\end{align}
we have
 \begin{align*}
 g(\boldsymbol\theta|\widehat{\boldsymbol\theta}^{(m-1)}) \leq  \ell(\boldsymbol\theta) ~~\mbox{for all}~~ \boldsymbol\theta.
\end{align*}
   }
   
Proposition 1 shows that  $g(\boldsymbol\theta|\widehat{\boldsymbol\theta}^{(m-1)})$ serves as a minority surrogate function of $\ell(\boldsymbol\theta)$. Thus, the resulting estimates $\widehat{\boldsymbol\theta}^{(m)}$ from the MMSA ensure the ascent property
\begin{align*}
\ell(\widehat{\boldsymbol\theta}^{(m)})\geq g(\widehat{\boldsymbol\theta}^{(m)}|\widehat{\boldsymbol\theta}^{(m-1)}) \geq g(\widehat{\boldsymbol\theta}^{(m-1)}|\widehat{\boldsymbol\theta}^{(m-1)})=\ell(\widehat{\boldsymbol\theta}^{(m-1)}), %\label{eq:ascent}
\end{align*}
and serve as refinements of the previous step.

Proposition 1 also helps clarify when a small learning rate is needed and provides
insights into the choice of norm for \eqref{eq:SD} in practical implementations. For example, for
classical gradient-based procedures, the updates at each iteration are computed based
on gradient information only. To ensure the ascent property for such gradient-based algorithms, the learning rate 
$\nu$ is required to satisfy
$
\lambda_{max}\left(\{- \triangledown^2 \ell(\boldsymbol{\widetilde\theta})\}\right) < 1/\nu.
$
This may explain the previous finding that  the learning rate $\nu$ is typically chosen to be sufficiently small for  classical gradient boosting to ensure better predictive and estimation accuracy. However,
a small value of $\nu$ requires a large number of boosting iterations and thus more computing
time, especially when the condition numbers of the observed information matrix are large as exemplified in the estimation of time-varying effects.  In contrast, the proposed MMSA is less sensitive
to the choice of learning rate and
substantially
improves the computational efficiency.
As shown in Figure 3,  compared with gradient-based procedure, the proposed method  achieves much computational efficiency and more accurate updates.

%The numerical convergence of the MMSA algorithm is summarized below.

\textbf{Proposition 2 (Numerical Convergence)}
{\it
 Assume the same condition on the learning rate as in Proposition 1. Suppose Conditions (A) and (B) hold. Then every cluster point of the iterates $\boldsymbol{\widehat\theta}^{(m)}=M(\boldsymbol{\widehat\theta}^{(m-1)})$ generated by the iteration map $M(\boldsymbol\theta)$ of the MMSA algorithm is a stationary point of $\ell(\boldsymbol\theta)$. Furthermore, the set of stationary points $\mathcal{F}$ is closed and the limit of the distance function is zero:
   \begin{align*}
 \lim_{m\rightarrow\infty}\inf_{\boldsymbol\theta \in \mathcal{F}} ||\boldsymbol{\widehat\theta}^{(m)} - \boldsymbol\theta||_2=0.
   \end{align*}
Moreover, if the observed information matrix $-\triangledown^2 \ell(\boldsymbol\theta)$ is positive definite in the super-level set 
defined in Condition (A),
any sequence of $\boldsymbol{\widehat\theta}^{(m)}$ possesses a limit, and this limit is a stationary point of $\ell(\boldsymbol\theta)$.
}

\subsection{Testing for Time-Varying Effects}

To assign p-values to determine whether the covariate effects are time-varying,
we explore the following property of B-splines: if
$\theta_{p1}=\cdots=\theta_{pK}=\theta, $
the corresponding covariate effect is time-independent; e.g.,
\begin{eqnarray}
\beta_p(t)=\sum_{k=1}^K\theta_{pk} B_k(t)=\theta.  \nonumber
\end{eqnarray}
 Specify a matrix $\bC_p$ such that
 $\bC_p\boldsymbol\theta=\textbf{0}$ corresponds to the contrast that $
 \theta_{p1}=\cdots=\theta_{pK}$. A Wald test can then be constructed by
 \begin{eqnarray}
 (\bC_p\boldsymbol{\widehat\theta})^T (\bC_p (- \triangledown^2 \ell(\boldsymbol{\widehat\theta}))^{-1}\bC_p^T)^{-1}(\bC_p\boldsymbol{\widehat\theta}),
   \nonumber
\end{eqnarray}
 where $\boldsymbol{\widehat\theta}$ is obtained through the proposed MMSA.

 In the kidney transplant database, however, computation of the observed information matrix outpowers a computer with an 32G memory.
 Numerically, the gradients are much easier to compute. Therefore, we consider the following test
\begin{align*}
 S_p=(\bC_p\boldsymbol{\widehat\theta})^T (\bC_p \bV^{-1}(\boldsymbol{\widehat\theta})\bC_p^T)^{-1}(\bC_p\boldsymbol{\widehat\theta}),
\end{align*}
 where
   \begin{align*}
 \bV(\boldsymbol{\widehat\theta})=\sum_{j=1}^J\sum_{i=1}^{n_j}\Psi_{ij} (\boldsymbol{\widehat\theta}) \Psi_{ij} (\boldsymbol{\widehat\theta})^T
   \end{align*}
   is an approximation of the empirical information matrix, with
\begin{align*}
\Psi_{ij} (\boldsymbol{\widehat\theta})= \delta_{ij} \left \{ \bX_{ij}- \overline{\textbf{Z}}_{ij} (\widehat{\boldsymbol\theta}, T_{ij}) \right \} \otimes \textbf{B}(T_{ij}).
\end{align*}
Here $\otimes$ is the Kronecker product and
\begin{align*}
\overline{\textbf{Z}}_{ij}(\widehat{\boldsymbol\theta}, T_{ij})= \frac{\sum_{i' \in R_{ij}}
 \textbf{X}_{i' j}\exp\{\textbf{X}_{i' j}^T\widehat{\boldsymbol\Theta} \textbf{B}(T_{ij})\}}{\sum_{i'
 \in R_{ij}}  \exp\{\textbf{X}_{i' j}^T\widehat{\boldsymbol\Theta} \textbf{B}(T_{ij})\}}.
\end{align*}
Under the null hypothesis that the covariate effect is time-independent, the statistics $S_p$ is asymptotically chi-square
 distributed with $K-1$ degree of freedom.

%Our numerical experiments indicate that a sufficiently small learning rate (e.g. 0.001) is typically needed for gradient-based method  to correctly model the time-varying effects, which requires a large number of  iterations and thus more computing time. In contrast, with the proposed MMSA, convergence is usually achieved with $\nu=0.05$. Moreover, our result in the Appendix helps clarify when a small learning rate is needed and provides insights into its choice in practical implementations.

\section{Simulations}
\label{s:Simulation}

We consider the following simulation settings:

\begin{itemize}
\item [Setting 1:] Death times were generated from an exponential model with a baseline hazard
0.5.  Censoring times were generated from the Uniform
(0,3) distribution. Continuous predictors were generated from a multivariate normal distribution, where the covariance followed an AR1 with an auto-correlation parameter 0.6.
We varied the number of predictors from 5, 20 to 50.  We chose $\beta_2(t)=sin3(\pi t/4)$ and $\beta_4(t)=-(t/3)^2 \exp(t/2)$ to represent time-varying effects. The remaining covariate effects  were set as constants (e.g. time-independent effects). 

\item [Setting 2:]   To mimic the motivating real data example, binary covariates were generated with specified frequencies between 0.05 and 0.2.  The remaining simulation set-ups were the same as Setting 1.

\item [Setting 3:]   Two continuous covariates were generated with coefficients $\beta_1=1$ and $\beta_2(t)=\gamma\cdot\mbox{sin}(3(\pi t/4))$,
where $\gamma$ varied between 0 and 3, representing the magnitude of the time-varying effects. The remaining simulation set-ups were the same as Setting 1.
\end{itemize}

%Moreover, to separate variables with time-varying effects and those with time-independent effects, a relaxed fitting is implemented such that the MMSA w on the set of selected variables.

%It has been found empirically that stability selection is often substantially better than approaches that do not use the additional sub-sampling procedures.

%In our setting, it is likely that not all non-null covariate effects are time-varying. Therefore,  simultaneous selection between varying effects and time-independent effects in addition to selection between nonzero and zero coefficient presents an additional challenge. Moreover,

\subsection{Evaluation of Computation Speed}
 Table 1 compares the computation time for the proposed MMSA and the Kronecker product-based Newton method (implemented in Rcpp through R package {\it RcppArmadillo}) with step size determined by backtracking line search \cite{r1}, quasi-Newton method (He et al. \cite{r44}; implemented by R function {\it optim}), and the likelihood-based boosting (termed as L-Boost and implemented by R package {\it COX$_{flex}$Boost}).  The simulation set-up was based on Setting 1 with various combinations of sample sizes and numbers of predictors. The convergence criterion was chosen as the maximal absolute change of $\boldsymbol\theta$ or the change of $\ell(\boldsymbol\theta)$ less than $10^{-6}$.  These timings were taken on a HP workstation with 4-core 3.50-GHz Intel Core E5-1620v3 processor and 32GB RAM. When the sample size is very large, no results are reported for the competing methods due to the computation exceeding the computer's maximum memory capacity.

\begin{table}[htbp] %h=here, t=top, b=bottom, p=page
\caption{Computation Time: n=sample size; J=number of centers; P=number of covariates;  NA for L-Boost: no results are reported due to its intensive computation; NA for Newton or quasi-Newton: the computation exceeds the computer's max memory capacity; based on Setting 1.}
\begin{center}
%\vspace{3mm}
\begin{tabular}{rrrcccc}
\hline  \hline
 n & J&  P   &  Newton & quasi-Newton & L-Boost & MMSA  \\
 \hline  \hline
   1,000 & 1 & 10 & 0.08  minutes & 15.43  minutes  & 10.36 hours   & 0.34 minutes \\
      \hline
   10,000 & 10  & 20  &  1.12 minutes & 1.93 hours  & NA    & 7.15 minutes  \\
        \hline

   347,668 & 293  & 164  &  NA & NA  & NA    & 11.64 hours  \\
  \hline  \hline
\end{tabular}
\end{center}
\end{table}

\begin{figure}[h]
 \caption{Estimated coefficients  (solid lines) and 95\% confidence interval (dashed lines) for simulation setting 1; $n=1,000$, $p=10$; $\beta_2(t)=sin3(\pi t/4)$ and $\beta_4(t)=-(t/3)^2 \exp(t/2)$ are time-varying effects; $\beta_1=1$ and  $\beta_3=-1$ are constant effects; K: number of B-spline basis functions; based on Setting 1.}
\centering
\subfloat[Newton (K=5)]{
\includegraphics[width=1.35in, height=1.65in]{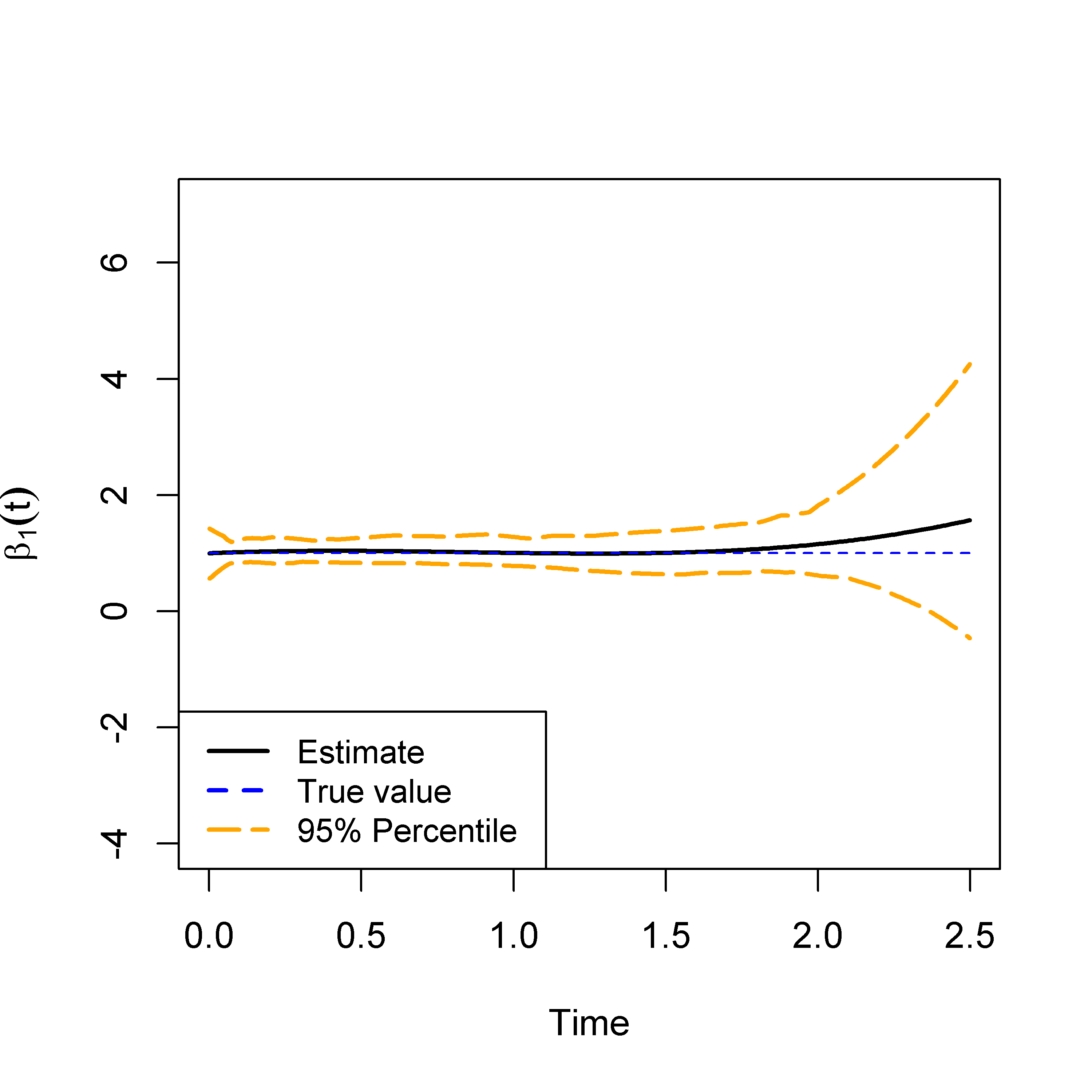}
  }
    \centering
\hspace{0.1pt}
\subfloat[Newton (K=10)]{
\includegraphics[width=1.35in, height=1.65in]{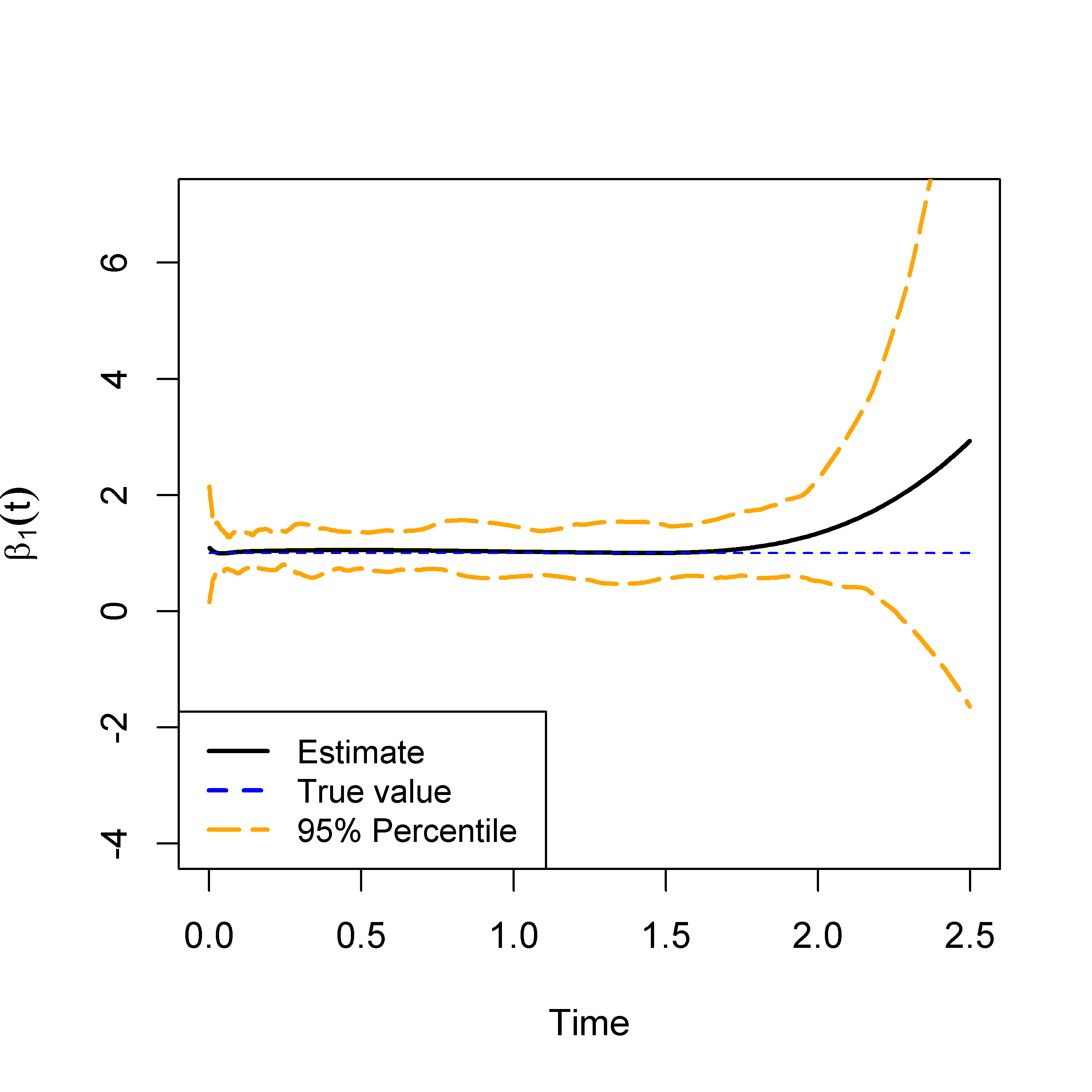}
  }
     \centering
\hspace{0.1pt}
\subfloat[MMSA (K=5)]{
\includegraphics[width=1.35in, height=1.65in]{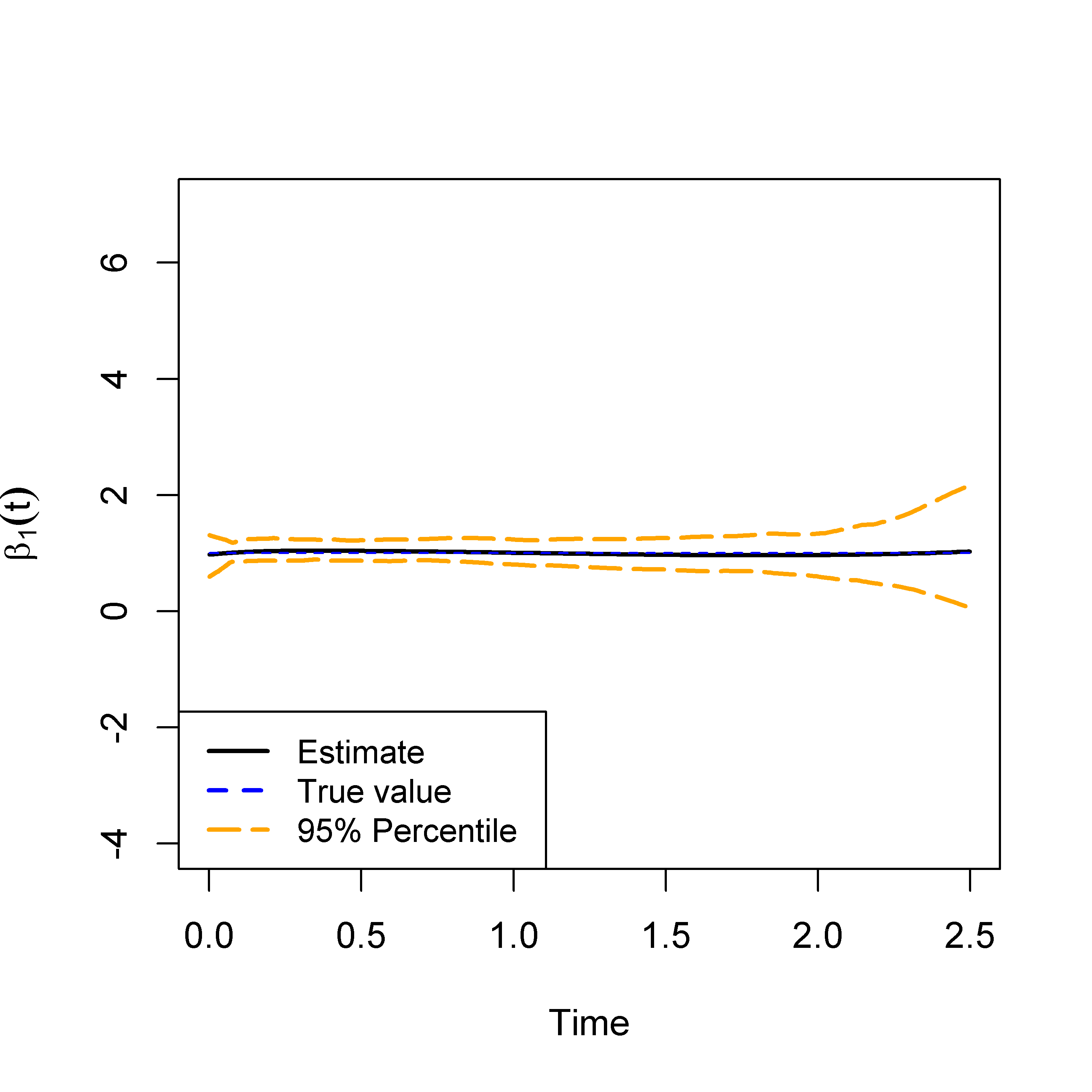}
  }
    \centering
\hspace{0.1pt}
\subfloat[MMSA (K=10)]{
\includegraphics[width=1.35in, height=1.65in]{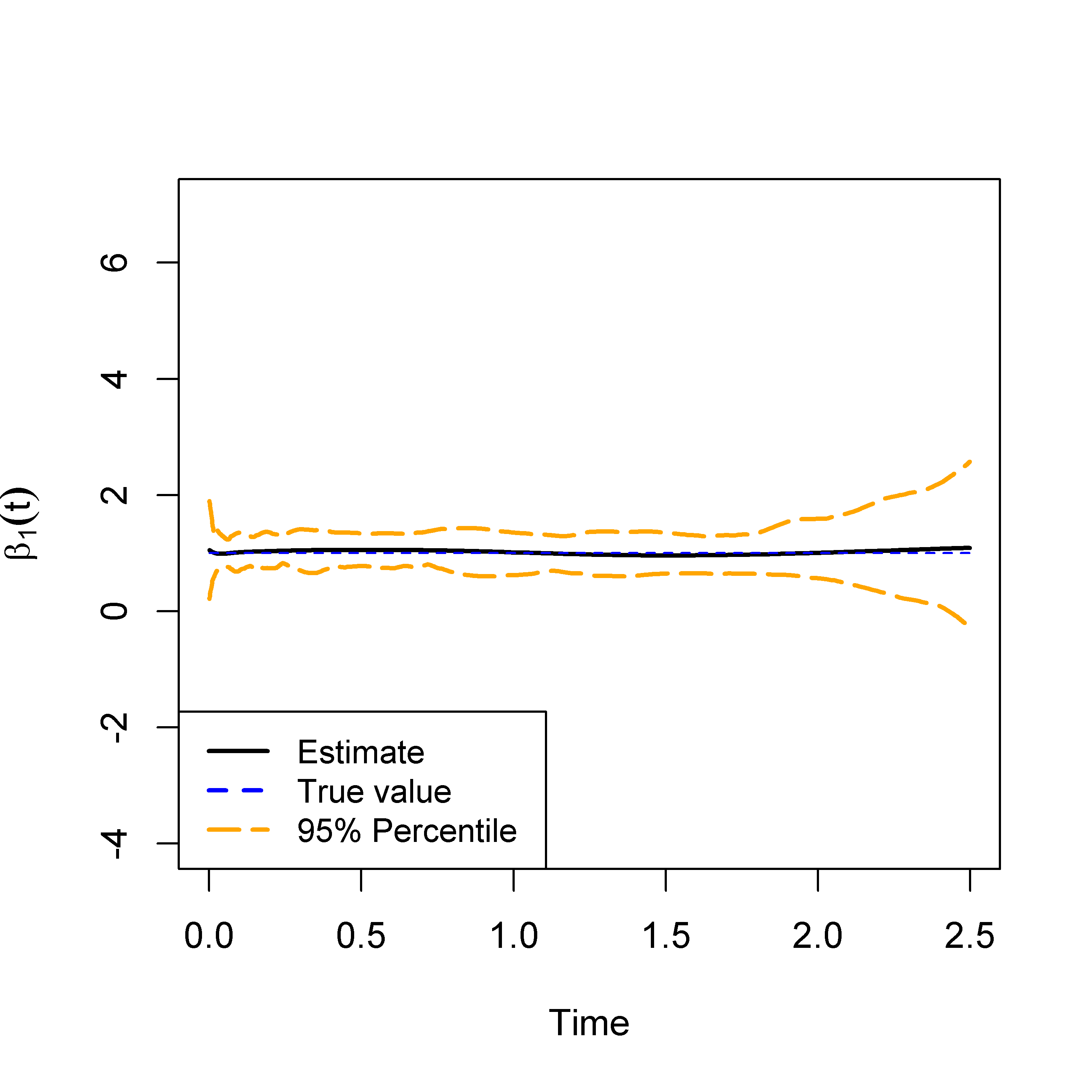}
  }
      \centering
     \vspace{1pt}
\subfloat[Newton (K=5)]{
\includegraphics[width=1.35in, height=1.65in]{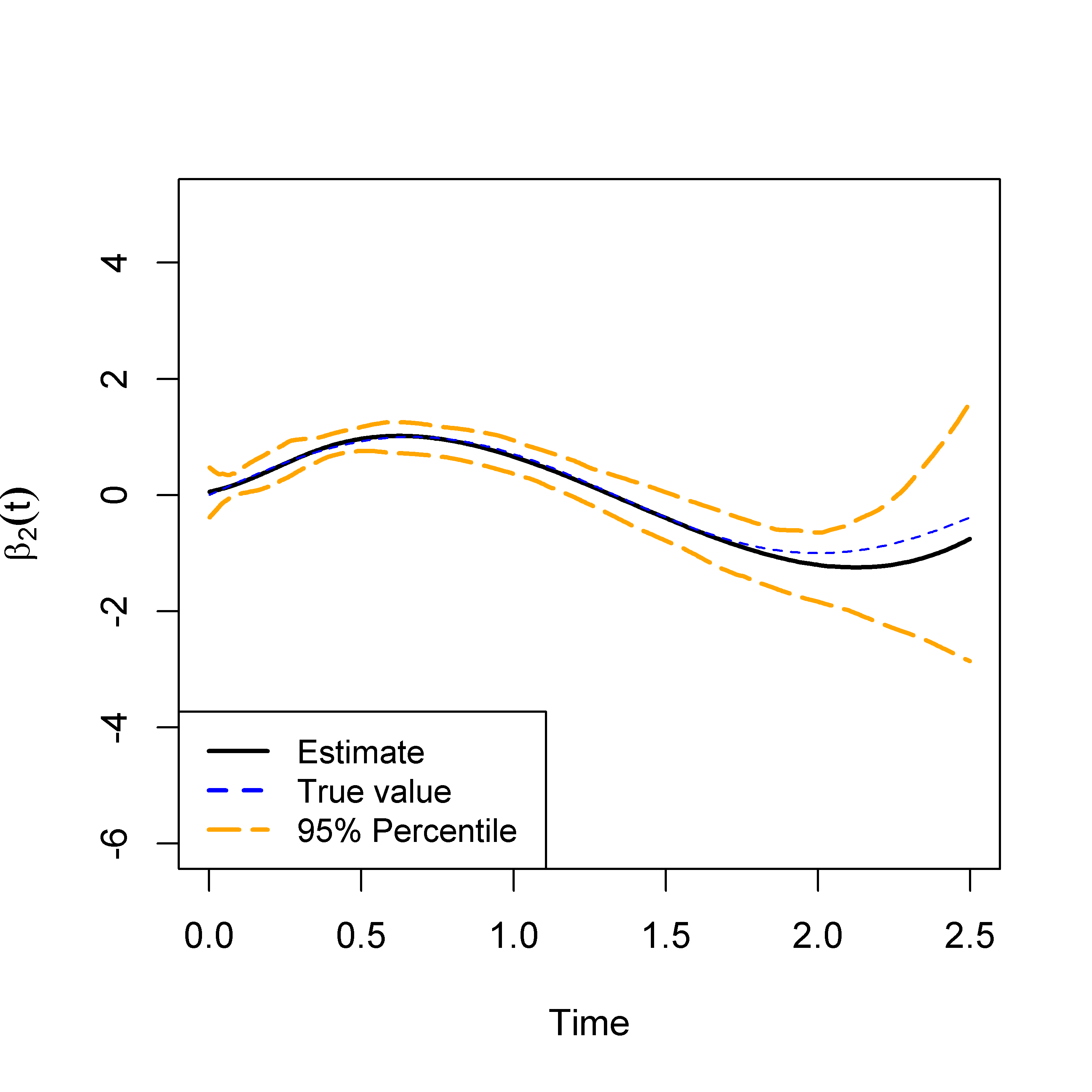}
  }      \centering
\hspace{0.1pt}
\subfloat[Newton (K=10)]{
\includegraphics[width=1.35in, height=1.65in]{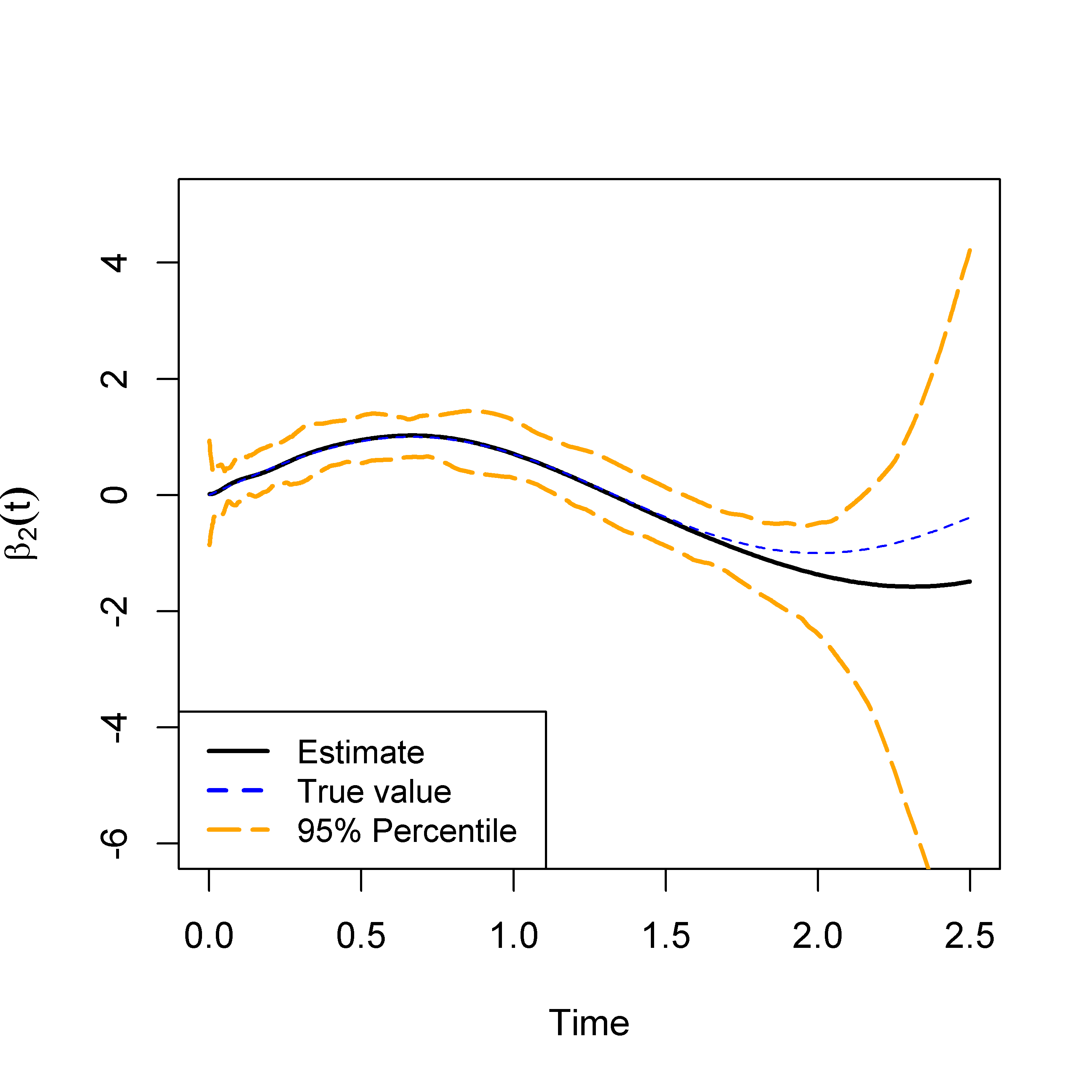}
  }
        \centering
\hspace{0.1pt}
\subfloat[MMSA (K=5)]{
\includegraphics[width=1.35in, height=1.65in]{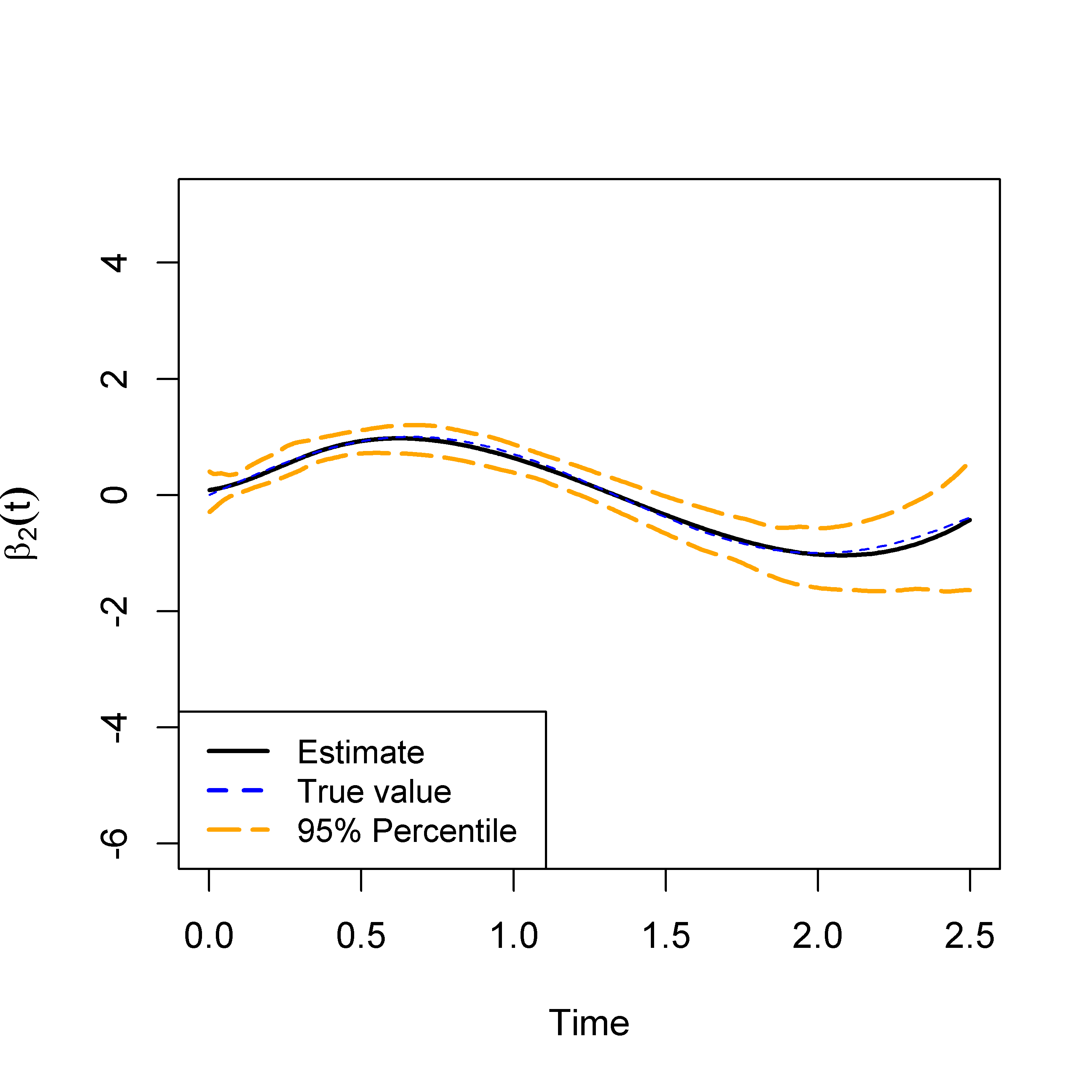}
  }      \centering
\hspace{0.1pt}
\subfloat[MMSA (K=10)]{
\includegraphics[width=1.35in, height=1.65in]{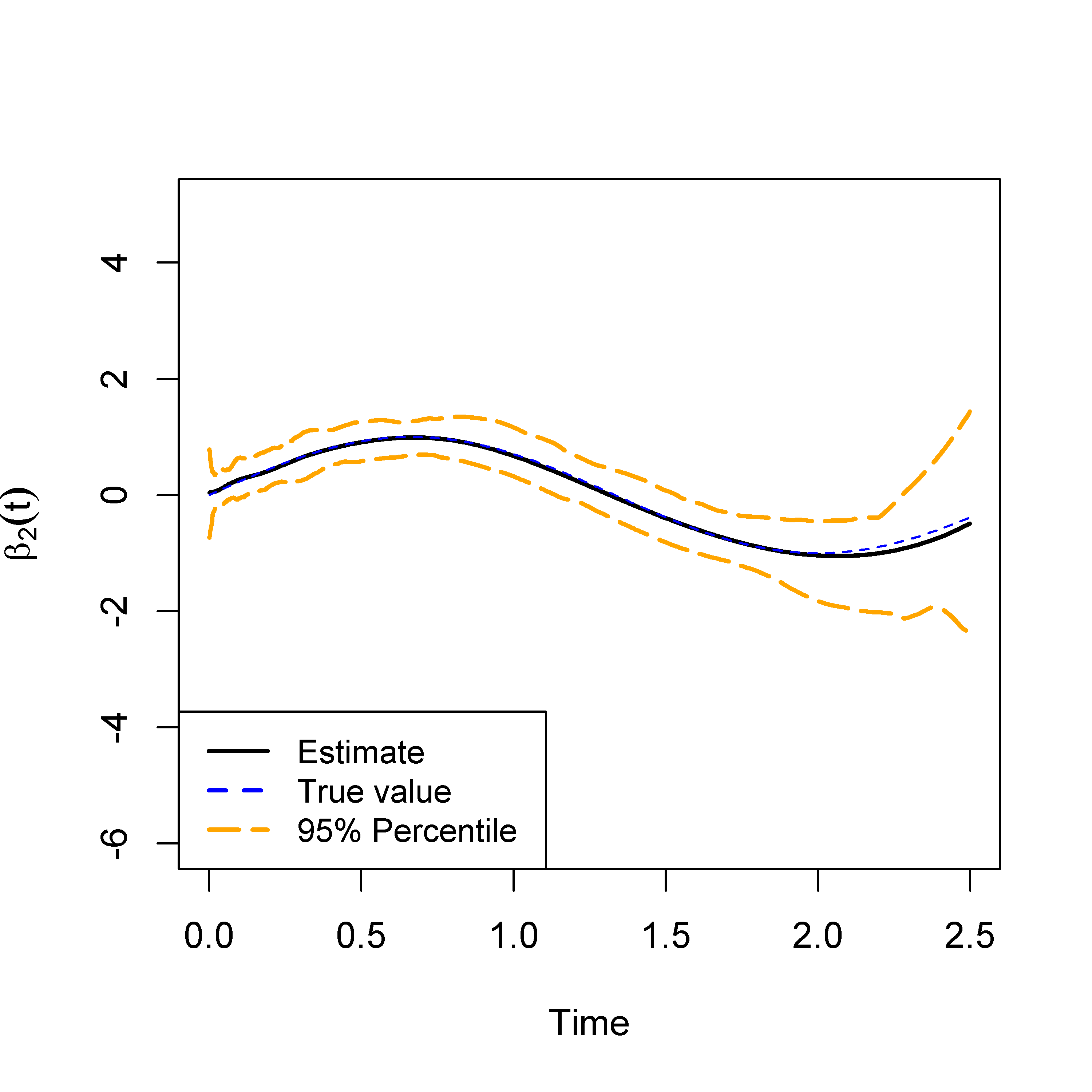}
  }
  \centering
  \vspace{1pt}
\subfloat[Newton (K=5)]{
\includegraphics[width=1.35in, height=1.65in]{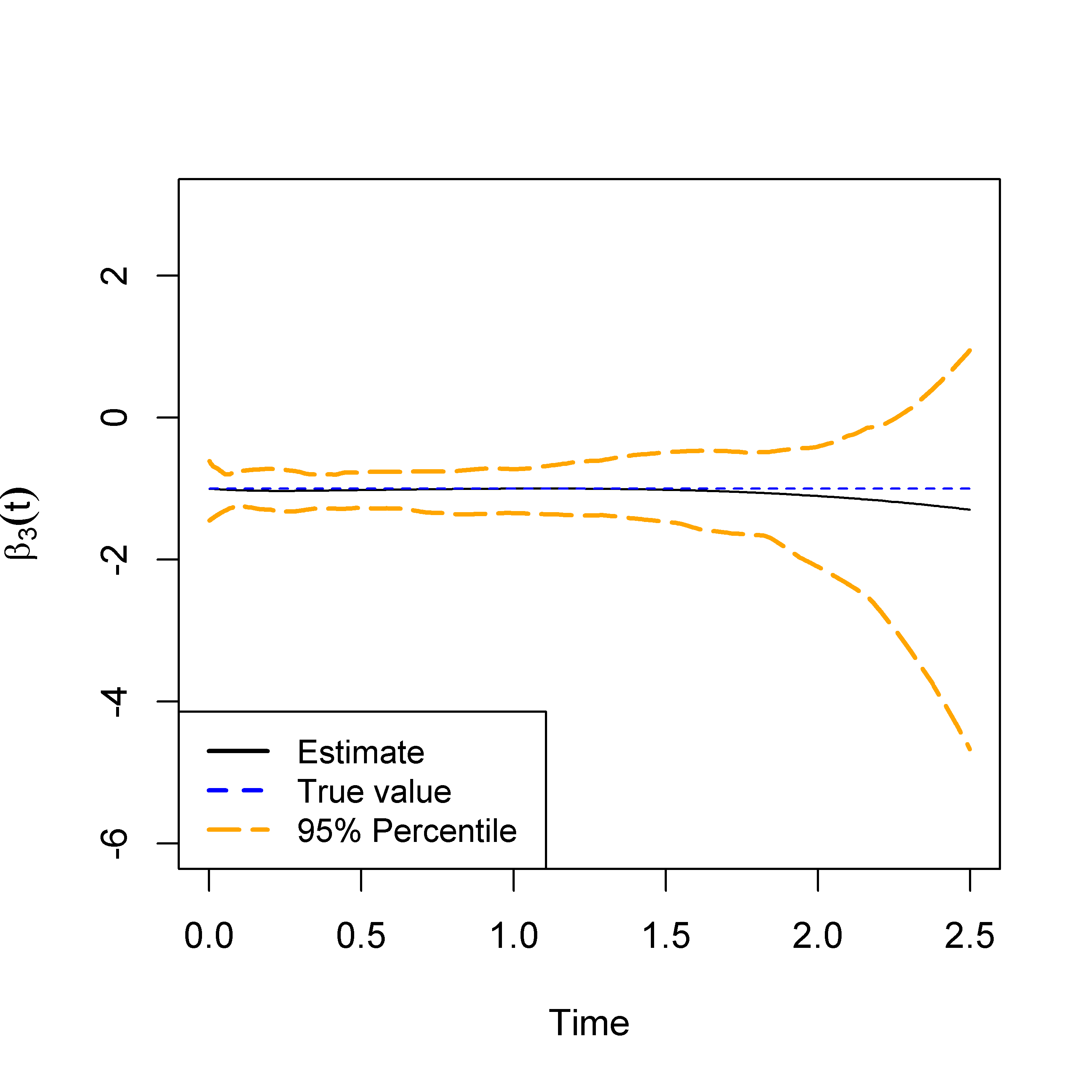}
  }
    \centering
\hspace{0.1pt}
\subfloat[Newton (K=10)]{
\includegraphics[width=1.35in, height=1.65in]{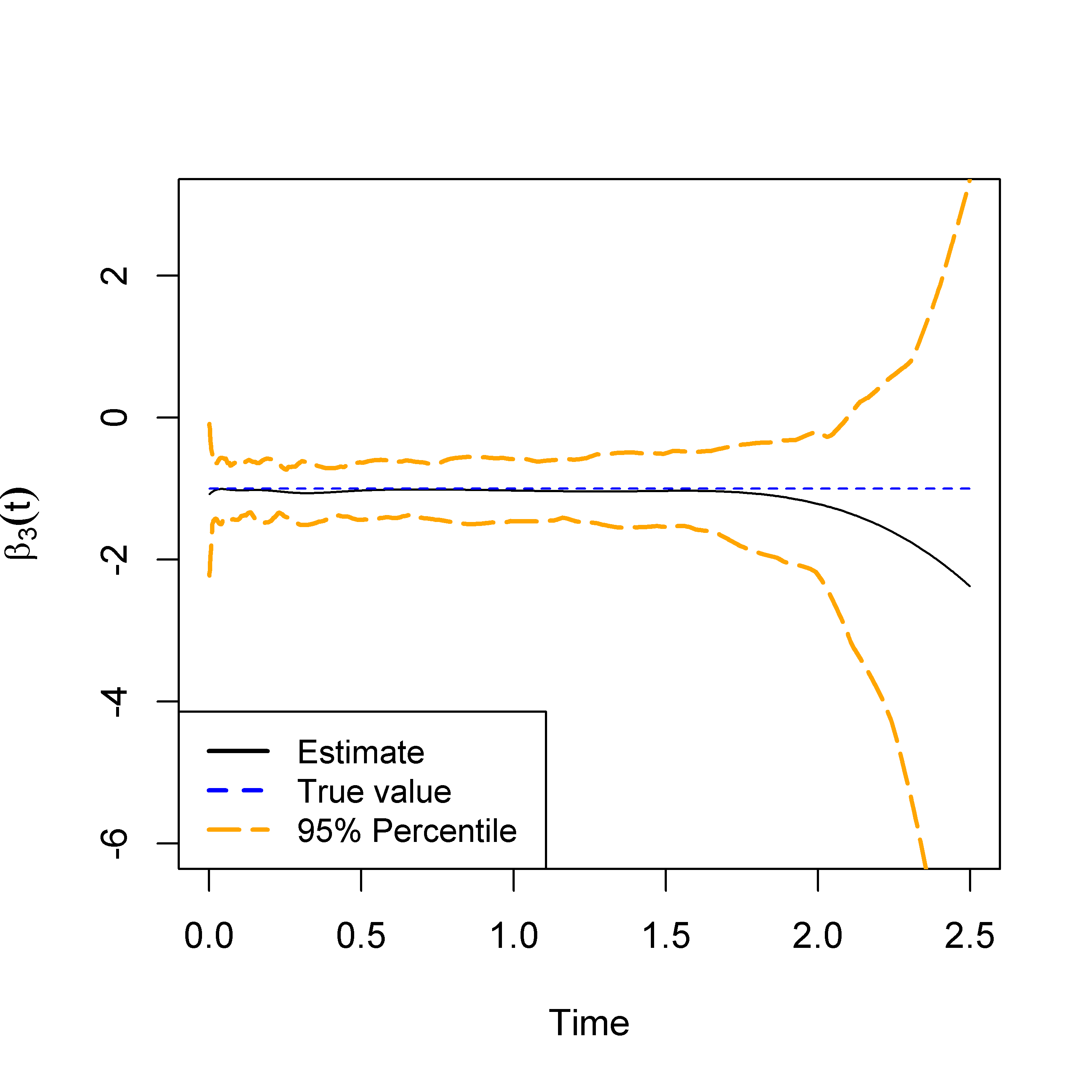}
  }
     \centering
 \hspace{0.1pt}
\subfloat[MMSA (K=5)]{
\includegraphics[width=1.35in, height=1.65in]{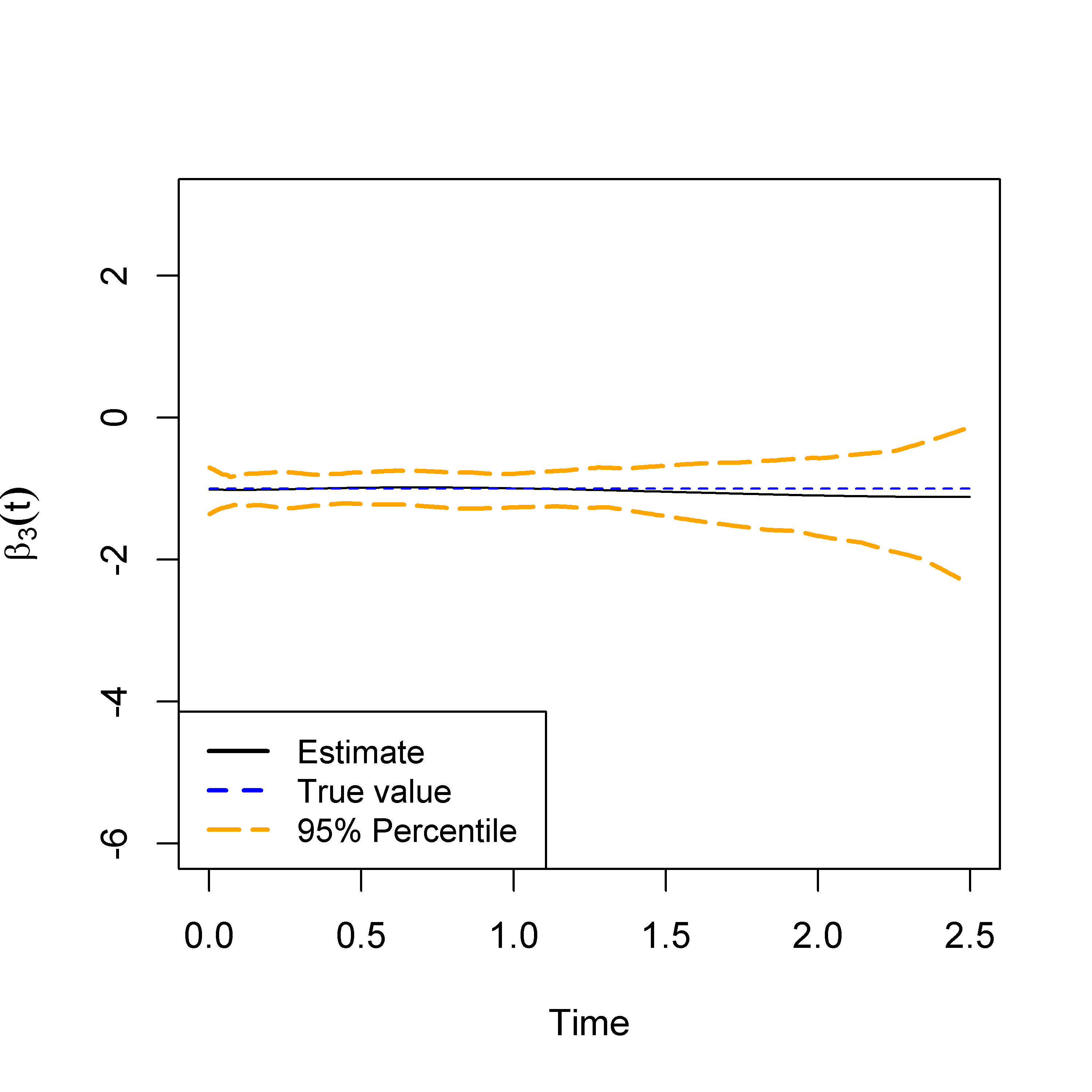}
  }
    \centering
\hspace{0.1pt}
\subfloat[MMSA (K=10)]{
\includegraphics[width=1.35in, height=1.65in]{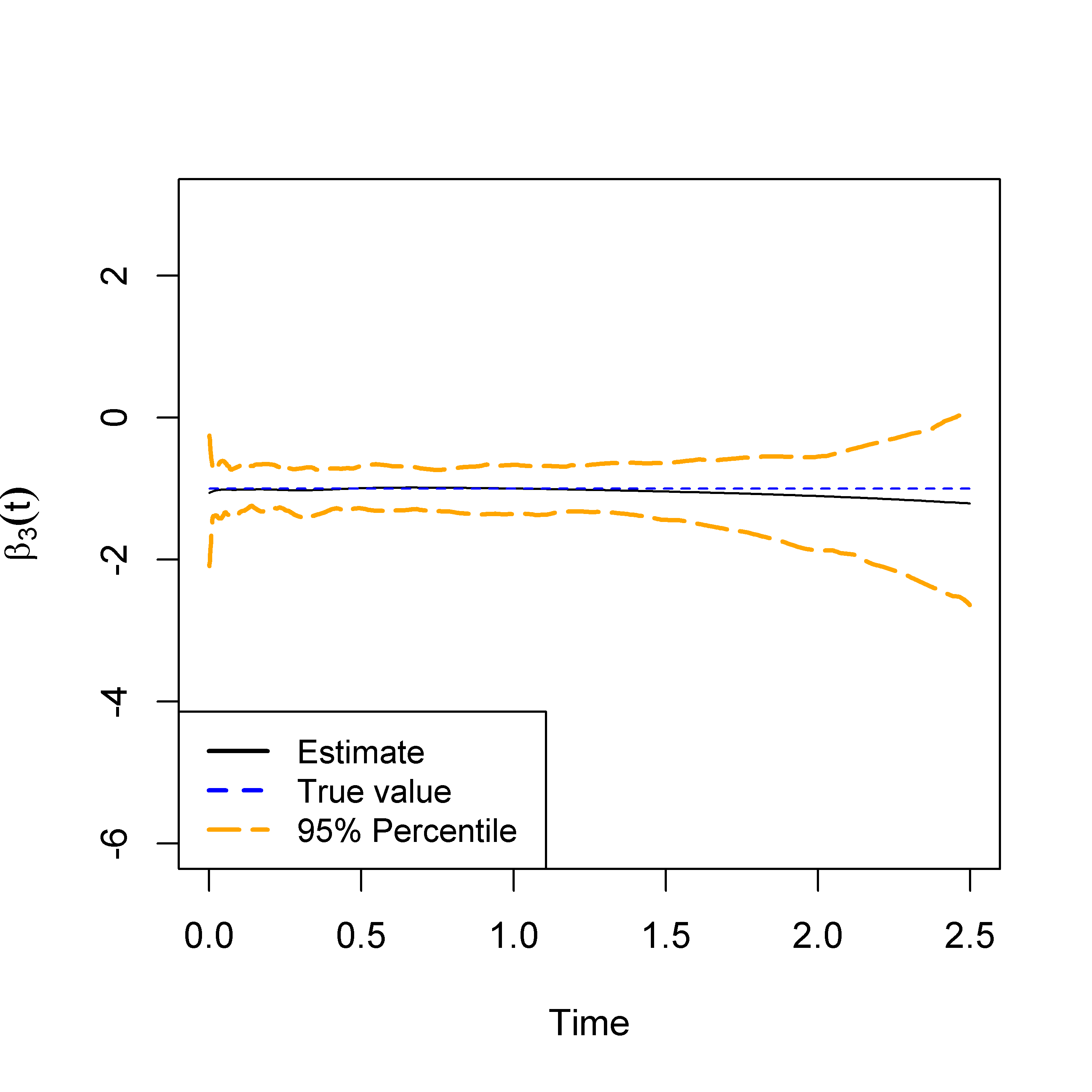}
  }
      \centering
 \vspace{1pt}
\subfloat[Newton (K=5)]{
\includegraphics[width=1.35in, height=1.65in]{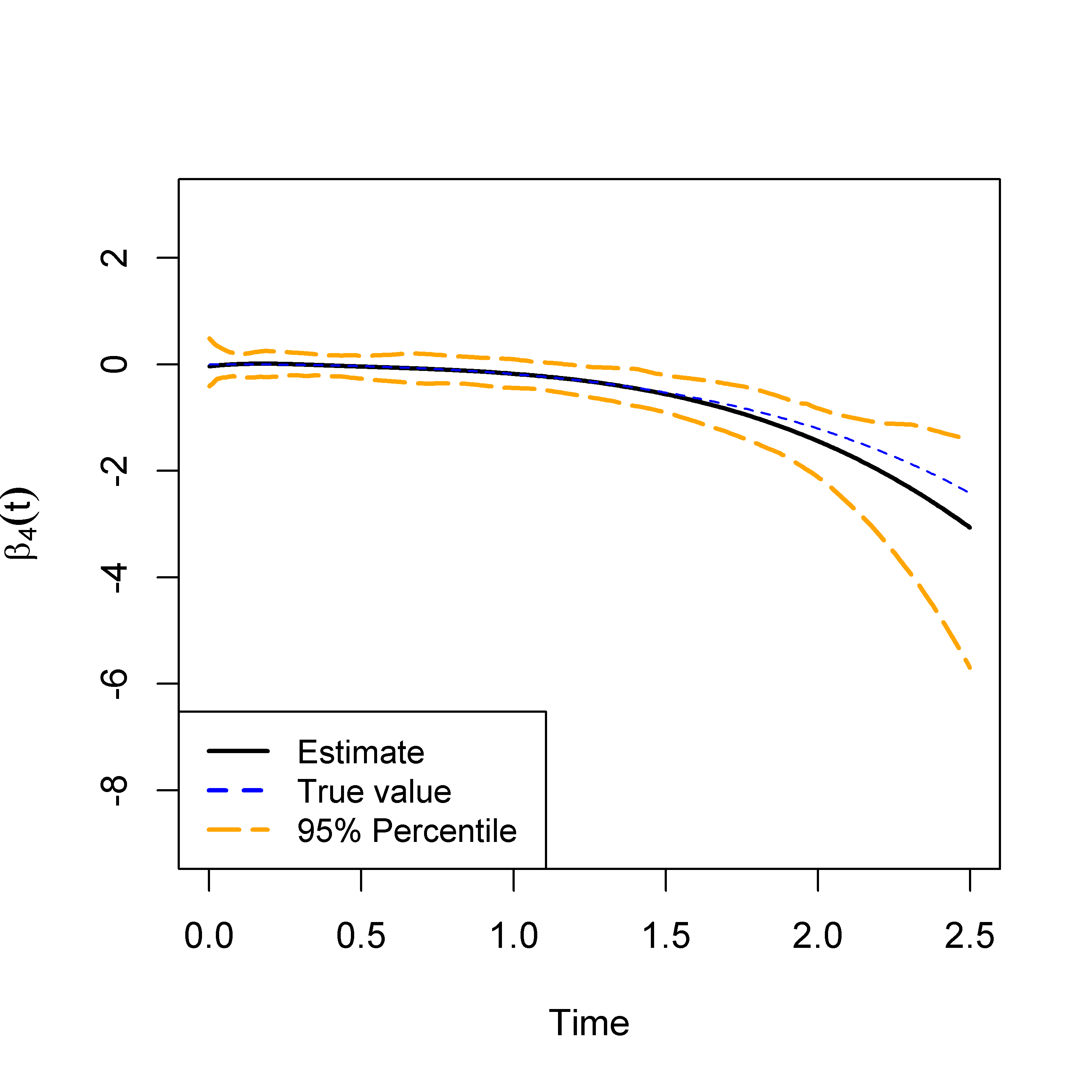}
  }
   \centering
     \hspace{0.1pt}
\subfloat[Newton (K=10)]{
\includegraphics[width=1.35in, height=1.65in]{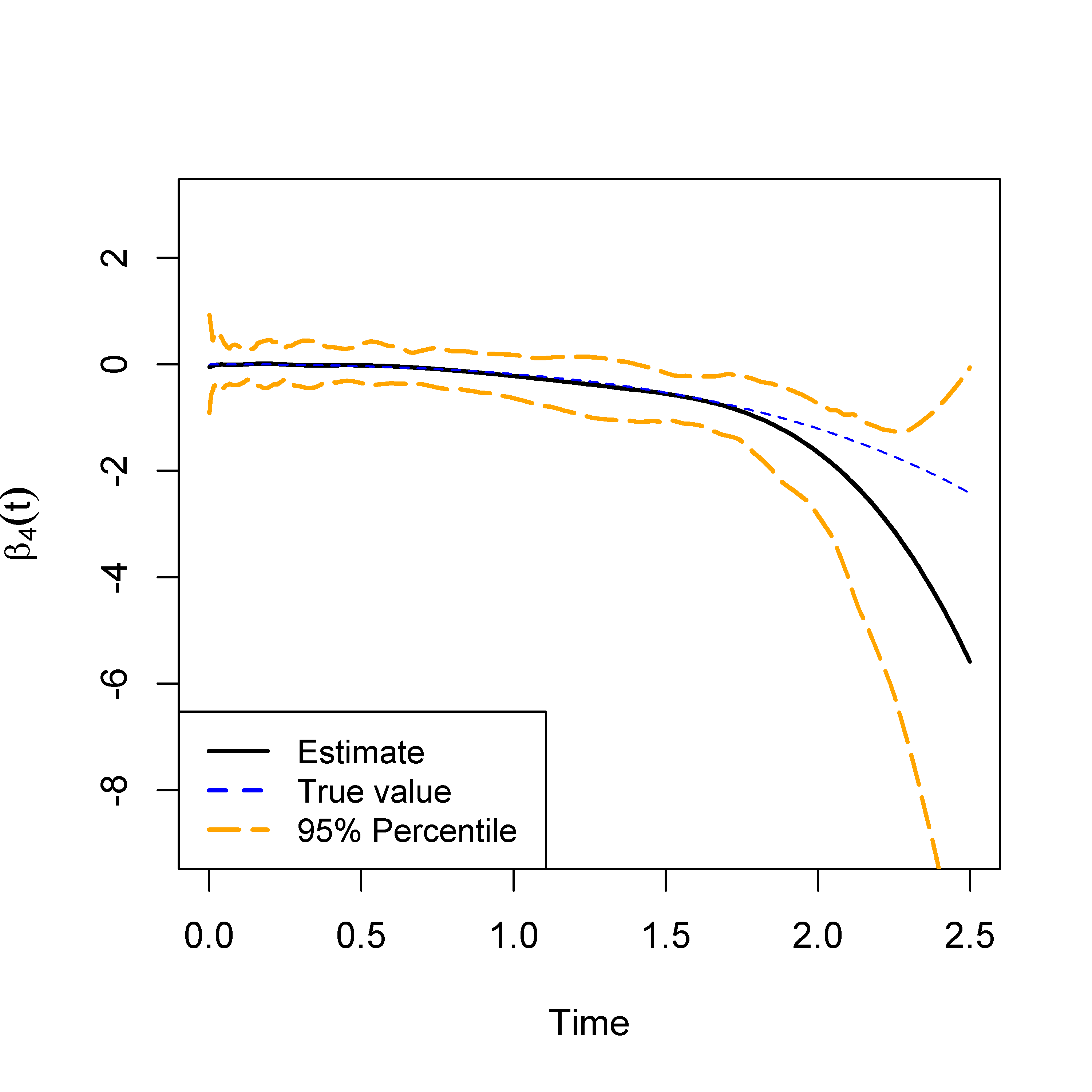}
  }
        \centering
\hspace{0.1pt}
\subfloat[MMSA (K=5)]{
\includegraphics[width=1.35in, height=1.65in]{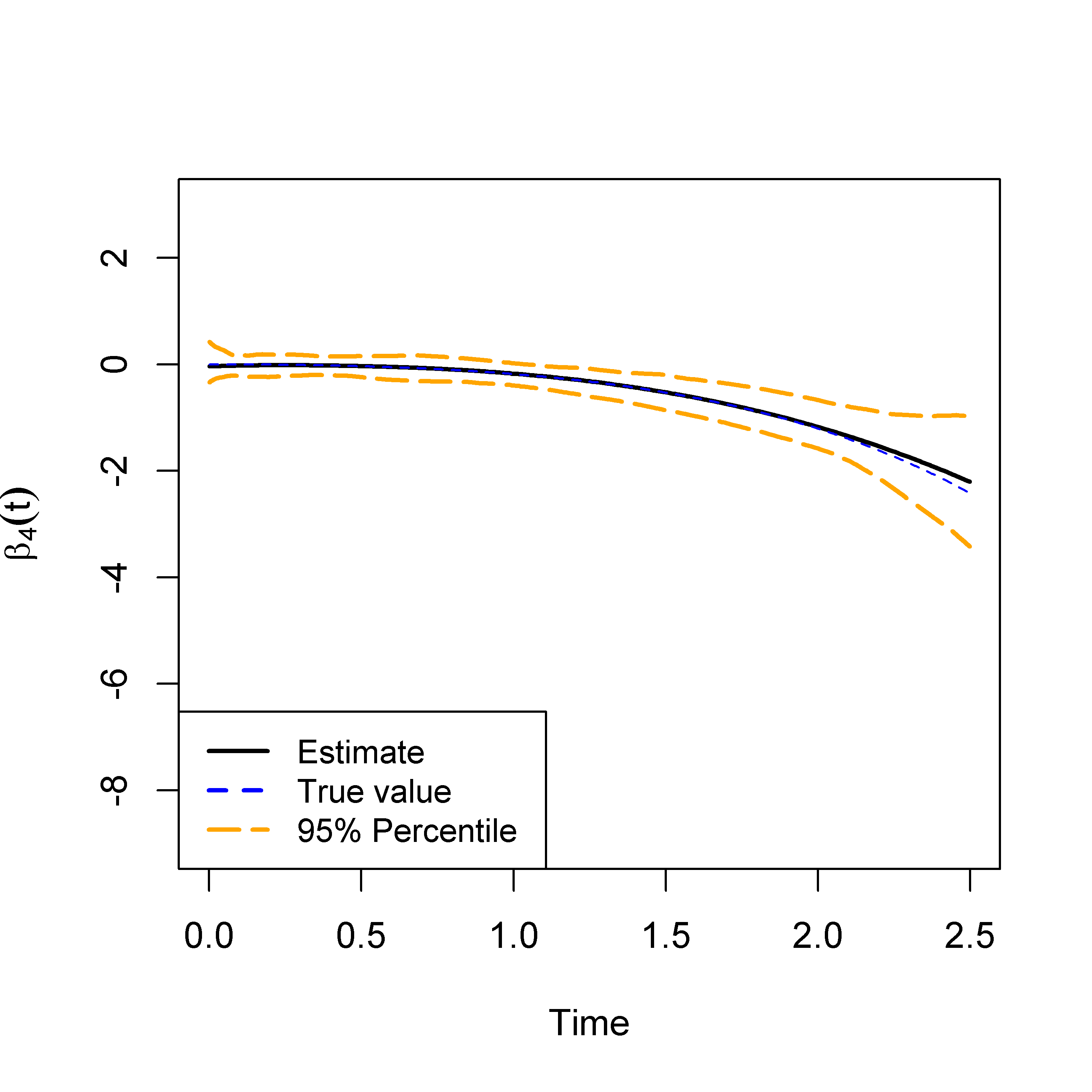}
  }      \centering
 \hspace{0.1pt}
\subfloat[MMSA (K=10)]{
\includegraphics[width=1.35in, height=1.65in]{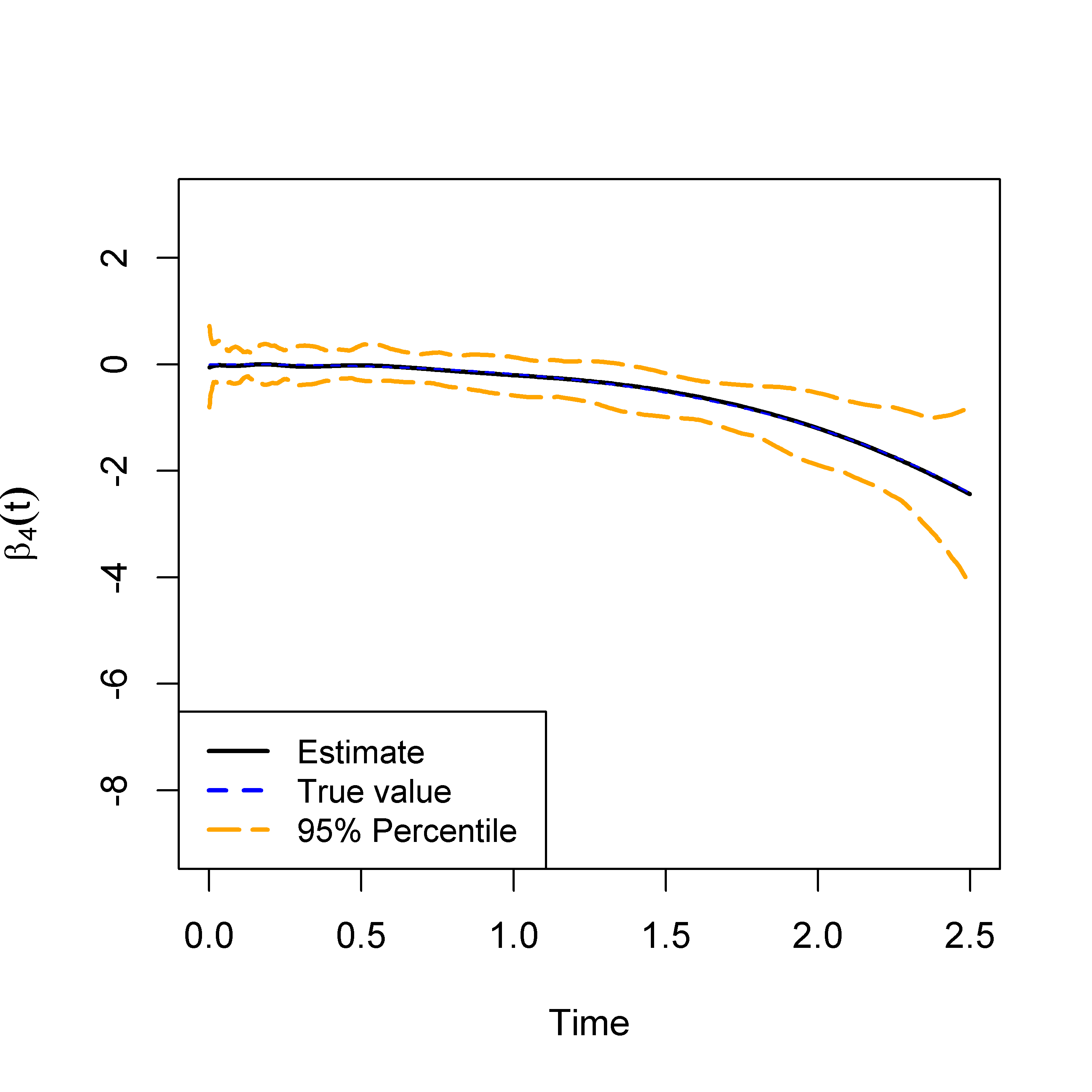}
  }
\end{figure}

\subsection{Estimation of Time-Varying Effects}

%\textbf{Setting 1: Moderate Sample Size (N=1,000 and=100)}

%We first considered a simulation setting  with moderate sample size. Empirically, the best strategy appears to be a two-stage approach. In the first stage, the estimation of the optimal stopping iteration, $M_{stop}$, was determined by cross-validation. Early stopping of MMSA procedure plays an important role in this stage to prevent overfitting and control false positives. Then, in the second stage, to separate variables with time-varying effects and those with time-independent effects, only the subset of covariates selected from stage 1 are considered, and the MMSA with larger iterations is used on the selected subset to improve estimations. Such a strategy is therefore intimately connected with the relaxed Lasso \cite{Meinshausen2007}. It is also worth pointing out that in stage 2 computing the relaxed estimators is often very quick due to the dimension deduction. Figure 1 depicts the estimates for both time-varying and time-independent effects. The proposed algorithm works well in this setting.

\begin{table}[htbp] %h=here, t=top, b=bottom, p=page
	\caption{Average computation time (in seconds), average estimation error (Bias) and average integrated mean square error (IMSE)
	for various methods; based on Setting 2.}
	\begin{center}
		%\vspace{3mm}
		\begin{tabular}{cclrrr}
			\hline  \hline
	P & &   Method   &  Time &  Bias &  IMSE   \\
			\hline  \hline
	\multirow{5}{*}{5}  &&	Newton & 0.22    & 0.646 & 0.592\\
 &&	Coordinate Ascent & 356.09    & 0.810 & 0.632\\
  &&	Gradient Ascent & 169.33    & 0.249 & 0.202\\
   &&	Stochastic Gradient Ascent & 183.51    & 0.290 & 0.256\\
    &&	MMSA & 25.60    & 0.156 & 0.136\\
			\hline \hline 
	\multirow{5}{*}{20}  &&	Newton & 1.10    & 0.305 & 0.169\\
 &&	Coordinate Ascent & 1260.24    & 0.184 & 0.186\\
  &&	Gradient Ascent &687.15    & 0.136 & 0.058 \\
   &&	Stochastic Gradient Ascent & 415.43   & 0.140 & 0.070\\
    &&	MMSA &43.48    & 0.075 & 0.055\\
			\hline \hline
	\multirow{5}{*}{50}  &&	Newton & 9.05    & 0.147 & 0.086\\
 &&	Coordinate Ascent & 2760.26    & 0.807 & 0.224\\
  &&	Gradient Ascent & 1620.21    &0.150 & 0.050\\
   &&	Stochastic Gradient Ascent & 757.07    & 0.118 & 0.038\\
    &&	MMSA & 95.27    & 0.064 & 0.030\\
			\hline \hline			
		\end{tabular}
	\end{center}
\end{table}
 
% We generated 10 variables from a multivariate normal distribution with a first-order auto-regressive (AR1) parameter of 0.6. The total sample size was 1,000. The remaining simulation set-up was similar to that in Section 3.1.

Table 2 compares the average computation time, average estimation errors and
average integrated mean square error (IMSE)
for the Newton approach, the coordinate ascent approach \cite{r8}, the gradient ascent, and the stochastic gradient ascent with  step size determined by Adagrad algorithm \cite{r32}. The reported bias and IMSE were the averages of point-wise estimates over simulated time points. The simulation set-up was based on Setting 2 with sample size 10,000 and various numbers of predictors. For each configuration, a total of 100 independent data were generated. As shown in Table 2, the Newton and coordinate ascent  approaches
suffer from large estimation biases and IMSE. %The block-wise steepest ascent  substantially improved the computational efficiency, but its estimation errors were unstable and sometimes larger than the Newton approach. 
The gradient ascent, and the stochastic gradient ascen substantially improve the estimation errors, but they %occasionally 
suffer from slow convergence.
In contrast, the proposed MMSA is computationally efficient
and achieved the smallest estimation biases in all scenarios. Figure 3 further compares the average estimated coefficients across various iterations of the proposed method and the gradient ascent. Compared with gradient-based procedure, the proposed method  achieves much computational efficiency and more accurate updates.
Figure 4 compares the average estimates and the $95\%$ empirical percentiles over 100 simulation replications for the conventional Newton approach and the MMSA algorithm.  We varied the number of  basis functions from 5 to 10. The simulation set-up was based on Setting 1 with 10 variables. The poor performance of the Newton can be explained in part as follows:
 in the late stage of the follow-up period, the at-risk set is small, causing unstable estimation.  In contrast, the proposed MMSA is
less sensitive to the number of basis functions,  achieving much more stable results. %Further simulation results for estimation error and computation time are provided in the Supplementary Material.

%\textbf{Setting 2:  Large Sample Size (N=10,000 and p=1,000)}
%We next considered a simulation setting  with larger sample size and number of predictors.

\subsection{Testing for Time-Varying Effects}

We next compared the proposed testing algorithm and the test based on the scaled Schoenfeld residuals (implemented by R {\it Survival} package).  Figures 5a and 5b compares  the empirical Type-I error and the empirical power based on Setting 3.
 The proposed testing
outperforms the traditional method with higher power to correctly identify the time-varying effects and smaller Type-I error for falsely identifying time-independent effects as time-varying (e.g. false
positive). In contrast, the Type-I error for the test based on the scaled Schoenfeld residuals increased with the magnitude of the time-varying effect.
One possible explanation is as follows: as noted in Grambsch and Therneau \cite{r14},
 the scaled Schoenfeld residuals are one-step Newton estimators
 with initial values fitted from the Cox proportional hazards
  model. When the magnitude of true time-varying effects is
 relatively large, the initial values are far away from the truth and hence, the one-step
 estimator may result in biased estimation. %In contrast, the proposed testing outperformed the competing method with higher power and smaller Type-I errors.

\begin{figure}[h]
 \caption{Testing for time-varying effects at significance level 0.05; Type-I error: empirical probability of falsely identifying time-independent effects as time-varying; Power: empirical probabilities of correctly identifying the time-varying effects; based on Setting 3.}
\centering
\subfloat[Type I Error]{
\includegraphics[width=2.5in, height=2.5in]{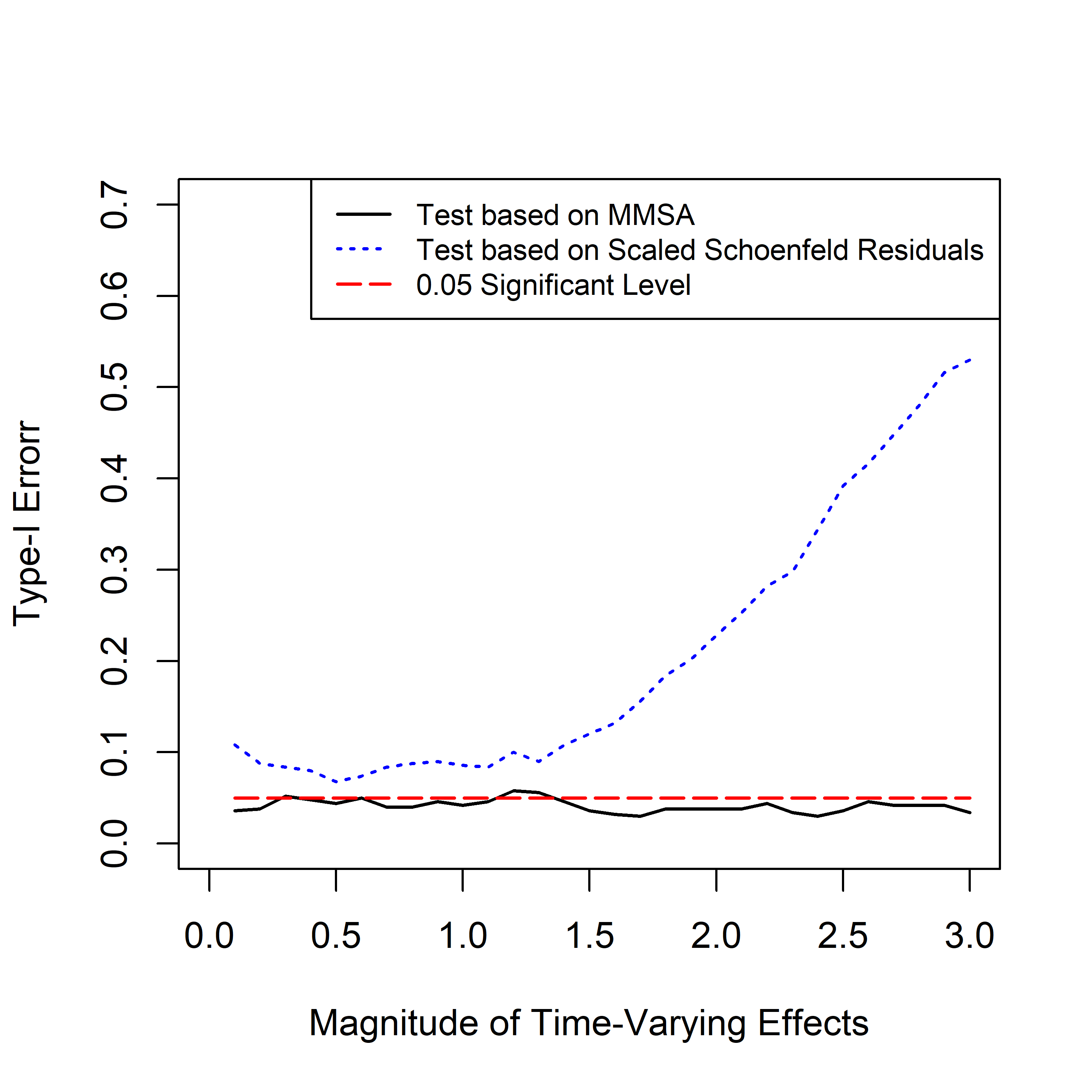}
  }
    \centering
\hspace{0.1pt}
\subfloat[Power]{
\includegraphics[width=2.5in, height=2.5in]{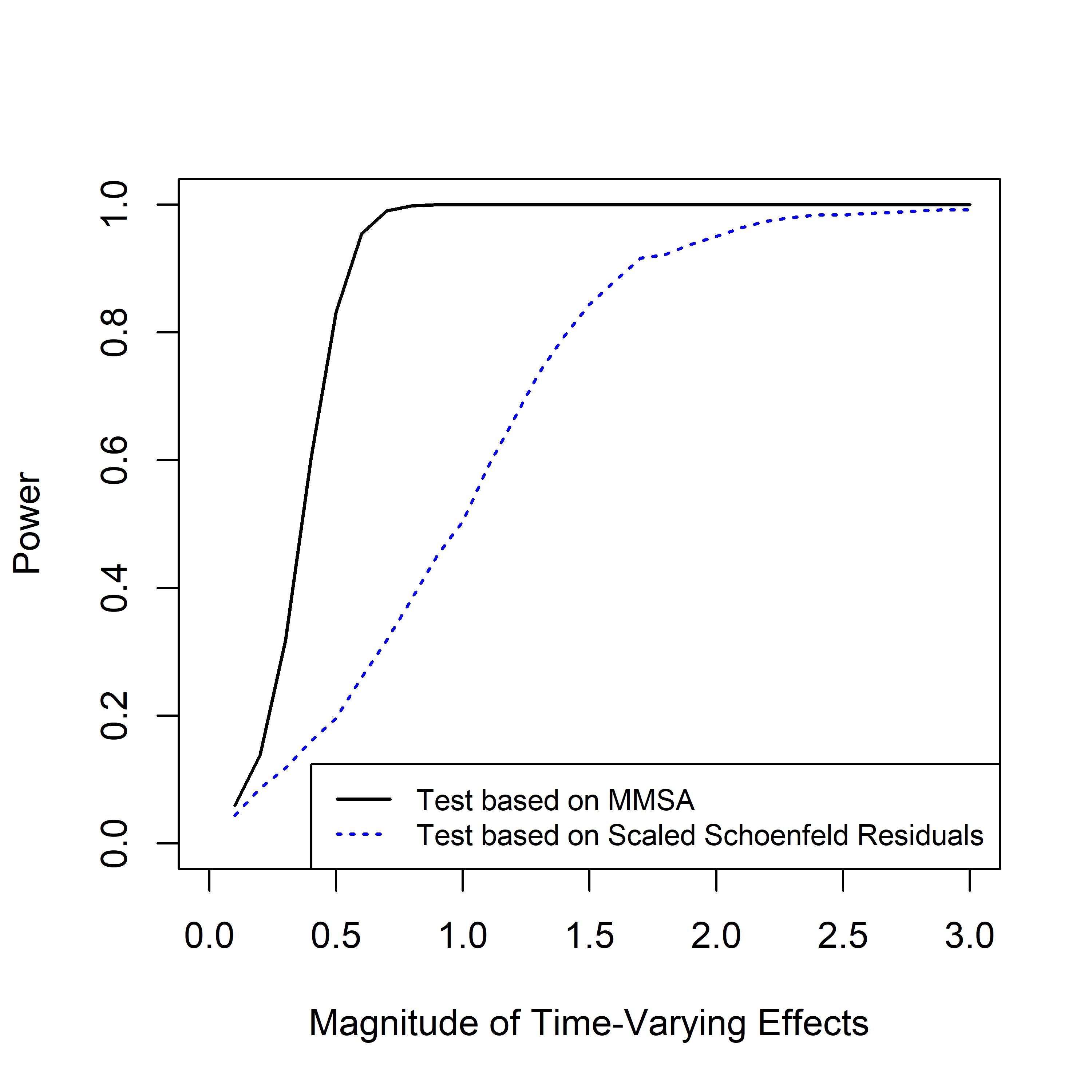}
  }
\end{figure}

\section{Application}
\label{s:Analysis}

\subsection{Kidney Transplant Dataset}
Data were obtained from the  U.S. Organ Procurement and Transplantation Network (OPTN).
%The graft lifetime estimates were based on graft failure, including death, reported graft failure and return to chronic dialysis, with censoring at the end of study.
Included in our analysis were $347, 668$ patients (from $293$ centers) who underwent kidney transplantation between January $1988$ and December $2012$. Patient survival was censored 10 year  post-transplant or at the end of study
(2012).  Failure time (recorded
in years) was defined as the time from transplantation to graft failure or death, whichever occurred first.
The overall censoring rate was $62\%$. Adjustment covariates ($P=164$) in this study included baseline recipient characteristics such as age, race,
gender,  BMI, time on dialysis,
indicator of previous kidney transplant,  immunosuppression, and cormorbidity conditions (e.g.
glomerulonephritis, polycystic kidney disease, diabetes, and hypertension), and donor characteristics such as blood type, cold ischemic time and type of donor kidney.
 Race was categorized as White, African American, Hispanic, Asian or other.  Cold ischemia times were categorized as Low (20 hours or less) or High (longer than 20 hours). Type of donor kidney was categorized as living, standard criteria donor, or expanded criteria donor (ECD). Waiting times on dialysis were categorized as Short (less than 5 years) or Long (greater than 5 years).

%Table 1 in the Supplementary Material gives the descriptive statistics for all covariates.

 \subsection{Assessing Time-Varying Effects}

To determine the number of basis functions, we performed
 5-fold cross-validation \cite{r36}. Ten basis functions were chosen for further analysis.
A total of 12 variables were identified
 with significant time-varying effects (p-values $< 0.001$).
 Figures 6-7 show the fitted coefficients (solid lines) and 95\% confidence intervals (dashed lines).
Figures 6a and 6b show that anti-viral therapies and anti-rejection immunosuppressant medications
had strong protective effects in the short term after
transplantation, with a weakening association over time.
One possible explanation is that these therapies prevent recipient's body from rejecting new kidney and declining rates of acute rejection have led to improvements
in short term kidney transplant survival \cite{r27}.
Figure 6c strongly supports previous finding \cite{r26} that
longer waiting times on dialysis (greater than 5 years) negatively impact post-transplant survival, especially in the short run.
%In other words, patients who reach end-stage renal disease should receive a renal transplant as early as possible in order to enhance their chances of survival.
Figure 6d indicates that the effects of stroke, the most frequent donor cause of death, varies over time, showing an increased risk of worse recipient outcomes in the short run, and then a slightly weakening association over time. One possible explanation is as follows: although stroke is a predictor for worse survival for kidney transplantation,
it is also associated with lower rates of rejection immediately after the renal transplantation \cite{r13}, which may lead to varying association in the short run.
%Human leukocyte antigens (HLA) play a central role in the cellular and humoral immune responses that determine the outcome of a transplant. The increasing demand for organ donors to supply the increasing number of patients on kidney waiting lists has led to the question of whether HLA matching should be used in organ allocation.
Figure 6e suggests that the effect
of Human leukocyte antigens (HLA) matching varies over time, resulting in an eventually weakening association.
Thus, special care (such as pre-transplant antiviral therapy) must be dedicated to improve the long-term results.
%Hypertension is one of the most common clinical problems in renal transplant recipients.
Figure 6f suggests that blood pressure management in the kidney transplant recipient reduces the likelihood of graft failure.  Given the greater cardiovascular burden in the kidney transplant recipient,  more effective control of blood pressure may further reduce cardiovascular-related death.

%The results regarding gender differences should be interpreted with caution.
Figure 7a shows that male recipients is a protective factor immediately after the renal transplantation and then becomes a risk factor in the long run. 
Regarding racial disparities, Figure 7b indicates that survival outcomes for African Americans continue to lag behind non-African Americans.
%This disparity  indicates that the impact of nonimmunologic variables, such as behavioral, social, environmental, and occupational factors, can be key issues.
Our results for previous Kidney Transplant and high cold ischemic time (Figures 7c and 7d) show that they
are risk factors for mortality in the short run. Thus, special care should be dedicated to improve the short-term results.
As shown in Figure 7e,  donors with higher height and hence larger adjusted cortical volume has a protective effect in the short-term after transplantation, which suggests that larger adjusted cortical volumes are more likely to achieve better glomerular filtration rate than those with smaller cortical volumes.
Finally, polycystic kidney
disease (PKD) is the most common
genetic kidney disease, which accounts for 2\% to
9\% of patients with end-stage renal failure \cite{r31}.
There are conflicting reports of differing renal
allograft outcomes for PKD patients \cite{r17}. %\citet{Roozbeh2008} reported that short-term survivals were slightly better in the PKD population than other subjects; however, the result was not statistically significant. Other studies \cite{r17} have reported complications occur more commonly following renal transplantation among PKD patients.
In our analysis, the time-varying coefficient (Figure 7f) suggests
that there is a varying association of PKD.
Thus, accounting for time-varying effects provides valuable clinical information that could be missed otherwise.

%Finally, we implemented the time-dependent ROC curve to assess how well the developed time-varying risk scores distinguish between subjects who are likely to die and subjects who are likely to survive. We random split the study samples into two disjointed sub-data sets of nearly equal sizes and implement the proposed MMSA algorithm and the he Cox proportional hazards model using the first half of samples. Figure 6 compares the ROC curves at selected time points using the validation samples. The results suggest a good short-term discriminating ability for the time-varying scores to separate early cases from early controls.
%Moreover, the accuracy of the time-varying scores gradually decreases
%with increasing temporal distance from the baseline. Declining prognostic accuracy is
%not surprising, particularly because the patient health status is actually time-varying
%and the transplant recipient may die for reasons unrelated to the transplant procedure
%during the long-term follow-up.

\begin{figure}[htbp]
 \caption{Transplant data: estimated coefficients (solid lines) and 95\% confidence interval (dashed lines) for time-varying effects.}
   \centering
\subfloat[Anti-viral Therapies]{
\includegraphics[width=2.85in, height=2.45in]{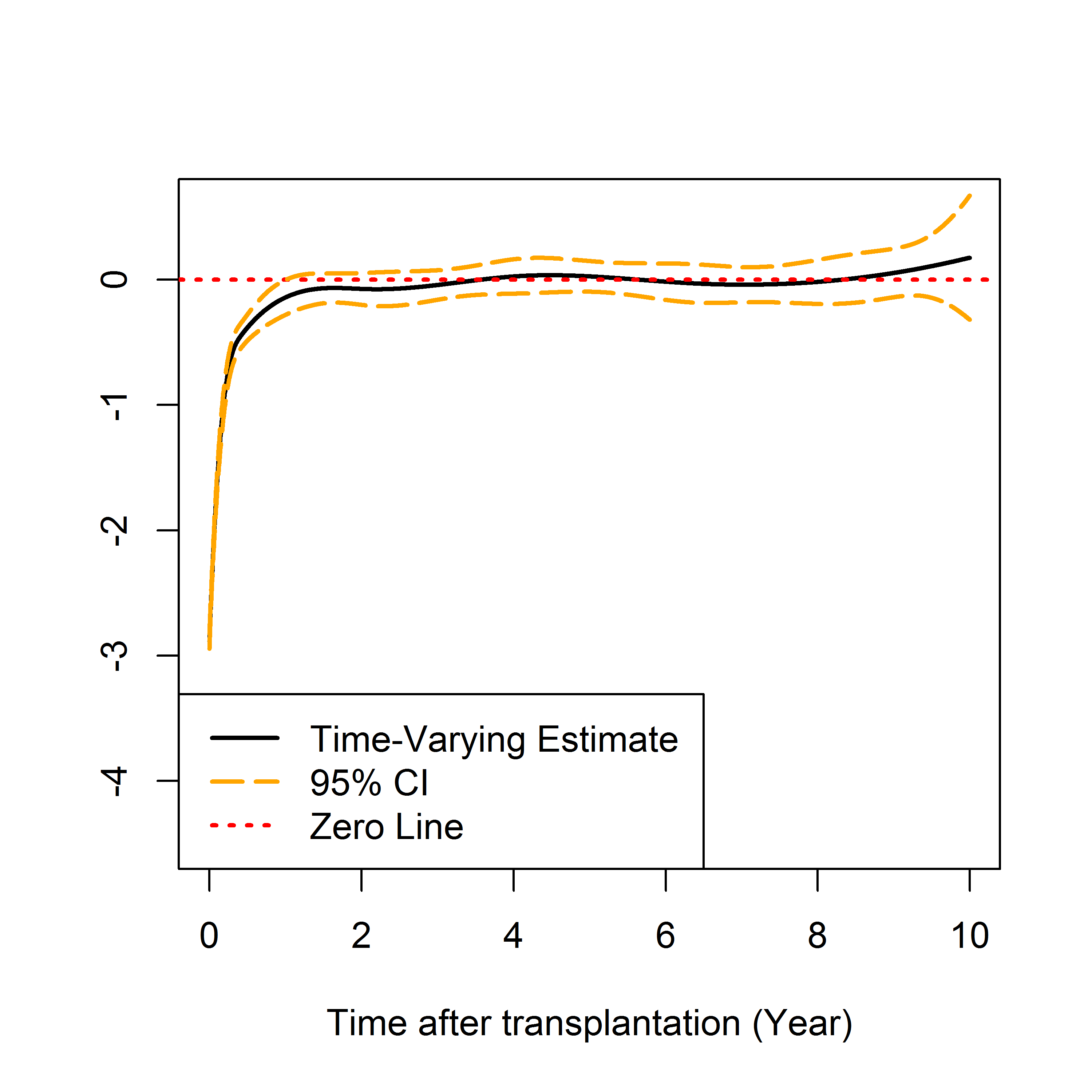}
  }
     \centering
\hspace{1pt}
\subfloat[Immunosuppressant Medications]{
\includegraphics[width=2.85in, height=2.45in]{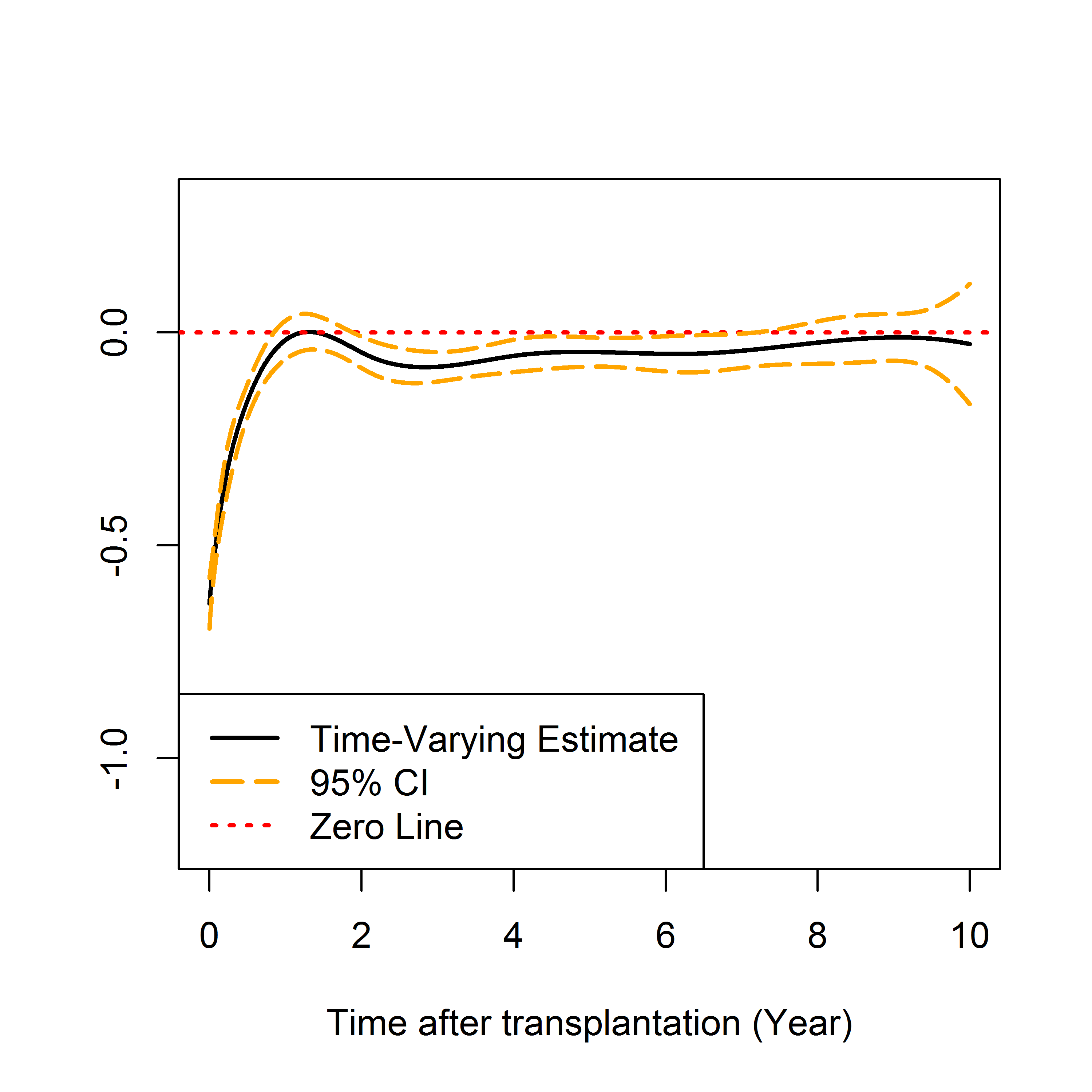}
  }
\centering
  \vspace{1pt}
\subfloat[Long Waiting Times]{
\includegraphics[width=2.85in, height=2.45in]{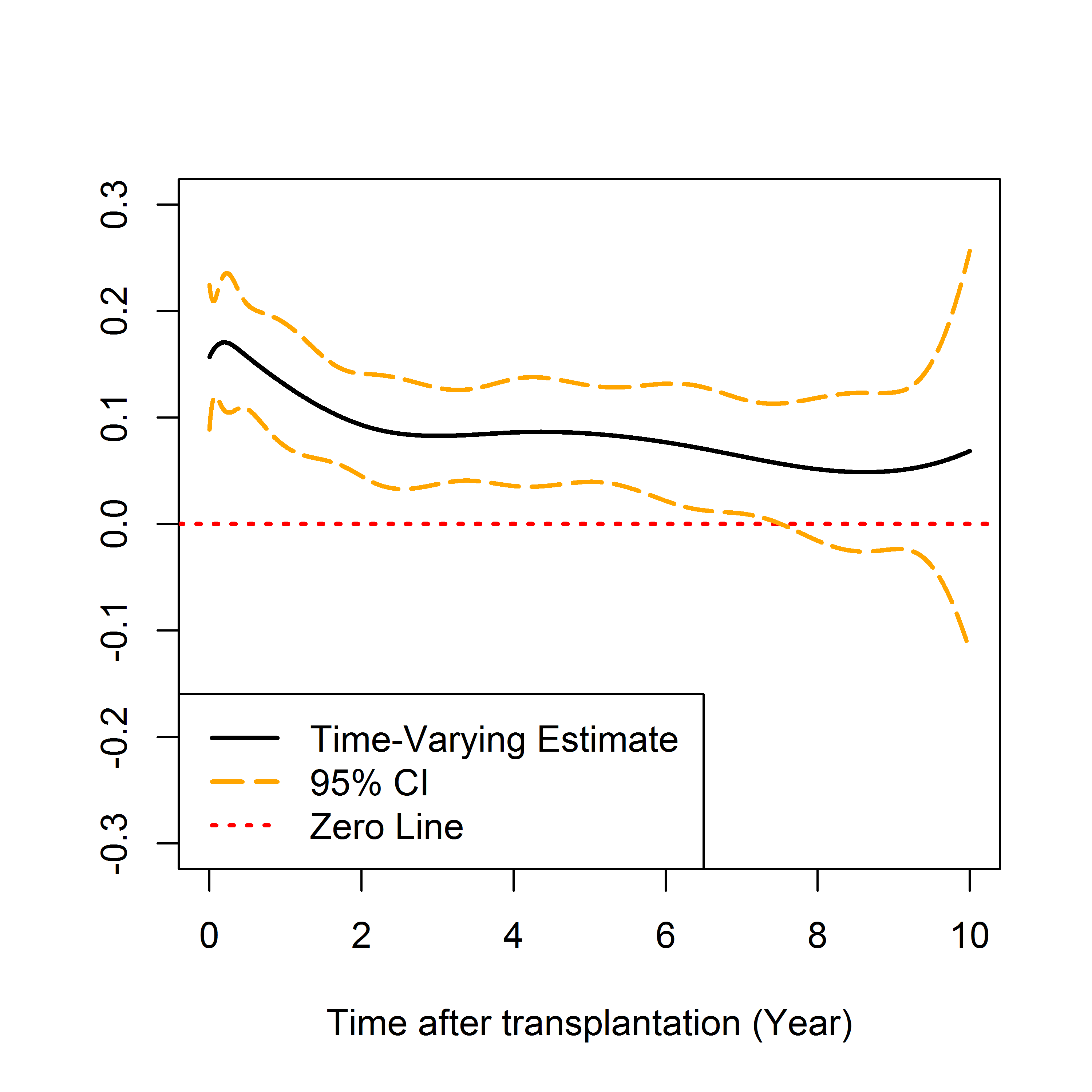}
  }
  \centering
  \hspace{1pt}
\subfloat[Stroke]{
\includegraphics[width=2.85in, height=2.45in]{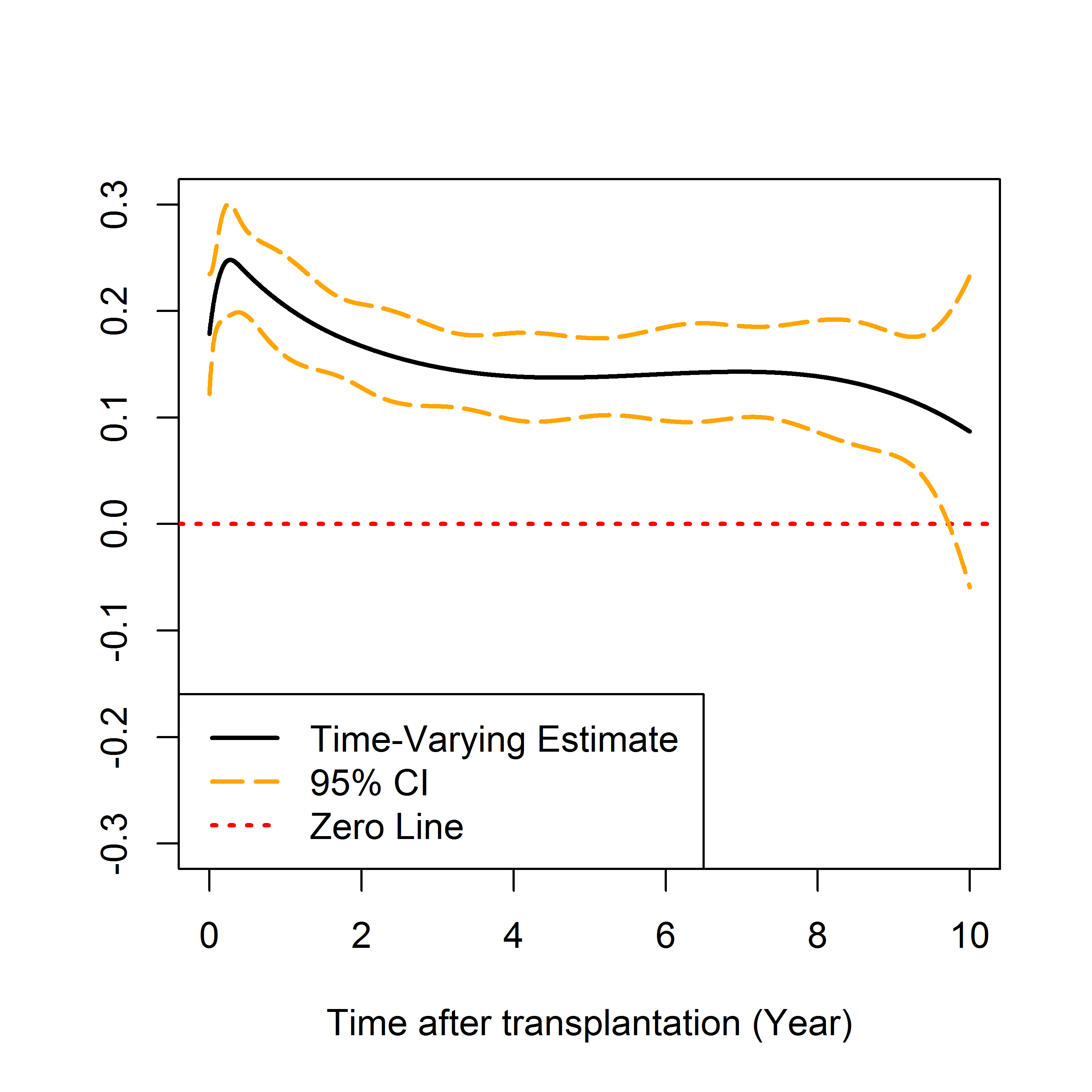}
  }
  \centering
\vspace{1pt}
\subfloat[HLA Match]{
\includegraphics[width=2.85in, height=2.45in]{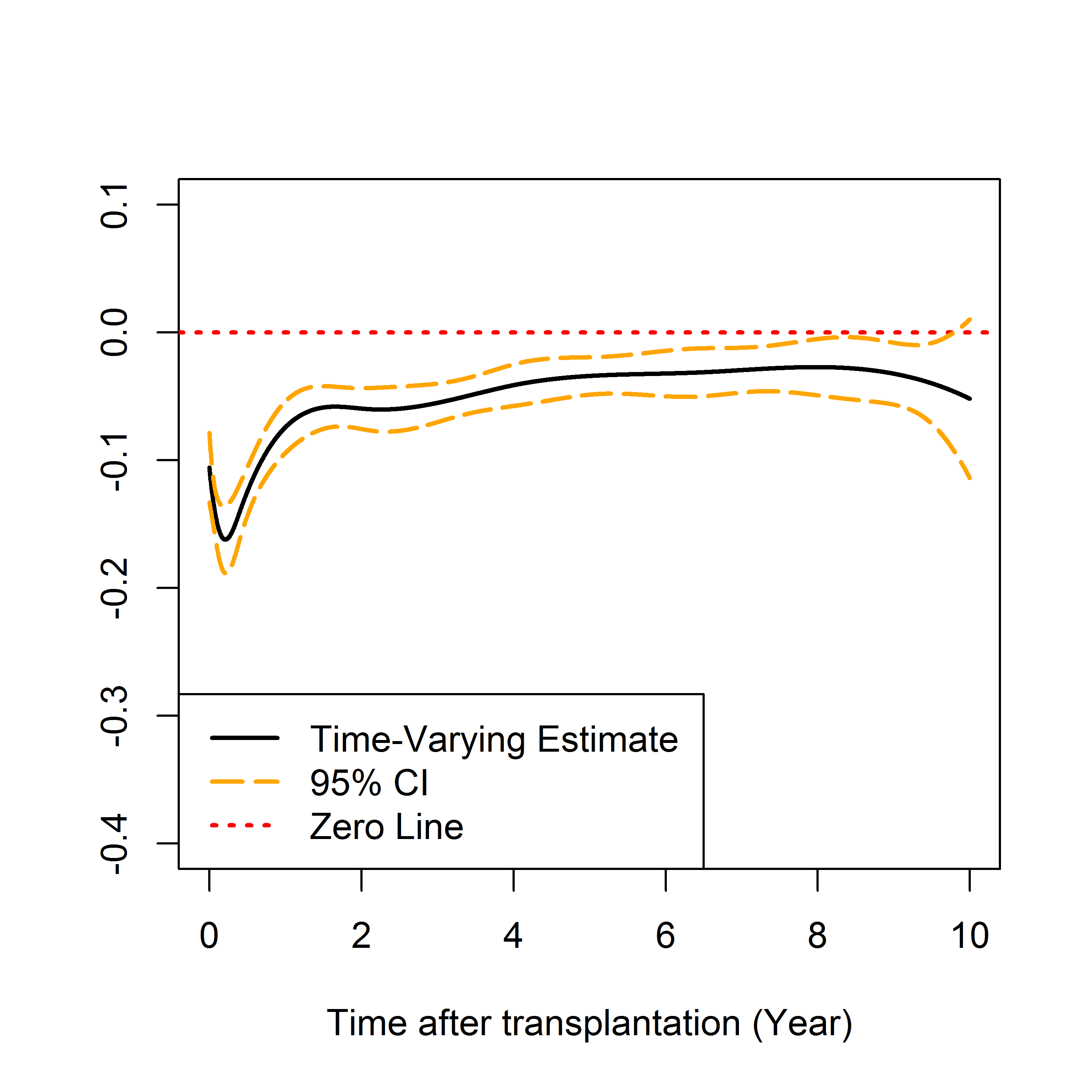}
  }
         \centering
  \hspace{1pt}
  \subfloat[Drug Treated Hypertension]{
\includegraphics[width=2.85in, height=2.45in]{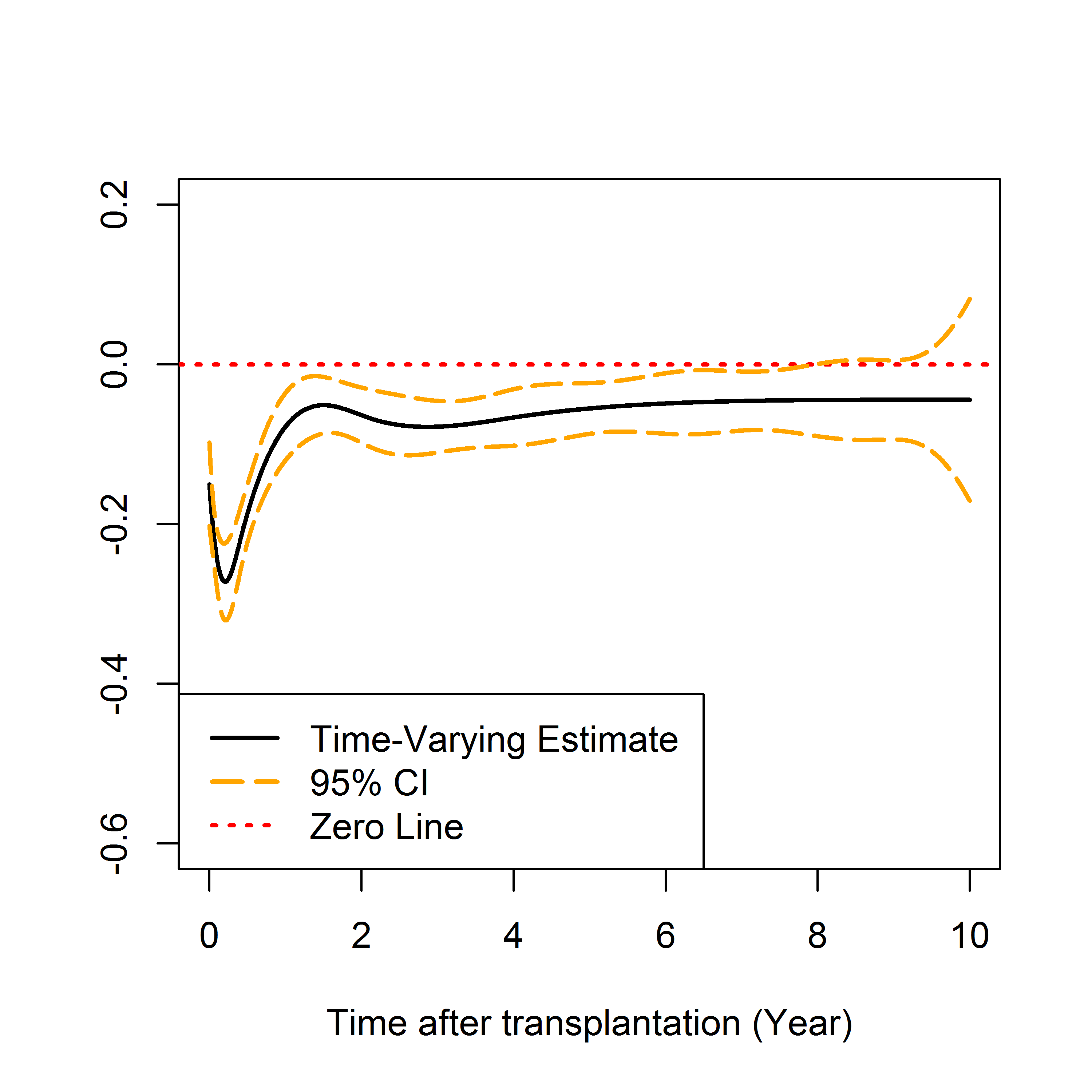}
  }
  \end{figure}

  \begin{figure}[htbp]
  \caption{Transplant data: estimated coefficients  (solid lines) and 95\% confidence interval (dashed lines) for time-varying effects.}
    \centering
\subfloat[Male]{
\includegraphics[width=2.85in, height=2.45in]{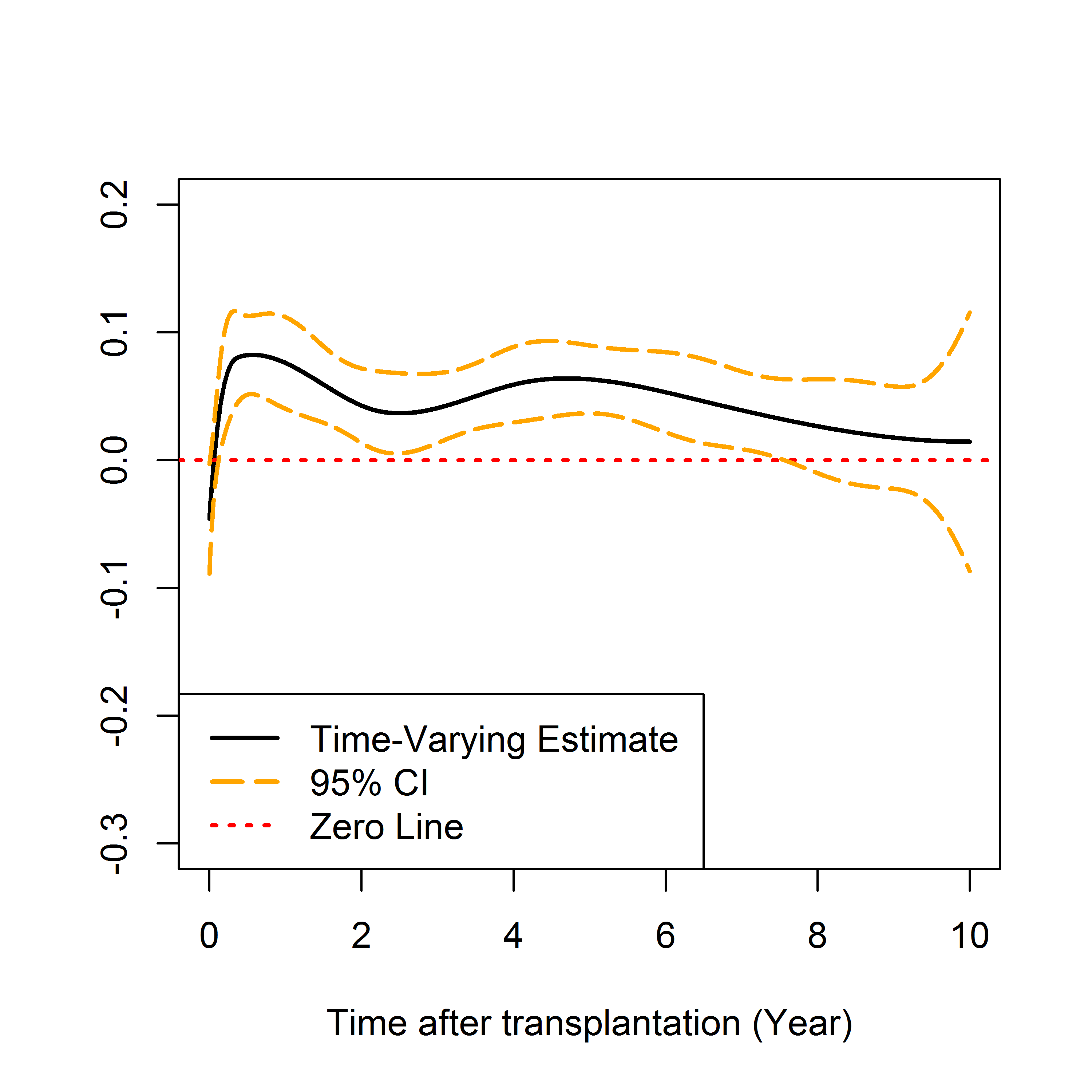}
  }
   \centering
\hspace{1pt}
\subfloat[African American]{
\includegraphics[width=2.85in, height=2.45in]{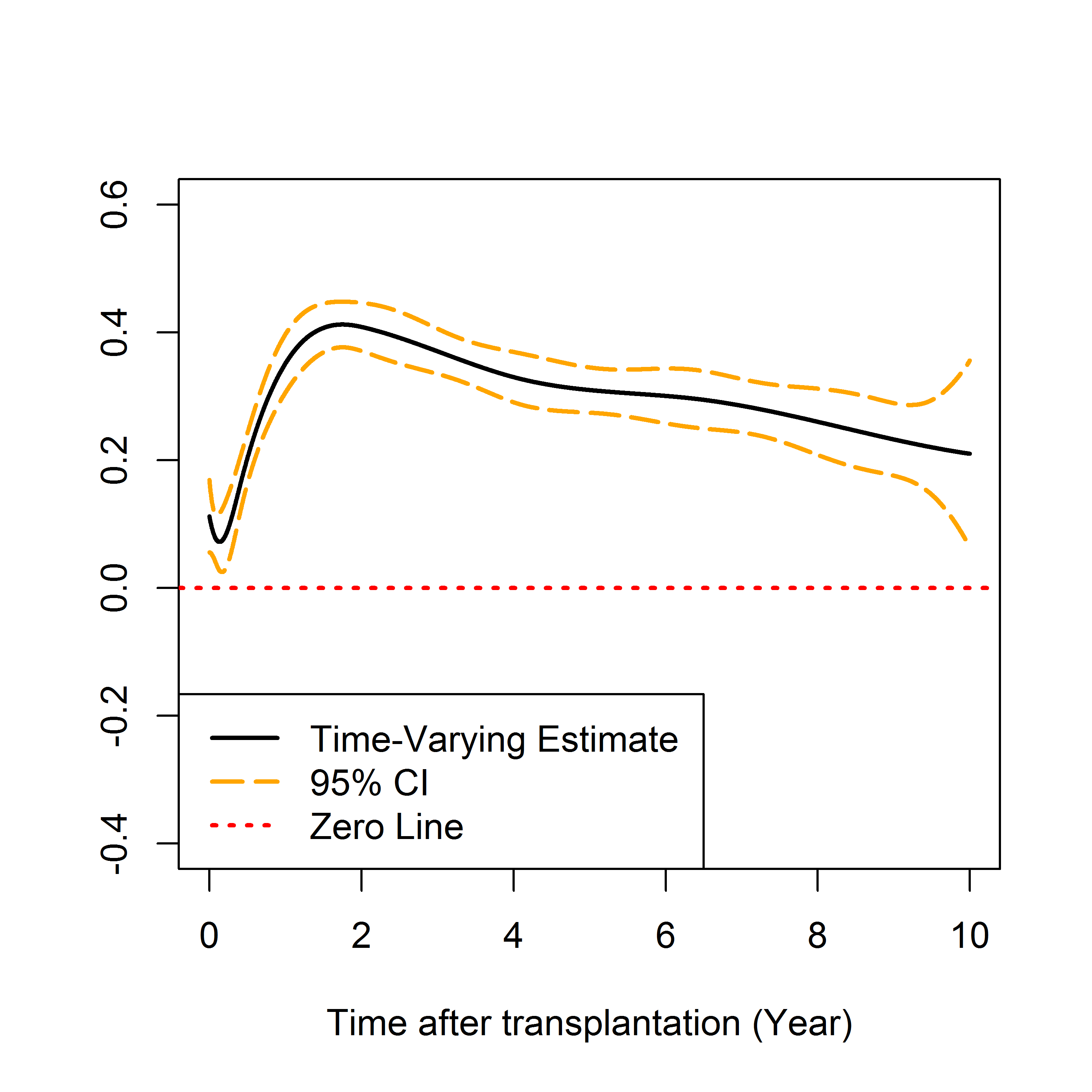}
  }
     \centering
     \vspace{1pt}
\subfloat[Recipient Other Therapies]{
\includegraphics[width=2.85in, height=2.45in]{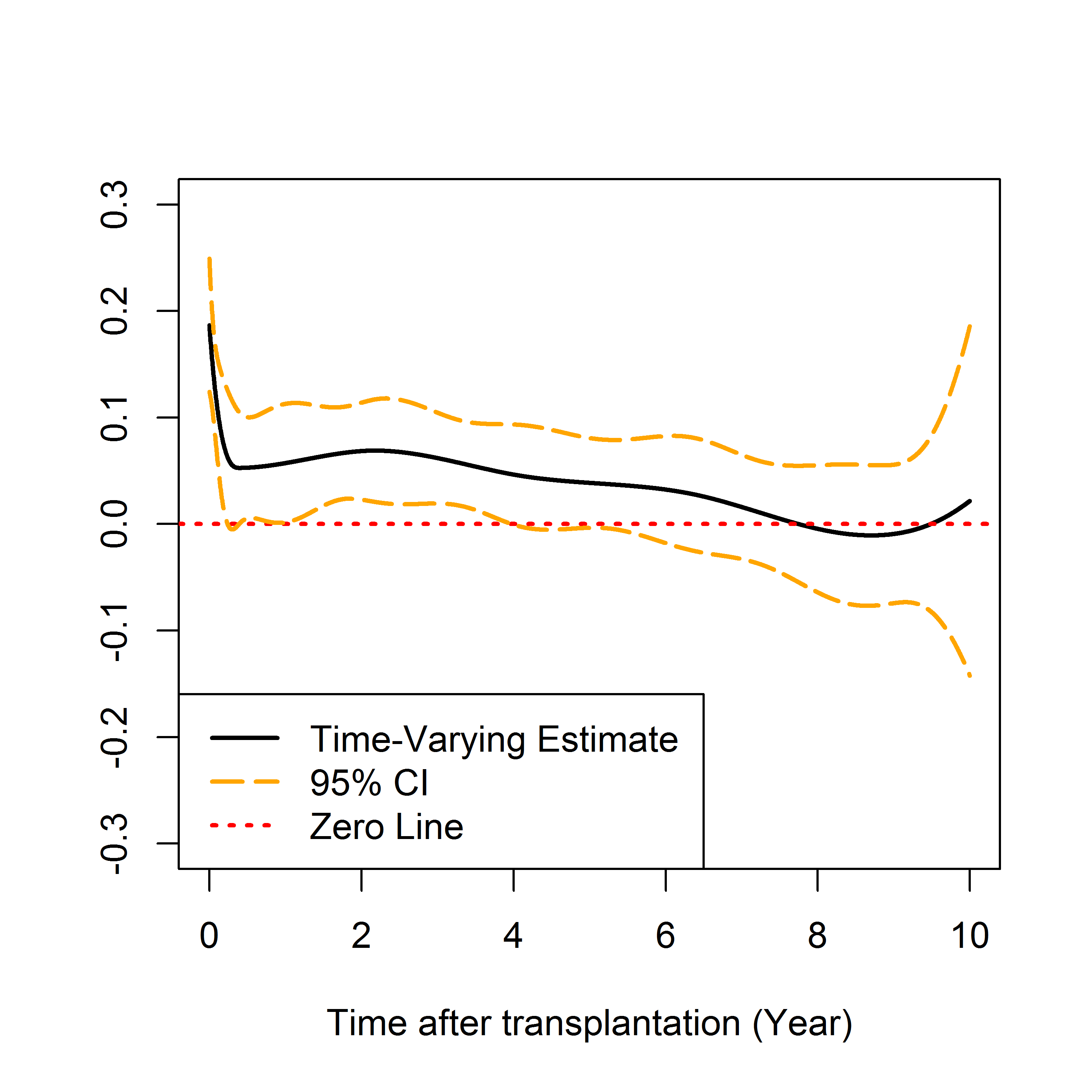}
  }
     \centering
  \hspace{1pt}
  \subfloat[High Cold Ischemic Time]{
\includegraphics[width=2.85in, height=2.45in]{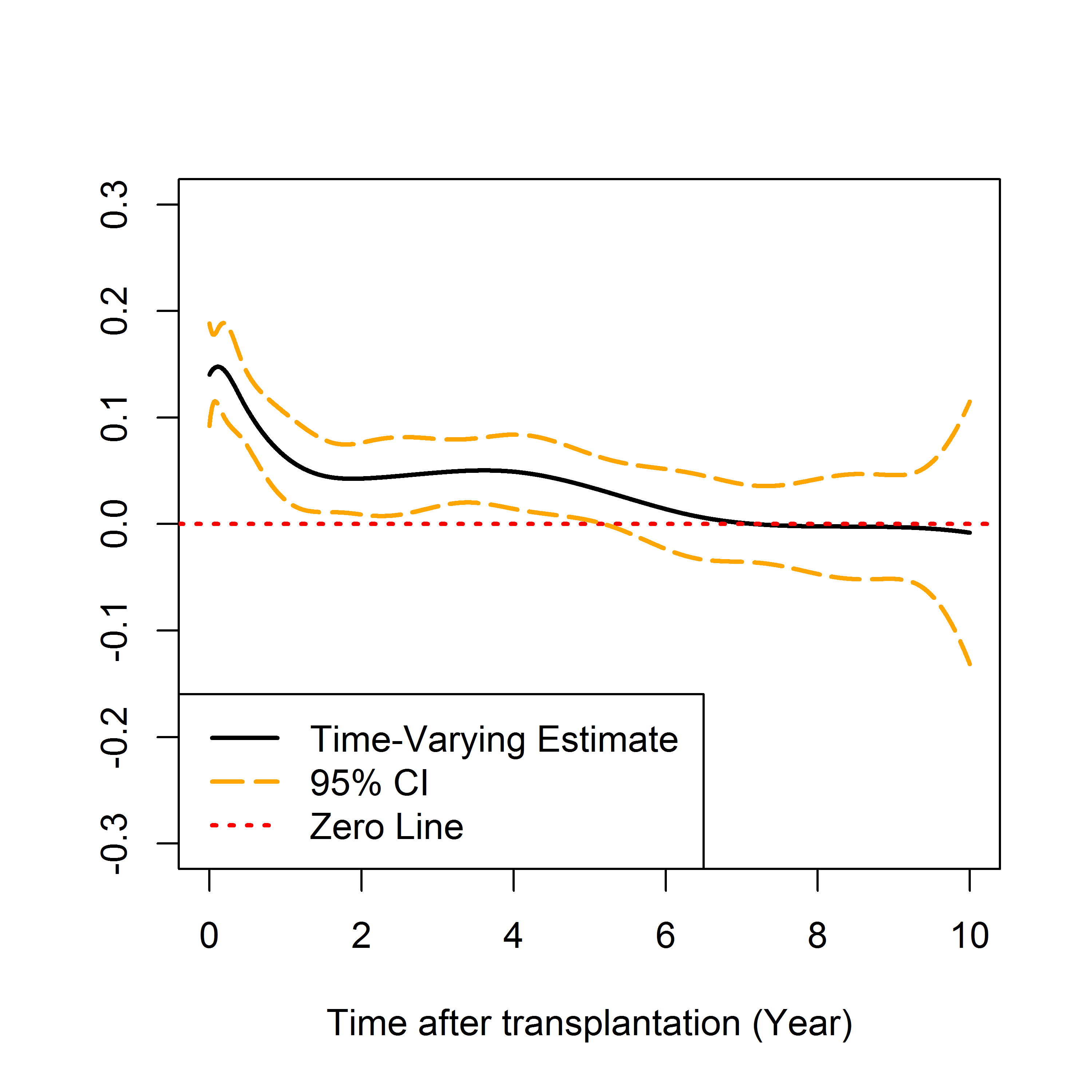}
  }
       \centering
\vspace{1pt}
\subfloat[Donor Height]{
\includegraphics[width=2.85in, height=2.45in]{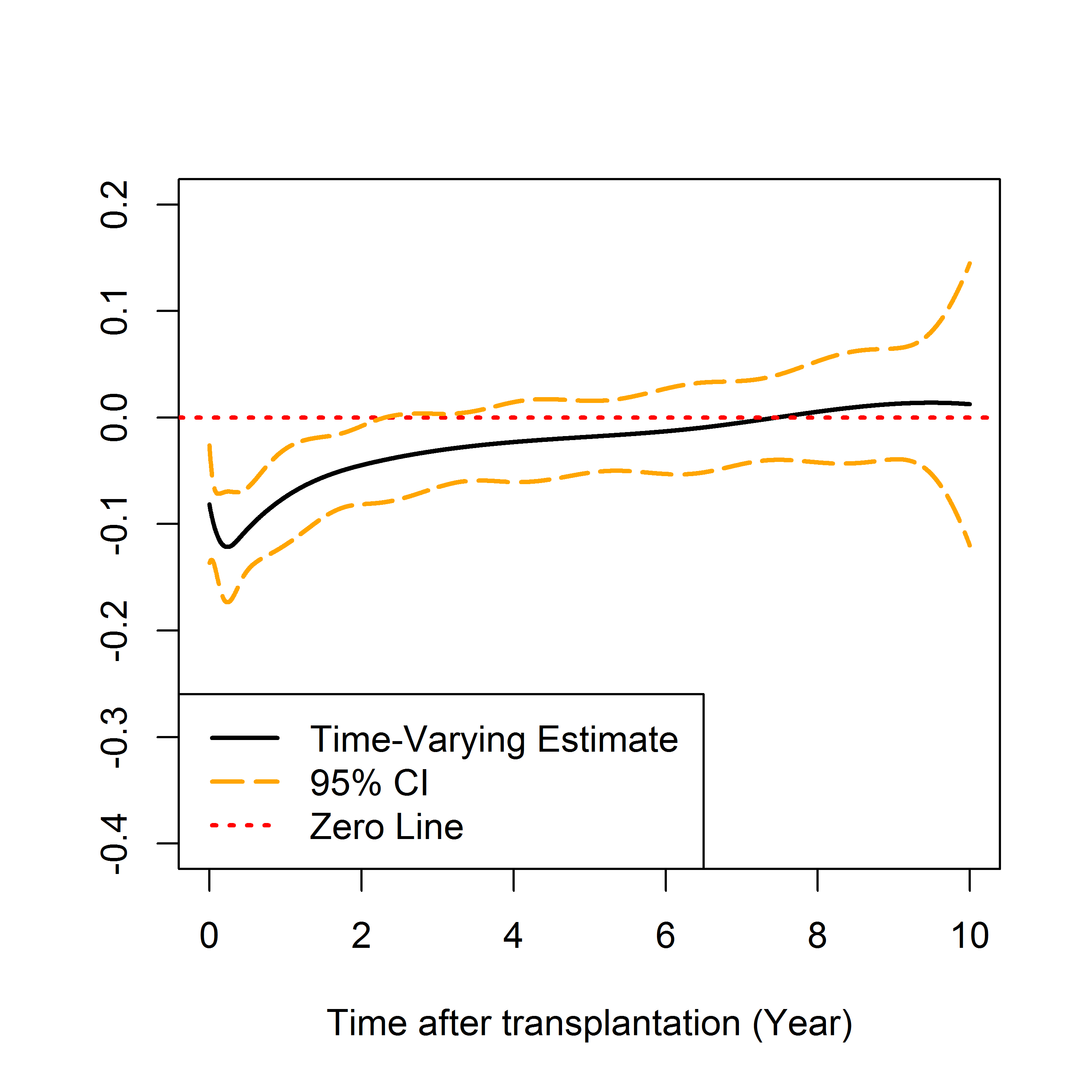}
  }
   \centering
   \hspace{1pt}
\subfloat[Polycystic Kidney Disease]{
\includegraphics[width=2.85in, height=2.45in]{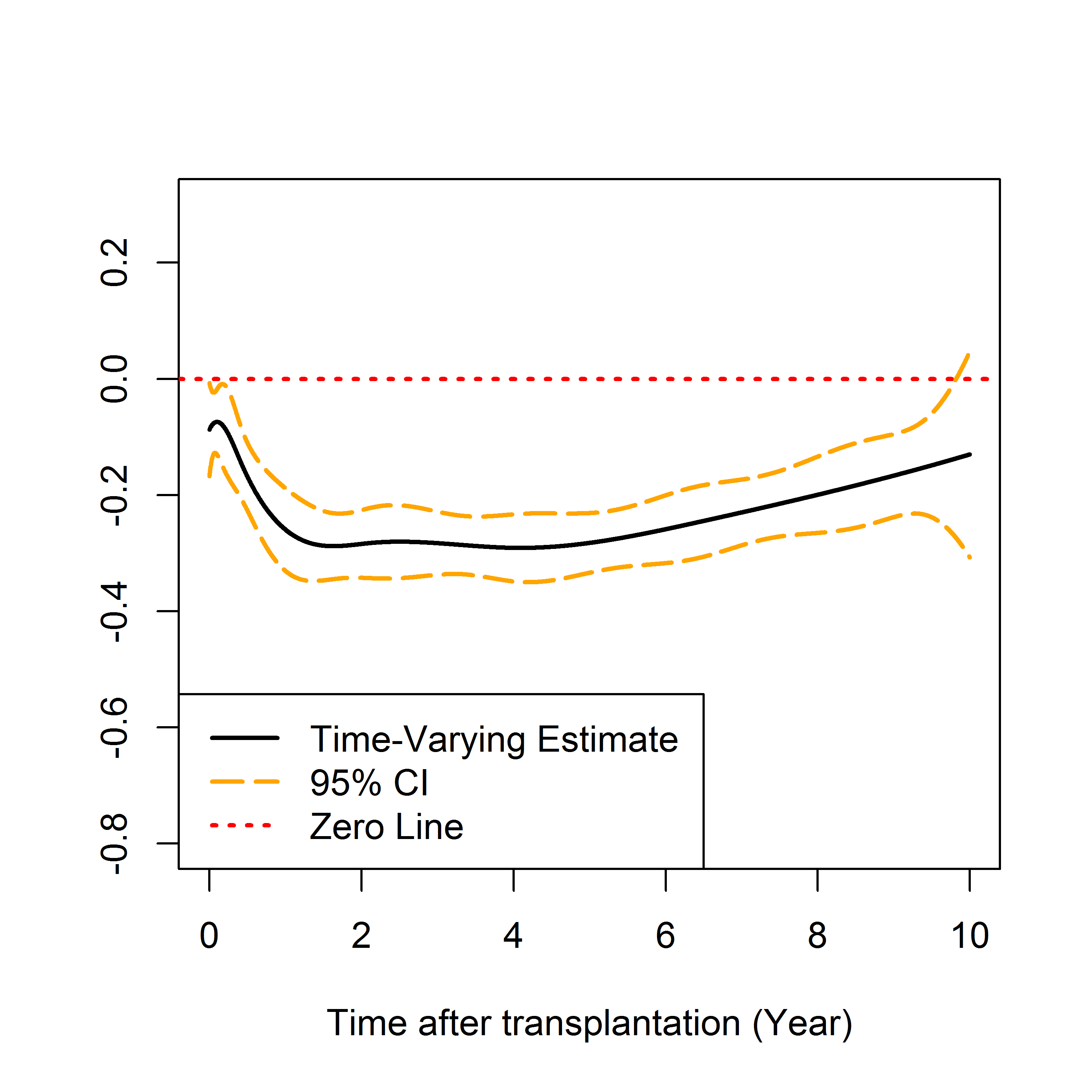}
  }
\end{figure}

\section{Discussion}
\label{s:discuss}

%Newton method requires iterative calculations and inversion of the Hessian matrix, which may have unreasonable costs or may even be impractical. Even the standard penalized procedures are plagued due to the computational complexity. This might have explained why only relative small data sets were considered in existing literatures to estimate time-varying effects in survival analysis.

%Thus, one of the main challenges is the limited availability of computationally feasible methods for large-scale datasets.

%This functional form selection problem is fundamentally important, as the validity of a fitted model heavily depends on the correct model structure.Our endeavor here is motivated by the study of large-scale national kidney transplant data. 
Detecting and accounting for time-varying effects are particularly important in the context of clinical studies, as non-proportional hazards have already been reported in the clinical literature \cite{r6,r7}. However, in survival analysis, 
the computational burden to model time-varying effects increases quickly as the sample size or the number of predictors grows.
In this report, we propose a Minorization-Maximization-based steepest ascent
method.
Our procedure iteratively updates the optimal block-wise direction along which the directional derivative is
maximized and, hence, the approximate increase in log-partial likelihood is greatest. Our approach is a computationally simple technique, which extends existing boosting methods to estimate time-varying effects for time-to-event data.
Numerical studies suggest that the proposed method provides feasible and accurate estimates for large-scale settings.

The proposed method can be extended to high-dimensional settings with the number of predictors larger than the sample size.
Since only one variable is updated at each MMSA iteration, variable selection can be achieved by using a finite number of boosting iterations, which can be determined by cross-validation.
Compared with penalized methods, the MMSA is
flexible and easily implemented without the need to apply constrained optimizations. %The parallel computing algorithms can also be easily optimized to improve the computational speed. 
The resulting approach simultaneously selects and automatically determines potential time-varying effects in high-dimensional time-to-event data. We will report this work in another report.

\appendix

\section{}\label{app}

\subsection*{Proof of Proposition 1}

 Consider a second-order Taylor expansion,
\begin{align*}
\ell(\boldsymbol\theta)=\ell(\boldsymbol{\widehat\theta})+ \triangledown \ell(\widehat{\boldsymbol\theta})^T (\boldsymbol\theta-\boldsymbol{\widehat\theta})+\frac{1}{2}(\boldsymbol\theta-\boldsymbol{\widehat\theta})^T \triangledown^2 \ell(\widetilde{\boldsymbol\theta})(\boldsymbol\theta-\boldsymbol{\widehat\theta}),
\end{align*}
where $\widetilde{\boldsymbol\theta}$ lies between $\boldsymbol\theta$ and $\boldsymbol{\widehat\theta}$.
Assuming $\bH(\boldsymbol\theta)$ is positive definite,
we have,
\begin{align*}
\lambda_{max}\left(\{\bH(\boldsymbol{\widehat\theta})\}^{-1/2}\{- \triangledown^2 \ell(\boldsymbol{\widetilde\theta})\}\{\bH(\boldsymbol{\widehat\theta})\}^{-1/2}\right)=\lambda_{max}\left(\{\bH(\boldsymbol{\widehat\theta})\}^{-1}\{- \triangledown^2 \ell(\boldsymbol{\widetilde\theta})\} \right),
\end{align*}
and for all $\boldsymbol\theta \neq \boldsymbol{\widehat\theta}$
\begin{align*}
\lambda_{max}\left(\{\bH(\boldsymbol{\widehat\theta})\}^{-1}\{- \triangledown^2 \ell(\boldsymbol{\widetilde\theta})\}\right) \geq \frac{(\boldsymbol\theta-\boldsymbol{\widehat\theta})^T(-\triangledown^2 \ell(\boldsymbol{\widetilde\theta}))(\boldsymbol\theta-\boldsymbol{\widehat\theta})}{(\boldsymbol\theta-\boldsymbol{\widehat\theta})^T \bH(\boldsymbol{\widehat\theta})(\boldsymbol\theta-\boldsymbol{\widehat\theta})}.
\end{align*}
Thus, if
$
\lambda_{max}\left(\{\bH(\boldsymbol{\widehat\theta})\}^{-1/2}\{- \triangledown^2 \ell(\boldsymbol{\widetilde\theta})\}\{\bH(\boldsymbol{\widehat\theta})\}^{-1/2}\right) < 1/\nu,
$
we have
\begin{align*}
&{}(\boldsymbol\theta-\boldsymbol{\widehat\theta})^T(-\triangledown^2 \ell(\boldsymbol{\widetilde\theta}))(\boldsymbol\theta-\boldsymbol{\widehat\theta})
<  \frac{1}{\nu}(\boldsymbol\theta-\boldsymbol{\widehat\theta})^T \bH(\boldsymbol{\widehat\theta})(\boldsymbol\theta-\boldsymbol{\widehat\theta}).
\end{align*}
It follows that
$\ell(\boldsymbol\theta) \geq g(\boldsymbol\theta|\widehat{\boldsymbol\theta})$ for all $\boldsymbol\theta$.

%\section*{Acknowledgements}
%And this is an acknowledgements section with a heading that was produced by the $\backslash$section* command. Thank you all for helping me writing this \LaTeX\ sample file. See \ref{suppA} for the supplementary material example.

%\begin{supplement}
%\sname{Supplement A}\label{suppA}
%\stitle{Title of the Supplement A}
%\slink[url]{http://www.e-publications.org/ims/support/dowload/imsart-ims.zip}
%\sdescription{Dum esset rex in
%accubitu suo, nardus mea dedit odorem suavitatis. Quoniam confortavit seras portarum tuarum, benedixit filiis tuis in te. Qui posuit fines tuos}
%\end{supplement}

\subsection*{Proof of Proposition 2}

To assess the numerical convergence of the MMSA algorithm, we  follow the strategy in Lange \cite{r23}
and  define the following concepts:
\begin{enumerate}
 \item [(1)] A point $\boldsymbol\theta$ is a cluster point of a sequence $\boldsymbol{\widehat\theta}^{(m)}$ if every neighborhood of $\boldsymbol\theta$ (i.e., a subset that includes an open set containing  $\boldsymbol\theta$) contains infinitely many $\boldsymbol{\widehat\theta}^{(m)}$.
 \item[(2)] A point $\boldsymbol\theta$ is a fixed point if $\boldsymbol\theta=M(\boldsymbol\theta)$, where $M(\boldsymbol\theta)$ is the iteration map generated by the MMSA algorithm.
  \end{enumerate}

The following steps provide the ground for the proof of Proposition 2.

 \textbf{Step 1}
{\it
The fixed points of iteration map $M(\boldsymbol\theta)$ generated by the MMSA algorithm, and the stationary points of the objective function $\ell(\boldsymbol\theta)$ coincide.
}

Recall $\ell(\boldsymbol\theta)\geq g(\boldsymbol\theta|\boldsymbol{\widehat\theta}^{(m-1)})$ for all $\boldsymbol\theta$, and
$g(\boldsymbol{\widehat\theta}^{(m-1)}|\boldsymbol{\widehat\theta}^{(m-1)})=\ell(\boldsymbol{\widehat\theta}^{(m-1)})$.
 Thus,
$\boldsymbol{\widehat\theta}^{(m-1)}$ is a stationary point of the difference $g(\boldsymbol\theta|\boldsymbol{\widehat\theta}^{(m-1)})-\ell(\boldsymbol\theta)$. The following gradient identity holds
\begin{align*}
\nabla g(\boldsymbol{\widehat\theta}^{(m-1)}|\boldsymbol{\widehat\theta}^{(m-1)}) = \nabla \ell(\boldsymbol{\widehat\theta}^{(m-1)}).
\end{align*}
By Condition (A), $M(\boldsymbol{\widehat\theta}^{(m-1)})=\boldsymbol{\widehat\theta}^{(m-1)}$ if and only if
$
\nabla g(\boldsymbol{\widehat\theta}^{(m-1)}|\boldsymbol{\widehat\theta}^{(m-1)})= \nabla \ell(\boldsymbol{\widehat\theta}^{(m-1)})=0.
$
Therefore, the fixed points of iteration map $M(\boldsymbol\theta)$  and the stationary points of $\ell(\boldsymbol\theta)$ coincide.

 \textbf{Step 2}
{\it
 Every cluster point of iterates $\boldsymbol{\widehat\theta}^{(m)}=M(\boldsymbol{\widehat\theta}^{(m-1)})$ generated by the iteration map $M(\boldsymbol\theta)$ of the MMSA algorithm is a stationary point of $\ell(\boldsymbol\theta)$. Furthermore, the set of stationary points $\mathcal{F}$ is closed and the limit of the following distance function is zero:
   \begin{align*}
 \lim_{m\rightarrow\infty}\inf_{\boldsymbol\theta \in \mathcal{F}} ||\boldsymbol{\widehat\theta}^{(m)} - \boldsymbol\theta||_2=0.
   \end{align*}
}

By the coercive condition assumed in Condition (B), the sequence $\boldsymbol{\widehat\theta}^{(m)}$ is contained within the compact super-level set
$
\{\boldsymbol\theta \in \mathcal{U}: \ell(\boldsymbol\theta) \geq \ell(\boldsymbol\theta_0)\},
$
for some $\boldsymbol\theta_0 \in \mathcal{U}$.
Existence of a cluster point is then guaranteed by the compactness of the super-level set. Consider a cluster point
$\boldsymbol{\widetilde\theta}=\lim_{r \rightarrow\infty}\boldsymbol{\widehat\theta}^{(m_r)}$.
 The ascent property in Proposition 1
 guarantees that $\lim_{r \rightarrow\infty}\ell(\boldsymbol{\widehat\theta}^{(m_r)})$ exists. Moreover, the continuity of $M(\boldsymbol\theta)$ and $\ell(\boldsymbol\theta)$ imply
 \begin{eqnarray}
 \lim_{r \rightarrow\infty} \ell(\boldsymbol{\widehat\theta}^{(m_r)})=\ell(\lim_{r \rightarrow\infty} \boldsymbol{\widehat\theta}^{(m_r)}).
   \nonumber
\end{eqnarray}
Using the ascent property once again,
\begin{align*}
\ell(M(\boldsymbol{\widetilde\theta}))\geq g(M(\boldsymbol{\widetilde\theta})|\boldsymbol{\widetilde\theta}) \geq g(\boldsymbol{\widetilde\theta}|\boldsymbol{\widetilde\theta})=\ell(\boldsymbol{\widetilde\theta}).
\end{align*}
Note that the ascent property and the fact that
$\ell(\boldsymbol{\widetilde\theta})$ is the supremum of $\ell(\boldsymbol{\widehat\theta}^{(m_r)})$ imply that $\ell(\boldsymbol{\widetilde\theta})=\ell(M(\boldsymbol{\widetilde\theta}))$. Thus, equality holds throughout the inequality above and hence
\begin{align*}
g(M(\boldsymbol{\widetilde\theta})|\boldsymbol{\widetilde\theta}) = g(\boldsymbol{\widetilde\theta}|\boldsymbol{\widetilde\theta}).
\end{align*}
  By Condition (A), %$g(M(\boldsymbol{\widetilde\theta})|\boldsymbol{\widetilde\theta}) > g(\boldsymbol{\widetilde\theta}|\boldsymbol{\widetilde\theta})$ unless $M(\boldsymbol{\widetilde\theta})|\boldsymbol{\widetilde\theta} = \boldsymbol{\widetilde\theta}|\boldsymbol{\widetilde\theta}$. Therefore,
 $M(\boldsymbol{\widetilde\theta})=\boldsymbol{\widetilde\theta}$, e.g. the cluster point $\boldsymbol{\widetilde\theta}$ is also a fixed point.
 Results in step 1  then imply that the fixed point $\boldsymbol{\widetilde\theta}$ is also a stationary point of $\ell(\boldsymbol\theta)$.

 To show that the set of stationary points $\mathcal{F}$ is closed, suppose there exists a sub-sequence $\boldsymbol{\widehat\theta}^{(m_r)} \in \mathcal{F}$ and $\lim_{r \rightarrow\infty} \boldsymbol{\widehat\theta}^{(m_r)}=\boldsymbol{\widetilde\theta}$. By the continuity of $M(\boldsymbol\theta)$, we have $\lim_{r \rightarrow\infty} M(\boldsymbol{\widehat\theta}^{(m_r)})=M(\boldsymbol{\widetilde\theta})$, where $M(\boldsymbol{\widehat\theta}^{(m_r)})=\boldsymbol{\widehat\theta}^{(m_r)}$ by the definition of $\mathcal{F}$. It follows that $\lim_{r \rightarrow\infty} \boldsymbol{\widehat\theta}^{(m_r)}=M(\boldsymbol{\widetilde\theta})$ and $M(\boldsymbol{\widetilde\theta})=\boldsymbol{\widetilde\theta}$. Thus, the set of stationary points $\mathcal{F}$ is closed. To show
    \begin{align*}
 \lim_{m\rightarrow\infty}\inf_{\boldsymbol\theta \in \mathcal{F}} ||\boldsymbol{\widehat\theta}^{(m)} - \boldsymbol\theta||_2=0,
   \end{align*}
we assume on the contrary there exist an $\varepsilon>0$ and a sequence $\boldsymbol{\widehat\theta}^{(m_r)}$ with $||\boldsymbol{\widehat\theta}^{(m_r)} - \boldsymbol\theta||_2 \geq \varepsilon$ for all $m_r$. By the compactness and taking a convergent subsequence, we have a cluster point outside of $\mathcal{F}$, which contradicts the definition of $\mathcal{F}$.

  % \textbf{Step 3}  {\it If all stationary points of $\ell(\boldsymbol\theta)$ are isolated, any sequence of $\boldsymbol{\widehat\theta}^{(m)}$ possesses a limit, and this limit is a stationary point of $\ell(\boldsymbol\theta)$.}The isolated condition implies there are only a finite number of stationary points in the compact super-level set $\{\boldsymbol\theta \in U: \ell(\boldsymbol\theta) \geq \ell(\boldsymbol\theta_0)\}$. An infinite number of stationary points would admit a convergent sequence whose limit contradicts the isolated condition. By results in Step 2, the set of cluster points  of $M(\boldsymbol\theta)$ is a subset of this finite set of stationary point. According to Lange \cite{r23}, such a set of cluster points is also connected. Because a finite set with more than one point is disconnected, there can only be one cluster point. Therefore, the iteration map $M(\boldsymbol\theta)$ of the MMSA algorithm possess a limit, and that limit is a stationary point of the $\ell(\boldsymbol\theta)$. By the convexity, such a stationary point is a maximum point.
  Finally, if the observed information matrix $-\triangledown^2 \ell(\boldsymbol\theta)$ is positive definite in the super-level set, any sequence of $\boldsymbol{\widehat\theta}^{(m)}$ possesses a limit, and this limit is a stationary point of $\ell(\boldsymbol\theta)$.

\section*{Acknowledgements}
The authors would like to thank Dr. Abhijit Naik for helpful discussion
and comments. The authors also thank Dr. Kirsten Herold at the UM-SPH Writing lab  for her helpful suggestions.

\bibliographystyle{unsrt}  


\begin{thebibliography}{1}

\bibitem{r1}
\textsc{Boyd, S.} and \textsc{Vandenberghe, L.} (2004). \textit{Convex Optimization},
Cambridge University Press, New York.



\bibitem{r2}
\textsc{Breiman, L.} 
(1999).
Prediction games and arcing algorithms. 
\textit{Neural Computation}
\textbf{11} 1493--1517.


\bibitem{r3}
\textsc{B\"{u}hlmann, P.} and \textsc{Yu, B.}
(2003).
Boosting with the $L2$ loss: Regression and classification. 
\textit{Journal of the American Statistical Association}
\textbf{98(462)} 324--339.

\bibitem{r4}
\textsc{B\"{u}hlmann, P.} and \textsc{Yu, B.}
(2006).
Boosting for high-dimensional linear models. 
\textit{Annals of Statistics}
\textbf{34} 559--583.



\bibitem{r5}
\textsc{Cox, D.R.} 
(1972).
Regression models and life tables (with Discussion). 
\textit{Journal of the Royal Statistical Society, Series B}
\textbf{34} 187--200.


\bibitem{r6}
\textsc{Dekker, F.W.} and \textsc{Mutsert, R.} and \textsc{Dijk, P.C}. and \textsc{Zoccali, C.} and \textsc{Jager, K.J.}
(2008).
Survival analysis: time-dependent effects and time-varying risk factors. 
\textit{Kidney International},
\textbf{74} 994--997.


\bibitem{r7}
\textsc{Englesbe, M.J.} and \textsc{Lee, J.S.} and \textsc{Lee, J.S.} and \textsc{He, K.} 
and \textsc{Fan, L.} 
and \textsc{Schaubel, D.E.} 
and \textsc{Schaubel, D.E.} 
 and \textsc{Sheetz,
  K.H.} and \textsc{Harbaugh, C.M.} and \textsc{Holcombe, S.A.} and \textsc{Campbell, D.A.} and \textsc{Sonnenday, C.J.} and \textsc{Wang, S.C.}
(2012).
Analytic morphomics, core muscle size, and surgical outcomes. \textit{Annals of Surgery}
\textbf{256(2)} 255--261.

\bibitem{r8}
\textsc{Friedman, J.} and 
\textsc{Hastie, T.} and 
\textsc{Tibshirani, R.}
(2010).
Regularization paths for generalized linear models via coordinate descent.
\textit{Journal of Statistical Software}
\textbf{33(1)} 1--22.

%\bibitem{r4}
%\textsc{Simon, N.} and \textsc{Friedman, J.} and \textsc{Hastie, T.} and \textsc{Tibshirani, R.} 
%(2011).
%Regularization paths for Cox's proportional hazards model via coordinate descent. 
%\textit{Journal of Statistical Software}
%\textbf{39(5)} 1--13.

\bibitem{r9}
\textsc{Freund, Y.} and \textsc{Schapire, R.}
(1996).
Experiments with a new boosting algorithm. 
\textit{Machine Learning: Proceedings of the Thirteenth International Conference, Morgan Kauffman, San Francisco},
\textbf{74} 148--156.

\bibitem{r10}
\textsc{Friedman, J.H.} and \textsc{Hastie, T.} and \textsc{Tibshirani, R.}
(2000).
Additive logistic regression: A statistical view of boosting
(with discussion). 
\textit{Annals of Statistics}
\textbf{28(2)} 337--407.

\bibitem{r11}
\textsc{Friedman, J.H.} 
(2001).
Greedy function approximation: A gradient boosting machine. 
\textit{Annals of Statistics}
\textbf{29(5)} 1189--1232.

\bibitem{r12}
\textsc{Friedman, J.H.} 
(2002).
Stochastic Gradient Boosting. 
\textit{Computational Statistics and Data Analysis}
\textbf{38(4)} 367--378.

\bibitem{r13}
\textsc{Frohnert, P.P.} and \textsc{Donadio, J.V.Jr} and \textsc{Velosa, J.A.} and \textsc{Holley, K.E.} and \textsc{Sterioff, S.} 
(1997).
The fate of
renal transplants in patients with IgA nephropathy. 
\textit{Clinical Transplant}
\textbf{11(2)} 127--133.

\bibitem{r14}
\textsc{Grambsch, P.}  and  \textsc{Therneau, T.} 
(1994).
Proportional hazards tests and diagnostics based on weighted residuals. 
\textit{Biometrika}
\textbf{81} 515--526.

\bibitem{r15}
\textsc{Gray, R.J.} 
(1992).
Flexible methods for analyzing survival data using splines, with applications to breast cancer prognosis. 
\textit{Journal of the American Statistical Association}
\textbf{87(420)} 942--951.

\bibitem{r16}
\textsc{Gray, R.J.} 
(1994).
Spline-based tests in survival analysis. 
\textit{Biometrics}
\textbf{50(3)} 640--652.


\bibitem{r17}
\textsc{Hadimeri, H.} and  \textsc{Norden, G.} and  \textsc{Friman, S.} and  \textsc{Nyberg,
G.} 
(1997).
Autosomal dominant polycystic kidney
disease in a kidney transplant population. 
\textit{Nephrol Dial Transplant}
\textbf{12} 1431--1436.

\bibitem{r18}
\textsc{Hastie, T.} and \textsc{Tibshirani, R.} 
(1993).
Varying-coefficient models. 
\textit{Journal of the Royal Statistical Society, Series B}
\textbf{55} 757--796.


\bibitem{r43}
\textsc{He, K.} and \textsc{Li, Y.M.} and \textsc{Wei, Q.Y.} and \textsc{Li, Y.} 
(2017).
Computationally efficient approach for modeling complex and big survival data.
\textit{Big and Complex Data Analysis: Statistical Methodologies and Applications, Edited volume by Springer}
 193-207.
 
 \bibitem{r44}
\textsc{He, K.} and \textsc{Yang, Y.} and \textsc{Li, Y.M.} and \textsc{Zhu, J.} and \textsc{Li, Y.} 
(2017).
Modeling time-varying effects with large-scale survival data: an efficient quasi-Newton approach.
\textit{Journal of Computational and Graphical Statistics}
\textbf{26(3)} 635-645.

 \bibitem{r45}
\textsc{He, K.} and \textsc{Li, Y.M.} and \textsc{Zhu, J.} and \textsc{Liu, H.L.} and \textsc{Lee, J.E.} and \textsc{Amos, C.I.} and \textsc{Hyslop, T.} and \textsc{Jin, J.S.} and \textsc{Lin, H.Z.} and \textsc{Wei, Q.Y.}  and \textsc{Li, Y.}
(2016).
Component-wise gradient boosting and false discovery control in survival analysis with high-dimensional covariates.
\textit{Bioinformatics}
\textbf{32(1)} 50-57.

\bibitem{r19}
\textsc{Hofner, B.} and \textsc{Hothorn, T.} and \textsc{Kneib, T.} 
(2008).
Variable selection and model choice in survival models with time-varying effects.
\textit{Technical Report}
Department of Statistics, University of Munich.

\bibitem{r20}
\textsc{Honda, T.} and \textsc{H\"{a}rdle, W.K.} 
(2014).
Variable selection in Cox regression models with varying coefficients. 
\textit{Journal of Statistical Planning and Inference}
\textbf{148} 67--81.

\bibitem{r21}
\textsc{Kalantar-Zadeh, K.} 
(2005).
Causes and consequences of the reverse epidemiology
of body mass index in dialysis patients. 
\textit{Journal of Renal Nutrition}
\textbf{15} 142--147.

%\bibitem{r22}
%\textsc{Kokado, Y.} and \textsc{Takahara, S.} and %\textsc{Kameoka, H.} 
%(1996).
%Hypertension in renal transplant recipients and its effect on long term allograft survival. 
%\textit{Transplant Proc}
%\textbf{28} 1600--1602.


\bibitem{r22}
\textsc{Kalbfleisch, J.D.} and \textsc{Wolfe, R.A.} 
(2013).
On Monitoring Outcomes of Medical Providers. 
\textit{Statistics in Biosciences}
\textbf{2} 286--302.


\bibitem{r23}
\textsc{Lange, K.} (2012). \textit{Optimization}, 2nd ed.
Springer, New York.

%\bibitem{r4}
%\textsc{Li, H.} and \textsc{Luan, Y.} 
%(2005).
%Boosting proportional hazards models using smoothing splines, with applications to high-dimensional microarray data. 
%\textit{Bioinformatics}
%\textbf{21(10)} 2403--2409.

\bibitem{r24}
\textsc{Liu, M.} and \textsc{Lu, W.} and \textsc{Shore, R. E.} and \textsc{Zeleniuch-Jacquotte, A.} 
(2010).
Cox regression model with time-varying coefficients in nested case-control studies. 
\textit{Biostatistics}
\textbf{11(4)} 693--706.

\bibitem{r25}
\textsc{Mason, L.} and \textsc{Baxter, J.} and \textsc{Bartlett, P.}r and \textsc{Frean, M.} 
(1999).
Boosting algorithms as gradient descent in function space. 
\textit{In Advances in Large Margin Classifiers. MIT Press}.


\bibitem{r26}
\textsc{Meier-Kriesche, H.U.} and \textsc{Port, F.K.} and \textsc{Ojo, A.O.} and \textsc{Rudich, S.M.} and \textsc{Hanson, J.A.} and \textsc{Cibrik, D.M.} and \textsc{Leichtman,
A.B.} and \textsc{Kaplan, B.} 
(2000).
Effect of waiting time on renal transplant outcome. 
\textit{Kidney International}
\textbf{58(3)} 1311--1317.

\bibitem{r27}
\textsc{Muntean, A.} and \textsc{Lucan, M.} 
(2013).
Immunosuppression in kidney transplantation. 
\textit{Clujul Medical}
\textbf{86} 177--180.

\bibitem{r28}
\textsc{Perperoglou, A.} and  \textsc{le Cessie, S.} and  \textsc{van Houwelingen, H.C.} 
(2006).
A fast routine for fitting Cox models with time
varying effects of the covariates. 
\textit{Computer Methods and Programs in Biomedicine}
\textbf{25} 154--161.

\bibitem{r29}
\textsc{Phelan, P.J.} and \textsc{Shields, W.} and \textsc{O'Kelly, P.} and  \textsc{Pendergrass, M.} and \textsc{Holian, J.} and
 \textsc{Walshe, J.} and  \textsc{Magee, C.} and  \textsc{Little, D.} and \textsc{Hickey, D.} and \textsc{Conlon, P.J.} 
(2009).
Left versus right deceased donor renal allograft outcome. 
\textit{Transplant International}
\textbf{25} 1159--1163.


\bibitem{r30}
\textsc{Ridgeway, G.} 
(1999).
The State of Boosting. 
\textit{Computing Science and Statistics}
\textbf{31} 172--181.


\bibitem{r31}
\textsc{Rozanski, J.} and \textsc{Kozlowska, I.} and \textsc{Myslak, M.} 
(2005).
Pretransplant nephrectomy in patients
with autosomal dominant polycystic kidney
disease. 
\textit{Transplant Proc}
\textbf{37} 666--668.

\bibitem{r32}
\textsc{Ruder, S.} 
(2016).
An overview of gradient descent optimization algorithm. 
\textit{arXiv preprint}

\bibitem{r33}
\textsc{Saran, R.} and  \textsc{Robinson, B.} and  \textsc{Abbott, K.C. et al.} 
(2018).
US Renal Data System 2017 Annual
Data Report: Epidemiology of Kidney Disease in the United States. 
\textit{American Journal of
Kidney Diseases}
\textbf{71(3)} S1--S672.

\bibitem{r34}
\textsc{Tian, L.} and \textsc{Zucker, D.} and  \textsc{Wei, L.} 
(2005).
On the Cox Model with Time-Varying Regression Coefficients. 
\textit{Journal of the American Statistical Association}
\textbf{100(469)} 72--183.


\bibitem{r35}
\textsc{Therneau, T.M.} and \textsc{Grambsch, P.M.} (2000). \textit{Modeling Survival Data: Extending the Cox Model},
Springer, New York.

\bibitem{r36}
\textsc{Verweij, P.J.M.} and \textsc{van Houwelingen, H.C.} 
(1993).
Cross-validation in survival analysis. 
\textit{Statistics in Medicine}
\textbf{12} 2305--2314.


\bibitem{r37}
\textsc{Wolfson, J.} 
(2011).
EEBoost: A general method for prediction and variable selection based on estimating equations. 
\textit{Journal of the American Statistical Association}
\textbf{106(493)} 296--305.


\bibitem{r38}
\textsc{Xiao, W.} and \textsc{Lu, W.} and \textsc{Zhang, H.H.} 
(2016).
Joint structure selection and estimation in the time-varying coefficient Cox model. 
\textit{Statistica Sinica}
\textbf{26(2)} 547--567.

\bibitem{r39}
\textsc{Yan, J.} and \textsc{Huang, J.} 
(2012).
Model selection for Cox models with time-varying coefficients. 
\textit{Biometrics}
\textbf{68(2)} 419--428.

\bibitem{r40}
\textsc{Zou, H.} 
(2006).
The adaptive lasso and its oracle properties.
\textit{Journal of the American Statistical Association}
\textbf{101(476)} 1418--1429.

\bibitem{r41}
\textsc{Zhang, H.} and \textsc{Lu, W.} 
(2006).
Adaptive lasso for Cox's proportional hazards model.
\textit{Biometrika}
\textbf{94} 691--703.

\bibitem{r42}
\textsc{Zucker, D.M.} and   \textsc{Karr, A.F.} 
(1990).
Nonparametric survival analysis with time-dependent covariate effects: a penalized partial likelihood approach. 
\textit{Annals of Statistics}
\textbf{18(1)} 329--353.


\end{thebibliography}

\end{document}